# Multi-messenger Observations of a Binary Neutron Star Merger

LIGO Scientific Collaboration and Virgo Collaboration, Fermi GBM, INTEGRAL, IceCube Collaboration, AstroSat Cadmium Zinc Telluride Imager Team, IPN Collaboration, The Insight-Hxmt Collaboration, ANTARES Collaboration, The Swift Collaboration, AGILE Team, The 1M2H Team, The Dark Energy Camera GW-EM Collaboration and the DES Collaboration, The DLT40 Collaboration, GRAWITA: GRAvitational Wave Inaf TeAm, The Fermi Large Area Telescope Collaboration, ATCA: Australia Telescope Compact Array, ASKAP: Australian SKA Pathfinder, Las Cumbres Observatory Group, OzGrav, DWF (Deeper, Wider, Faster Program), AST3, and CAASTRO Collaborations, The VINROUGE Collaboration, MASTER Collaboration, J-GEM, GROWTH, JAGWAR, Caltech-NRAO, TTU-NRAO, and NuSTAR Collaborations, Pan-STARRS, The MAXI Team, TZAC Consortium, KU Collaboration, Nordic Optical Telescope, ePESSTO, GROND, Texas Tech University, SALT Group, TOROS: Transient Robotic Observatory of the South Collaboration, The BOOTES Collaboration, MWA: Murchison Widefield Array, The CALET Collaboration, IKI-GW Follow-up Collaboration, H.E.S.S. Collaboration, LOFAR Collaboration, LWA: Long Wavelength Array, HAWC Collaboration, The Pierre Auger Collaboration, ALMA Collaboration, Euro VLBI Team, Pi of the Sky Collaboration, The Chandra Team at McGill University, DFN: Desert Fireball Network, ATLAS, High Time Resolution Universe Survey, RIMAS and RATIR, and SKA South Africa/MeerKAT
(See the end matter for the full list of authors.)

## Abstract

On 2017 August 17 a binary neutron star coalescence candidate (later designated GW170817) with merger time 12:41:04 UTC was observed through gravitational waves by the Advanced LIGO and Advanced Virgo detectors. The *Fermi* Gamma-ray Burst Monitor independently detected a gamma-ray burst (GRB 170817A) with a time delay of $\sim$1.7 s with respect to the merger time. From the gravitational-wave signal, the source was initially localized to a sky region of 31 deg$^2$ at a luminosity distance of $40^{+8}_{-8}$ Mpc and with component masses consistent with neutron stars. The component masses were later measured to be in the range 0.86 to 2.26 $M_\odot$. An extensive observing campaign was launched across the electromagnetic spectrum leading to the discovery of a bright optical transient (SSS17a, now with the IAU identification of AT 2017gfo) in NGC 4993 (at $\sim$40 Mpc) less than 11 hours after the merger by the One-Meter, Two Hemisphere (1M2H) team using the 1 m Swope Telescope. The optical transient was independently detected by multiple teams within an hour. Subsequent observations targeted the object and its environment. Early ultraviolet observations revealed a blue transient that faded within 48 hours. Optical and infrared observations showed a redward evolution over $\sim$10 days. Following early non-detections, X-ray and radio emission were discovered at the transient's position $\sim$9 and $\sim$16 days, respectively, after the merger. Both the X-ray and radio emission likely arise from a physical process that is distinct from the one that generates the UV/optical/near-infrared emission. No ultra-high-energy gamma-rays and no neutrino candidates consistent with the source were found in follow-up searches. These observations support the hypothesis that GW170817 was produced by the merger of two neutron stars in NGC 4993 followed by a short gamma-ray burst (GRB 170817A) and a kilonova/macronova powered by the radioactive decay of *r*-process nuclei synthesized in the ejecta.

*Key words:* gravitational waves – stars: neutron

## 1. Introduction

Over 80 years ago Baade & Zwicky (1934) proposed the idea of neutron stars, and soon after, Oppenheimer & Volkoff (1939) carried out the first calculations of neutron star models. Neutron stars entered the realm of observational astronomy in the 1960s by providing a physical interpretation of X-ray emission from Scorpius X-1 (Giacconi et al. 1962; Shklovsky 1967) and of radio pulsars (Gold 1968; Hewish et al. 1968; Gold 1969).

The discovery of a radio pulsar in a double neutron star system by Hulse & Taylor (1975) led to a renewed interest in binary stars and compact-object astrophysics, including the development of a scenario for the formation of double neutron stars and the first population studies (Flannery & van den Heuvel 1975; Massevitch et al. 1976; Clark 1979; Clark et al. 1979; Dewey & Cordes 1987; Lipunov et al. 1987; for reviews see Kalogera et al. 2007; Postnov & Yungelson 2014). The Hulse-Taylor pulsar provided the first firm evidence (Taylor & Weisberg 1982) of the existence of gravitational waves (Einstein 1916, 1918) and sparked a renaissance of observational tests of general relativity (Damour & Taylor 1991, 1992; Taylor et al. 1992; Wex 2014). Merging binary neutron stars (BNSs) were quickly recognized to be promising sources of detectable gravitational waves, making them a primary target for ground-based interferometric detectors (see Abadie et al. 2010 for an overview). This motivated the development of accurate models for the two-body, general-relativistic dynamics (Blanchet et al. 1995; Buonanno & Damour 1999; Pretorius 2005; Baker et al. 2006; Campanelli et al. 2006; Blanchet 2014) that are critical for detecting and interpreting gravitational waves (Abbott et al. 2016c, 2016d, 2016e, 2017a, 2017c, 2017d).

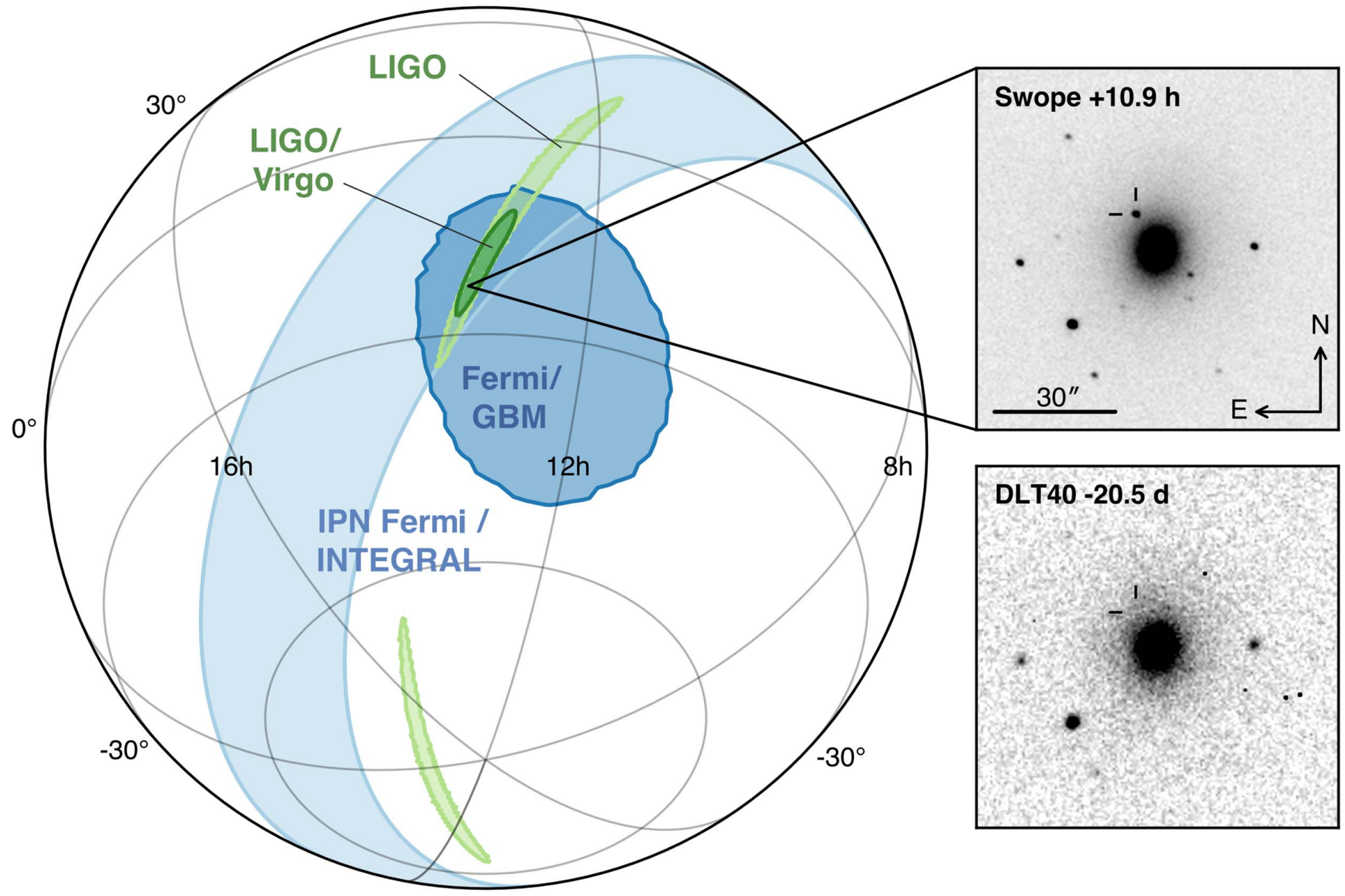

**Figure 1.** Localization of the gravitational-wave, gamma-ray, and optical signals. The left panel shows an orthographic projection of the 90% credible regions from LIGO (190 $deg^2$; light green), the initial LIGO-Virgo localization (31 $deg^2$; dark green), IPN triangulation from the time delay between *Fermi* and *INTEGRAL* (light blue), and *Fermi*-GBM (dark blue). The inset shows the location of the apparent host galaxy NGC 4993 in the Swope optical discovery image at 10.9 hr after the merger (top right) and the DLT40 pre-discovery image from 20.5 days prior to merger (bottom right). The reticle marks the position of the transient in both images.

In the mid-1960s, gamma-ray bursts (GRBs) were discovered by the Vela satellites, and their cosmic origin was first established by Klebesadel et al. (1973). GRBs are classified as *long* or *short*, based on their duration and spectral hardness (Dezalay et al. 1992; Kouveliotou et al. 1993). Uncovering the progenitors of GRBs has been one of the key challenges in high-energy astrophysics ever since (Lee & Ramirez-Ruiz 2007). It has long been suggested that short GRBs might be related to neutron star mergers (Goodman 1986; Paczynski 1986; Eichler et al. 1989; Narayan et al. 1992).

In 2005, the field of short gamma-ray burst (sGRB) studies experienced a breakthrough (for reviews see Nakar 2007; Berger 2014) with the identification of the first host galaxies of sGRBs and multi-wavelength observation (from X-ray to optical and radio) of their afterglows (Berger et al. 2005; Fox et al. 2005; Gehrels et al. 2005; Hjorth et al. 2005b; Villasenor et al. 2005). These observations provided strong hints that sGRBs might be associated with mergers of neutron stars with other neutron stars or with black holes. These hints included: (i) their association with both elliptical and star-forming galaxies (Barthelmy et al. 2005; Prochaska et al. 2006; Berger et al. 2007; Ofek et al. 2007; Troja et al. 2008; D'Avanzo et al. 2009; Fong et al. 2013), due to a very wide range of delay times, as predicted theoretically (Bagot et al. 1998; Fryer et al. 1999; Belczynski et al. 2002); (ii) a broad distribution of spatial offsets from host-galaxy centers (Berger 2010; Fong & Berger 2013; Tunnicliffe et al. 2014), which was predicted to arise from supernova kicks (Narayan et al. 1992; Bloom et al. 1999); and (iii) the absence of associated supernovae (Fox et al. 2005; Hjorth et al. 2005c, 2005a; Soderberg et al. 2006; Kocevski et al. 2010; Berger et al. 2013a). Despite these strong hints, proof that sGRBs were powered by neutron star mergers remained elusive, and interest intensified in following up gravitational-wave detections electromagnetically (Metzger & Berger 2012; Nissanke et al. 2013).

Evidence of beaming in some sGRBs was initially found by Soderberg et al. (2006) and Burrows et al. (2006) and confirmed by subsequent sGRB discoveries (see the compilation and analysis by Fong et al. 2015 and also Troja et al. 2016). Neutron star binary mergers are also expected, however, to produce isotropic electromagnetic signals, which include (i) early optical and infrared emission, a so-called kilonova/macronova (hereafter kilonova; Li & Paczyński 1998; Kulkarni 2005; Rosswog 2005; Metzger et al. 2010; Roberts et al. 2011; Barnes & Kasen 2013; Kasen et al. 2013; Tanaka & Hotokezaka 2013; Grossman et al. 2014; Barnes et al. 2016; Tanaka 2016; Metzger 2017) due to radioactive decay of rapid neutron-capture process (*r*-process) nuclei (Lattimer & Schramm 1974, 1976) synthesized in dynamical and accretion-disk-wind ejecta during the merger; and (ii) delayed radio emission from the interaction of the merger ejecta with the ambient medium (Nakar & Piran 2011; Piran et al. 2013; Hotokezaka & Piran 2015; Hotokezaka et al. 2016). The late-time infrared excess associated with GRB 130603B was interpreted as the signature of *r*-process nucleosynthesis (Berger et al. 2013b; Tanvir et al. 2013), and more candidates were identified later (for a compilation see Jin et al. 2016).

Here, we report on the global effort[958] that led to the first joint detection of gravitational and electromagnetic radiation from a single source. An $\sim 100$ s long gravitational-wave signal (GW170817) was followed by an sGRB (GRB 170817A) and an optical transient (SSS17a/AT 2017gfo) found in the host galaxy NGC 4993. The source was detected across the electromagnetic spectrum—in the X-ray, ultraviolet, optical, infrared, and radio bands—over hours, days, and weeks. These observations support the hypothesis that GW170817 was produced by the merger of two neutron stars in NGC4993, followed by an sGRB and a kilonova powered by the radioactive decay of *r*-process nuclei synthesized in the ejecta.

[958] A follow-up program established during initial LIGO-Virgo observations (Abadie et al. 2012) was greatly expanded in preparation for Advanced LIGO-Virgo observations. Partners have followed up binary black hole detections, starting with GW150914 (Abbott et al. 2016a), but have discovered no firm electromagnetic counterparts to those events.

## 2. A Multi-messenger Transient

On 2017 August 17 12:41:06 UTC the *Fermi* Gamma-ray Burst Monitor (GBM; Meegan et al. 2009) onboard flight software triggered on, classified, and localized a GRB. A Gamma-ray Coordinates Network (GCN) Notice (Fermi-GBM 2017) was issued at 12:41:20 UTC announcing the detection of the GRB, which was later designated GRB 170817A (von Kienlin et al. 2017). Approximately 6 minutes later, a gravitational-wave candidate (later designated GW170817) was registered in low latency (Cannon et al. 2012; Messick et al. 2017) based on a single-detector analysis of the Laser Interferometer Gravitational-wave Observatory (LIGO) Hanford data. The signal was consistent with a BNS coalescence with merger time, $t_c$, 12:41:04 UTC, less than 2 s before GRB 170817A. A GCN Notice was issued at 13:08:16 UTC. Single-detector gravitational-wave triggers had never been disseminated before in low latency. Given the temporal coincidence with the *Fermi*-GBM GRB, however, a GCN Circular was issued at 13:21:42 UTC (LIGO Scientific Collaboration & Virgo Collaboration et al. 2017a) reporting that a highly significant candidate event consistent with a BNS coalescence was associated with the time of the GRB[959]. An extensive observing campaign was launched across the electromagnetic spectrum in response to the *Fermi*-GBM and LIGO–Virgo detections, and especially the subsequent well-constrained, three-dimensional LIGO–Virgo localization. A bright optical transient (SSS17a, now with the IAU identification of AT 2017gfo) was discovered in NGC 4993 (at ∼40 Mpc) by the 1M2H team (August 18 01:05 UTC; Coulter et al. 2017a) less than 11 hr after the merger.

### 2.1. Gravitational-wave Observation

GW170817 was first detected online (Cannon et al. 2012; Messick et al. 2017) as a single-detector trigger and disseminated through a GCN Notice at 13:08:16 UTC and a GCN Circular at 13:21:42 UTC (LIGO Scientific Collaboration & Virgo Collaboration et al. 2017a). A rapid re-analysis (Nitz et al. 2017a, 2017b) of data from LIGO-Hanford, LIGO-Livingston, and Virgo confirmed a highly significant, coincident signal. These data were then combined to produce the first three-instrument skymap (Singer & Price 2016; Singer et al. 2016) at 17:54:51 UTC (LIGO Scientific Collaboration & Virgo Collaboration et al. 2017b), placing the source nearby, at a luminosity distance *initially* estimated to be $40^{+8}_{-8}$ Mpc in an elongated region of $\approx 31$ deg$^2$ (90% credibility), centered around R.A. $\alpha(\mathrm{J2000.0}) = 12^{\mathrm{h}}57^{\mathrm{m}}$ and decl. $\delta(\mathrm{J2000.0}) = -17°51'$. Soon after, a coherent analysis (Veitch et al. 2015) of the data from the detector network produced a skymap that was distributed at 23:54:40 UTC (LIGO Scientific Collaboration & Virgo Collaboration et al. 2017c), consistent with the initial one: a $\simeq 34$ deg$^2$ sky region at 90% credibility centered around $\alpha(\mathrm{J2000.0}) = 13^{\mathrm{h}}09^{\mathrm{m}}$ and $\delta(\mathrm{J2000.0}) = -25°37'$.

The offline gravitational-wave analysis of the LIGO-Hanford and LIGO-Livingston data identified GW170817 with a false-alarm rate of less than one per $8.0 \times 10^4$ (Abbott et al. 2017c). This analysis uses post-Newtonian waveform models (Blanchet et al. 1995, 2004, 2006; Bohé et al. 2013) to construct a matched-filter search (Sathyaprakash & Dhurandhar 1991; Cutler et al. 1993; Allen et al. 2012) for gravitational waves from the coalescence of compact-object binary systems in the (detector frame) total mass range 2–500 $M_\odot$. GW170817 lasted for ∼100 s in the detector sensitivity band. The signal reached Virgo first, then LIGO-Livingston 22 ms later, and after 3 ms more, it arrived at LIGO-Hanford. GW170817 was detected with a combined signal-to-noise ratio across the three-instrument network of 32.4. For comparison, GW150914 was observed with a signal-to-noise ratio of 24 (Abbott et al. 2016c).

The properties of the source that generated GW170817 (see Abbott et al. 2017c for full details; here, we report parameter ranges that span the 90% credible interval) were derived by employing a coherent Bayesian analysis (Veitch et al. 2015; Abbott et al. 2016b) of the three-instrument data, including marginalization over calibration uncertainties and assuming that the signal is described by waveform models of a binary system of compact objects in quasi-circular orbits (see Abbott et al. 2017c and references therein). The waveform models include the effects introduced by the objects' intrinsic rotation (spin) and tides. The source is located in a region of 28 deg$^2$ at a distance of $40^{+8}_{-14}$ Mpc, see Figure 1, consistent with the early estimates disseminated through GCN Circulars (LIGO Scientific Collaboration & Virgo Collaboration et al. 2017b, 2017c). The misalignment between the total angular momentum axis and the line of sight is $\leqslant 56°$.

The (source-frame[960]) masses of the primary and secondary components, $m_1$ and $m_2$, respectively, are in the range $m_1 \in (1.36\text{–}2.26)M_\odot$ and $m_2 \in (0.86\text{–}1.36)M_\odot$. The chirp mass,[961] $\mathcal{M}$, is the mass parameter that, at the leading order, drives the frequency evolution of gravitational radiation in the inspiral phase. This dominates the portion of GW170817 in the instruments' sensitivity band. As a consequence, it is the best measured mass parameter, $\mathcal{M} = 1.188^{+0.004}_{-0.002}\,M_\odot$. The total mass is $2.82^{+0.47}_{-0.09}M_\odot$, and the mass ratio $m_2/m_1$ is bound to the range 0.4–1.0. These results are consistent with a binary whose components are neutron stars. White dwarfs are ruled out since the gravitational-wave signal sweeps through 200 Hz in the instruments' sensitivity band, implying an orbit of size ∼100 km, which is smaller than the typical radius of a white dwarf by an order of magnitude (Shapiro & Teukolsky 1983). However, for this event gravitational-wave data *alone* cannot rule out objects more compact than neutron stars such as quark stars or black holes (Abbott et al. 2017c).

### 2.2. Prompt Gamma-Ray Burst Detection

The first announcement of GRB 170817A came from the GCN Notice (Fermi-GBM 2017) automatically generated by *Fermi*-GBM at 12:41:20 UTC, just 14 s after the detection of the GRB at T0 = 12:41:06 UTC. GRB 170817A was detected by the *International Gamma-Ray Astrophysics Laboratory* (*INTEGRAL*) spacecraft using the Anti-Coincidence Shield (von Kienlin et al. 2003) of the spectrometer on board *INTEGRAL* (SPI), through an offline search initiated by the LIGO-Virgo and *Fermi*-GBM reports. The final *Fermi*-GBM localization constrained GRB 170817A to a region with highest probability at $\alpha(\mathrm{J2000.0}) = 12^{\mathrm{h}}28^{\mathrm{m}}$ and $\delta(\mathrm{J2000.0}) = -30°$ and 90% probability region covering ∼1100 deg$^2$ (Goldstein et al. 2017a). The difference between the binary merger and the GRB is $T0 - t_c = 1.734 \pm 0.054$ s (Abbott et al. 2017g). Exploiting the difference in the arrival time of the gamma-ray signals at *Fermi*-GBM and *INTEGRAL* SPI-ACS (Svinkin et al. 2017c) provides additional significant constraints on the gamma-ray localization area (see Figure 1). The IPN localization capability will be especially important in the case of future gravitational-wave events that might be less well-localized by LIGO-Virgo.

[959] The trigger was recorded with LIGO-Virgo ID G298048, by which it is referred throughout the GCN Circulars.

[960] Any mass parameter $m^{(\mathrm{det})}$ derived from the observed signal is measured in the detector frame. It is related to the mass parameter, $m$, in the source frame by $m^{(\mathrm{det})} = (1 + z)m$, where $z$ is the source's redshift. Here, we always report source-frame mass parameters, assuming standard cosmology (Ade et al. 2016) and correcting for the motion of the solar Ssystem barycenter with respect to the cosmic microwave background (Fixsen 2009). From the gravitational-wave luminosity distance measurement, the redshift is determined to be $z = 0.008^{+0.002}_{-0.003}$. For full details see Abbott et al. (2016b, 2017c, 2017e).

[961] The binary's chirp mass is defined as $\mathcal{M} = (m_1 m_2)^{3/5}/(m_1 + m_2)^{1/5}$.

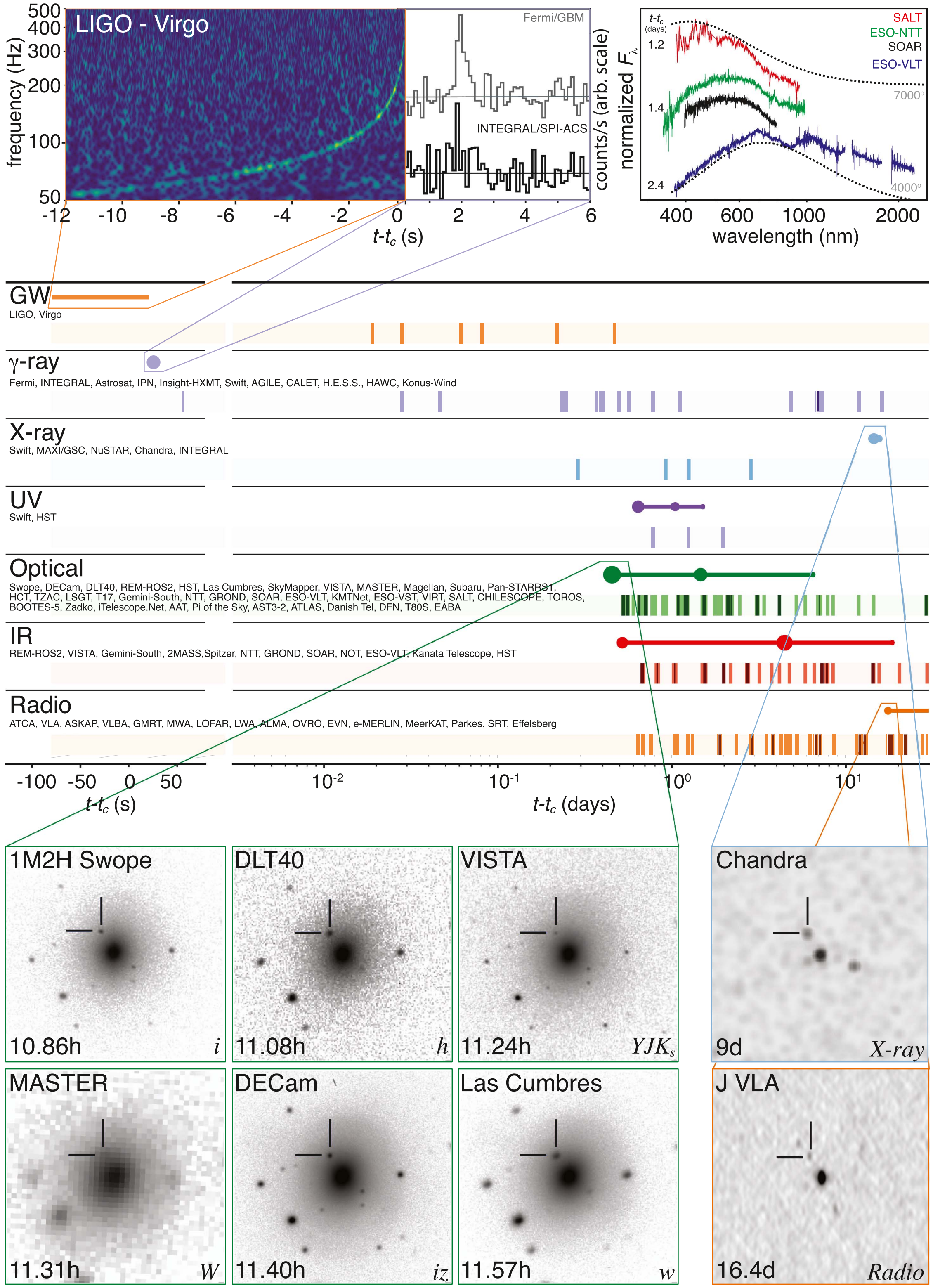

**Figure 2.** Timeline of the discovery of GW170817, GRB 170817A, SSS17a/AT 2017gfo, and the follow-up observations are shown by messenger and wavelength relative to the time $t_c$ of the gravitational-wave event. Two types of information are shown for each band/messenger. First, the shaded dashes represent the times when information was reported in a GCN Circular. The names of the relevant instruments, facilities, or observing teams are collected at the beginning of the row. Second, representative observations (see Table 1) in each band are shown as solid circles with their areas approximately scaled by brightness; the solid lines indicate when the source was detectable by at least one telescope. Magnification insets give a picture of the first detections in the gravitational-wave, gamma-ray, optical, X-ray, and radio bands. They are respectively illustrated by the combined spectrogram of the signals received by LIGO-Hanford and LIGO-Livingston (see Section 2.1), the *Fermi*-GBM and *INTEGRAL*/SPI-ACS lightcurves matched in time resolution and phase (see Section 2.2), $1\overset{'}{.}5 \times 1\overset{'}{.}5$ postage stamps extracted from the initial six observations of SSS17a/AT 2017gfo and four early spectra taken with the SALT (at $t_c + 1.2$ days; Buckley et al. 2017; McCully et al. 2017b), ESO-NTT (at $t_c + 1.4$ days; Smartt et al. 2017), the SOAR 4 m telescope (at $t_c + 1.4$ days; Nicholl et al. 2017d), and ESO-VLT-XShooter (at $t_c + 2.4$ days; Smartt et al. 2017) as described in Section 2.3, and the first X-ray and radio detections of the same source by *Chandra* (see Section 3.3) and JVLA (see Section 3.4). In order to show representative spectral energy distributions, each spectrum is normalized to its maximum and shifted arbitrarily along the linear *y*-axis (no absolute scale). The high background in the SALT spectrum below 4500 Å prevents the identification of spectral features in this band (for details McCully et al. 2017b).

Standard follow-up analyses (Goldstein et al. 2012; Paciesas et al. 2012; Gruber et al. 2014) of the *Fermi*-GBM trigger determined the burst duration to be $T_{90} = 2.0 \pm 0.5$ s, where $T_{90}$ is defined as the interval over which 90% of the burst fluence is accumulated in the energy range of 50–300 keV. From the *Fermi*-GBM $T_{90}$ measurement, GRB 170817A was classified as an sGRB with 3:1 odds over being a long GRB. The classification of GRB 170817A as an sGRB is further supported by incorporating the hardness ratio of the burst and comparing it to the *Fermi*-GBM catalog (Goldstein et al. 2017a). The SPI-ACS duration for GRB 170817A of 100 ms is consistent with an sGRB classification within the instrument's historic sample (Savchenko et al. 2012).

The GRB had a peak photon flux measured on a 64 ms timescale of $3.7 \pm 0.9$ photons $s^{-1}$ $cm^{-2}$ and a fluence over the $T_{90}$ interval of $(2.8 \pm 0.2) \times 10^{-7}$ erg $cm^{-2}$ (10–1000 keV; (Goldstein et al. 2017a). GRB 170817A is the closest sGRB with measured redshift. By usual measures, GRB 170817A is sub-luminous, a tantalizing observational result that is explored in Abbott et al. (2017g) and Goldstein et al. (2017a).

Detailed analysis of the *Fermi*-GBM data for GRB 170817A revealed two components to the burst: a main pulse encompassing the GRB trigger time from $T0 - 0.320$ s to $T0 + 0.256$ s followed by a weak tail starting at $T0 + 0.832$ s and extending to $T0 + 1.984$ s. The spectrum of the main pulse of GRB 170817A is best fit with a Comptonized function (a power law with an exponential cutoff) with a power-law photon index of $-0.62 \pm 0.40$, peak energy $E_{\rm peak} = 185 \pm 62$ keV, and time-averaged flux of $(3.1 \pm 0.7) \times 10^{-7}$ erg $cm^{-2}$ $s^{-1}$. The weak tail that follows the main pulse, when analyzed independently, has a localization consistent with both the main pulse and the gravitational-wave position. The weak tail, at 34% the fluence of the main pulse, extends the $T_{90}$ beyond the main pulse and has a softer, blackbody spectrum with $kT = 10.3 \pm 1.5$ keV (Goldstein et al. 2017a).

Using the *Fermi*-GBM spectral parameters of the main peak and $T_{90}$ interval, the integrated fluence measured by *INTEGRAL* SPI-ACS is $(1.4 \pm 0.4) \times 10^{-7}$ erg $cm^{-2}$ (75–2000 keV), compatible with the *Fermi*-GBM spectrum. Because SPI-ACS is most sensitive above 100 keV, it detects only the highest-energy part of the main peak near the start of the longer *Fermi*-GBM signal (Abbott et al. 2017f).

### 2.3. Discovery of the Optical Counterpart and Host Galaxy

The announcements of the *Fermi*-GBM and LIGO-Virgo detections, and especially the well-constrained, three-dimensional LIGO-Virgo localization, triggered a broadband observing campaign in search of electromagnetic counterparts. A large number of teams across the world were mobilized using ground- and space-based telescopes that could observe the region identified by the gravitational-wave detection. GW170817 was localized to the southern sky, setting in the early evening for the northern hemisphere telescopes, thus making it inaccessible to the majority of them. The LIGO-Virgo localization region (LIGO Scientific Collaboration & Virgo Collaboration et al. 2017b, 2017c) became observable to telescopes in Chile about 10 hr after the merger with an altitude above the horizon of about 45°.

The One-Meter, Two-Hemisphere (1M2H) team was the first to discover and announce (August 18 01:05 UTC; Coulter et al. 2017a) a bright optical transient in an *i*-band image acquired on August 17 at 23:33 UTC ($t_c + 10.87$ hr) with the 1 m Swope telescope at Las Campanas Observatory in Chile. The team used an observing strategy (Gehrels et al. 2016) that targeted known galaxies (from White et al. 2011b) in the three-dimensional LIGO-Virgo localization taking into account the galaxy stellar mass and star formation rate (Coulter et al. 2017). The transient, designated Swope Supernova Survey 2017a (SSS17a), was $i = 17.057 \pm 0.018$ mag[962] (August 17 23:33 UTC, $t_c + 10.87$ hr) and did not match any known asteroid or supernova. SSS17a (now with the IAU designation AT 2017gfo) was located at $\alpha$(J2000.0) = $13^{\rm h}09^{\rm m}48\overset{\rm s}{.}085 \pm 0.018$, $\delta$(J2000.0) = $-23°22'53\overset{''}{.}343 \pm 0.218$ at a projected distance of $10\overset{''}{.}6$ from the center of NGC 4993, an early-type galaxy in the ESO 508 group at a distance of $\simeq$40 Mpc (Tully–Fisher distance from Freedman et al. 2001), consistent with the gravitational-wave luminosity distance (LIGO Scientific Collaboration & Virgo Collaboration et al. 2017b).

Five other teams took images of the transient within an hour of the 1M2H image (and before the SSS17a announcement) using different observational strategies to search the LIGO-Virgo sky localization region. They reported their discovery of the same optical transient in a sequence of GCNs: the Dark Energy Camera (01:15 UTC; Allam et al. 2017), the Distance Less Than 40 Mpc survey (01:41 UTC; Yang et al. 2017a), Las Cumbres Observatory (LCO; 04:07 UTC; Arcavi et al. 2017a), the Visible and Infrared Survey Telescope for Astronomy (VISTA; 05:04 UTC; Tanvir et al. 2017a), and MASTER (05:38 UTC; Lipunov et al. 2017d). Independent searches were also carried out by the Rapid Eye Mount (REM-GRAWITA, optical, 02:00 UTC; Melandri et al. 2017a), *Swift* UVOT/XRT (utraviolet, 07:24 UTC; Evans et al. 2017a), and Gemini-South (infrared, 08:00 UT; Singer et al. 2017a).

The Distance Less Than 40 Mpc survey (DLT40; L. Tartaglia et al. 2017, in preparation) team independently detected SSS17a/AT 2017gfo, automatically designated DLT17ck (Yang et al. 2017a) in an image taken on August 17 23:50 UTC while carrying out high-priority observations of 51 galaxies (20 within the LIGO-Virgo localization and 31 within the wider *Fermi*-GBM localization region; Valenti et al. 2017, accepted). A confirmation image was taken on August 18 00:41 UTC after the observing program had cycled through all of the high-priority targets and found no other transients. The updated magnitudes for these two epochs are $r = 17.18 \pm 0.03$ and $17.28 \pm 0.04$ mag, respectively.

SSS17a/AT 2017gfo was also observed by the VISTA in the second of two 1.5 $deg^2$ fields targeted. The fields were chosen to be within the high-likelihood localization region of GW170817 and to contain a high density of potential host galaxies (32 of the 54 entries in the list of Cook et al. 2017a). Observations began during evening twilight and were repeated twice to give a short temporal baseline over which to search for variability (or proper motion of any candidates). The magnitudes of the transient source in the earliest images taken in the near-infrared were measured to be $K_s = 18.63 \pm 0.05$, $J = 17.88 \pm 0.03$, and $Y = 17.51 \pm 0.02$ mag.

[962] All apparent magnitudes are AB and corrected for the Galactic extinction in the direction of SSS17a ($E(B - V) = 0.109$ mag; Schlafly & Finkbeiner 2011).

On August 17 23:59 UTC, the MASTER-OAFA robotic telescope (Lipunov et al. 2010), covering the sky location of GW170817, recorded an image that included NGC 4993. The autodetection software identified MASTER OT J130948.10-232253.3, the bright optical transient with the unfiltered magnitude $W = 17.5 \pm 0.2$ mag, as part of an automated search performed by the MASTER Global Robotic Net (Lipunov et al. 2017a, 2017d).

The Dark Energy Camera (DECam; Flaugher et al. 2015) Survey team started observations of the GW170817 localization region on August 17 23:13 UTC. DECam covered 95% of the probability in the GW170817 localization area with a sensitivity sufficient to detect a source up to 100 times fainter than the observed optical transient. The transient was observed on 2017 August 18 at 00:05 UTC and independently detected at 00:42 UTC (Allam et al. 2017). The measured magnitudes of the transient source in the first images were $i = 17.30 \pm 0.02$, $z = 17.45 \pm 0.03$. A complete analysis of DECam data is presented in Soares-Santos et al. (2017).

Las Cumbres Observatory (LCO; Brown et al. 2013) surveys started their observations of individual galaxies with their global network of 1 and 2 m telescopes upon receipt of the initial *Fermi*-GBM localization. Approximately five hours later, when the LIGO-Virgo localization map was issued, the observations were switched to a prioritized list of galaxies (from Dalya et al. 2016) ranked by distance and luminosity (Arcavi et al. 2017, in preparation). In a 300 s $w$-band exposure beginning on August 18 00:15 UTC, a new transient, corresponding to AT 2017gfo/SSS17a/DLT17ck, was detected near NGC 4993 (Arcavi et al. 2017a). The transient was determined to have $w = 17.49 \pm 0.04$ mag (Arcavi et al. 2017e).

These early photometric measurements, from the optical to near-infrared, gave the first broadband spectral energy distribution of AT 2017gfo/SSS17a/DL17ck. They do not distinguish the transient from a young supernova, but they serve as reference values for subsequent observations that reveal the nature of the optical counterpart as described in Section 3.1. Images from the six earliest observations are shown in the inset of Figure 2.

## 3. Broadband Follow-up

While some of the first observations aimed to tile the error region of the GW170817 and GRB 170817A localization areas, including the use of galaxy targeting (White et al. 2011a; Dalya et al. 2016; D. Cook & M. Kasliwal 2017, in preparation; S. R. Kulkarni et al. 2017, in preparation), most groups focused their effort on the optical transient reported by Coulter et al. (2017) to define its nature and to rule out that it was a chance coincidence of an unrelated transient. The multi-wavelength evolution within the first 12–24 hr, and the subsequent discoveries of the X-ray and radio counterparts, proved key to scientific interpretation. This section summarizes the plethora of key observations that occurred in different wavebands, as well as searches for neutrino counterparts.

### *3.1. Ultraviolet, Optical, and Infrared*

The quick discovery in the first few hours of Chilean darkness, and the possibility of fast evolution, prompted the need for the ultraviolet–optical–infrared follow-up community to have access to both space-based and longitudinally separated ground-based facilities. Over the next two weeks, a network of ground-based telescopes, from 40 cm to 10 m, and space-based observatories spanning the ultraviolet (UV), optical (O), and near-infrared (IR) wavelengths followed up GW170817. These observations revealed an exceptional electromagnetic counterpart through careful monitoring of its spectral energy distribution. Here, we first consider photometric and then spectroscopic observations of the source.

Regarding photometric observations, at $t_c + 11.6$ hr, the *Magellan*-Clay and *Magellan*-Baade telescopes (Drout et al. 2017a; Simon et al. 2017) initiated follow-up observations of the transient discovered by the Swope Supernova Survey from the optical (g band) to NIR (Ks band). At $t_c + 12.7$ hr and $t_c + 12.8$ hr, the Rapid Eye Mount (REM)/ROS2 (Melandri et al. 2017b) detected the optical transient and the Gemini-South FLAMINGO2 instrument first detected near-infrared Ks-band emission constraining the early optical to infrared color (Kasliwal et al. 2017; Singer et al. 2017a), respectively. At $t_c + 15.3$ hr, the *Swift* satellite (Gehrels 2004) detected bright, ultraviolet emission, further constraining the effective temperature (Evans et al. 2017a, 2017b). The ultraviolet evolution continued to be monitored with the *Swift* satellite (Evans et al. 2017b) and the *Hubble Space Telescope* (*HST*; Adams et al. 2017; Cowperthwaite et al. 2017b; Kasliwal et al. 2017).

Over the course of the next two days, an extensive photometric campaign showed a rapid dimming of this initial UV–blue emission and an unusual brightening of the near-infrared emission. After roughly a week, the redder optical and near-infrared bands began to fade as well. Ground- and space-based facilities participating in this photometric monitoring effort include (in alphabetic order): CTIO1.3 m, DECam (Cowperthwaite et al. 2017b; Nicholl et al. 2017a, 2017d), IRSF, the Gemini-South FLAMINGO2 (Singer et al. 2017a, 2017b; Chornock et al. 2017b; Troja et al. 2017b, 2017d), Gemini-South GMOS (Troja et al. 2017b), GROND (Chen et al. 2017; Wiseman et al. 2017), *HST* (Cowperthwaite et al. 2017b; Levan & Tanvir 2017; Levan et al. 2017a; Tanvir & Levan 2017; Troja et al. 2017a), iTelescope.Net telescopes (Im et al. 2017a, 2017b), the Korea Microlensing Telescope Network (KMTNet; Im et al. 2017c, 2017d), LCO (Arcavi et al. 2017b, 2017c, 2017e), the Lee Sang Gak Telescope (LSGT)/SNUCAM-II, the *Magellan*-Baade and *Magellan*-Clay 6.5 m telescopes (Drout et al. 2017a; Simon et al. 2017), the Nordic Optical Telescope (Malesani et al. 2017a), Pan-STARRS1 (Chambers et al. 2017a, 2017b, 2017c, 2017d), REM/ROS2 and REM/REMIR (Melandri et al. 2017a, 2017c), SkyMapper (Wolf et al. 2017), Subaru Hyper Suprime-Cam (Yoshida et al. 2017a, 2017b, 2017c, 2017d; Tominaga et al. 2017), ESO-VISTA (Tanvir et al. 2017a), ESO-VST/OmegaCAM (Grado et al. 2017a, 2017b), and ESO-VLT/FORS2 (D'Avanzo et al. 2017).

One of the key properties of the transient that alerted the worldwide community to its unusual nature was the rapid luminosity decline. In bluer optical bands (i.e., in the $g$ band), the transient showed a fast decay between daily photometric measurements (Cowperthwaite et al. 2017b; Melandri et al. 2017c). Pan-STARRS (Chambers et al. 2017c) reported photometric measurements in the optical/infrared *izy* bands with the same cadence, showing fading by 0.6 mag per day, with reliable photometry from difference imaging using already existing sky images (Chambers et al. 2016; Cowperthwaite et al. 2017b). Observations taken every 8 hr by LCO showed an initial rise in the *w* band, followed by rapid fading in all optical bands (more than 1 mag per day in the blue) and reddening with time (Arcavi et al. 2017e). Accurate measurements from Subaru (Tominaga et al. 2017), LSGT/SNUCAM-II and KMTNet (Im et al. 2017c), ESO-VLT/FORS2 (D'Avanzo et al. 2017), and DECam (Cowperthwaite et al. 2017b; Nicholl et al. 2017b) indicated a similar rate of fading. On the contrary, the near-infrared monitoring reports by GROND and Gemini-South showed that the source faded more slowly in the infrared (Chornock et al. 2017b; Wiseman et al. 2017) and even showed a late-time plateau in the Ks band (Singer et al. 2017b). This evolution was recognized by the community as quite unprecedented for transients in the nearby (within 100 Mpc) universe (e.g., Siebert et al. 2017).

Table 1 reports a summary of the imaging observations, which include coverage of the entire gravitational-wave sky localization and follow-up of SSS17a/AT 2017gfo. Figure 2 shows these observations in graphical form.

Concerning spectroscopic observations, immediately after discovery of SSS17a/AT 2017gfo on the Swope 1 m telescope, the same team obtained the first spectroscopic observations of the optical transient with the LDSS-3 spectrograph on the 6.5 m *Magellan*-Clay telescope and the MagE spectrograph on the 6.5 m *Magellan*-Baade telescope at Las Campanas Observatory. The spectra, just 30 minutes after the first image, showed a blue and featureless continuum between 4000 and 10000 Å, consistent with a power law (Drout et al. 2017a; Shappee et al. 2017). The lack of features and blue continuum during the first few hours implied an unusual, but not unprecedented transient since such characteristics are common in cataclysmic–variable stars and young core-collapse supernovae (see, e.g., Li et al. 2011a, 2011b).

The next 24 hr of observation were critical in decreasing the likelihood of a chance coincidence between SSS17a/AT 2017gfo, GW170817, and GRB 170817A. The SALT-RSS spectrograph in South Africa (Buckley et al. 2017; McCully et al. 2017b; Shara et al. 2017), ePESSTO with the EFOSC2 instrument in spectroscopic mode at the ESO New Technology Telescope (NTT, in La Silla, Chile; Lyman et al. 2017), the X-shooter spectrograph on the ESO Very Large Telescope (Pian et al. 2017b) in Paranal, and the Goodman Spectrograph on the 4 m SOAR telescope (Nicholl et al. 2017c) obtained additional spectra. These groups reported a rapid fall off in the blue spectrum without any individual features identifiable with line absorption common in supernova-like transients (see, e.g., Lyman et al. 2017). This ruled out a young supernova of any type in NGC 4993, showing an exceptionally fast spectral evolution (Drout et al. 2017; Nicholl et al. 2017d). Figure 2 shows some representative early spectra (SALT spectrum is from Buckley et al. 2017; McCully et al. 2017b; ESO spectra from Smartt et al. 2017; SOAR spectrum from Nicholl et al. 2017d). These show rapid cooling, and the lack of commonly observed ions from elements abundant in supernova ejecta, indicating this object was unprecedented in its optical and near-infrared emission. Combined with the rapid fading, this was broadly indicative of a possible kilonova (e.g., Arcavi et al. 2017e; Cowperthwaite et al. 2017b; McCully et al. 2017b; Kasen et al. 2017; Kasliwal et al. 2017; Nicholl et al. 2017d; Smartt et al. 2017). This was confirmed by spectra taken at later times, such as with the Gemini Multi-Object Spectrograph (GMOS; Kasliwal et al. 2017; McCully et al. 2017b; Troja et al. 2017a, 2017b), the LDSS-3 spectrograph on the 6.5 m *Magellan*-Clay telescope at Las Campanas Observatory (Drout et al. 2017; Shappee et al. 2017), the LCO FLOYDS spectrograph at Faulkes Telescope South (McCully et al. 2017a, 2017b), and the AAOmega spectrograph on the 3.9 m Anglo-Australian Telescope (Andreoni et al. 2017), which did not show any significant emission or absorption lines over the red featureless continuum. The optical and near-infrared spectra over these few days provided convincing arguments that this transient was unlike any other discovered in extensive optical wide-field surveys over the past decade (see, e.g., Siebert et al. 2017).

The evolution of the spectral energy distribution, rapid fading, and emergence of broad spectral features indicated that the source had physical properties similar to models of kilonovae (e.g., Metzger et al. 2010; Kasen et al. 2013; Barnes & Kasen 2013; Tanaka & Hotokezaka 2013; Grossman et al. 2014; Metzger & Fernández 2014; Barnes et al. 2016; Tanaka 2016; Kasen et al. 2017; Metzger 2017). These show a very rapid shift of the spectral energy distribution from the optical to the near-infrared. The FLAMINGOS2 near-infrared spectrograph at Gemini-South (Chornock et al. 2017c; Kasliwal et al. 2017) shows the emergence of very broad features in qualitative agreement with kilonova models. The ESO-VLT/X-shooter spectra, which simultaneously cover the wavelength range 3200–24800 Å, were taken over 2 weeks with a close to daily sampling (Pian et al. 2017a; Smartt et al. 2017) and revealed signatures of the radioactive decay of *r*-process nucleosynthesis elements (Pian et al. 2017a). Three epochs of infrared grism spectroscopy with the *HST* (Cowperthwaite et al. 2017b; Levan & Tanvir 2017; Levan et al. 2017a; Tanvir & Levan 2017; Troja et al. 2017a)[963] identified features consistent with the production of lanthanides within the ejecta (Levan & Tanvir 2017; Tanvir & Levan 2017; Troja et al. 2017a).

The optical follow-up campaign also includes linear polarimetry measurements of SSS17a/AT 2017gfo by ESO-VLT/FORS2, showing no evidence of an asymmetric geometry of the emitting region and lanthanide-rich late kilonova emission (Covino et al. 2017). In addition, the study of the galaxy with the MUSE Integral Field Spectrograph on the ESO-VLT (Levan et al. 2017b) provides simultaneous spectra of the counterpart and the host galaxy, which show broad absorption features in the transient spectrum, combined with emission lines from the spiral arms of the host galaxy (Levan & Tanvir 2017; Tanvir & Levan 2017).

Table 2 reports the spectroscopic observations that have led to the conclusion that the source broadly matches kilonovae theoretical predictions.

### 3.2. *Gamma-Rays*

The fleet of ground- and space-based gamma-ray observatories provided broad temporal and spectral coverage of the source location. Observations spanned ∼10 orders of magnitude in energy and covered the position of SSS17a/AT 2017gfo from a few hundred seconds before the GRB 170817A trigger time (T0) to days afterward. Table 3 lists, in chronological order, the results reporting observation

[963] *HST* Program GO 14804 Levan, GO 14771 Tanvir, and GO 14850 Troja.

**Table 1**
A Partial Summary of Photometric Observations up to 2017 September 5 UTC with at Most Three Observations per Filter per Telescope/Group, i.e., the Earliest, the Peak, and the Latest in Each Case

| Telescope/Instrument | UT Date | Band | References |
|---|---|---|---|
| DFN/– | 2017 Aug 17 12:41:04 | visible | Hancock et al. (2017), |
| MASTER/– | 2017 Aug 17 17:06:47 | Clear | Lipunov et al. (2017a, 2017b) |
| PioftheSky/PioftheSkyNorth | 2017 Aug 17 21:46:28 | visible wide band | Cwiek et al. (2017); Batsch et al. (2017); Zadrozny et al. (2017) |
| MASTER/– | 2017 Aug 17 22:54:18 | Visible | Lipunov et al. (2017b, 2017a) |
| Swope/DirectCCD | 2017 Aug 17 23:33:17 | i | Coulter et al. (2017a, 2017b, 2017) |
| PROMPT5(DLT40)/– | 2017 Aug 17 23:49:00 | r | Yang et al. (2017a), Valenti et al. (submitted) |
| VISTA/VIRCAM | 2017 Aug 17 23:55:00 | K | Tanvir & Levan (2017) |
| MASTER/– | 2017 Aug 17 23:59:54 | Clear | Lipunov et al. (2017d, 2017a) |
| Blanco/DECam/– | 2017 Aug 18 00:04:24 | i | Cowperthwaite et al. (2017b); Soares-Santos et al. (2017) |
| Blanco/DECam/– | 2017 Aug 18 00:05:23 | z | Cowperthwaite et al. (2017b); Soares-Santos et al. (2017) |
| VISTA/VIRCAM | 2017 Aug 18 00:07:00 | J | Tanvir & Levan (2017) |
| *Magellan*-Clay/LDSS3-C | 2017 Aug 18 00:08:13 | g | Simon et al. (2017); Drout et al. (2017b) |
| *Magellan*-Baade/FourStar | 2017 Aug 18 00:12:19 | H | Drout et al. (2017b) |
| LasCumbres1-m/Sinistro | 2017 Aug 18 00:15:50 | w | Arcavi et al. (2017a, 2017e) |
| VISTA/VIRCAM | 2017 Aug 18 00:17:00 | Y | Tanvir & Levan (2017) |
| MASTER/– | 2017 Aug 18 00:19:05 | Clear | Lipunov et al. (2017d, 2017a) |
| *Magellan*-Baade/FourStar | 2017 Aug 18 00:25:51 | J | Drout et al. (2017b) |
| *Magellan*-Baade/FourStar | 2017 Aug 18 00:35:19 | Ks | Drout et al. (2017b) |
| PROMPT5(DLT40)/– | 2017 Aug 18 00:40:00 | r | Yang et al. (2017a), Valenti et al. (submitted) |
| REM/ROS2 | 2017 Aug 18 01:24:56 | g | Melandri et al. (2017a); Pian et al. (2017a) |
| REM/ROS2 | 2017 Aug 18 01:24:56 | i | Melandri et al. (2017a); Pian et al. (2017a) |
| REM/ROS2 | 2017 Aug 18 01:24:56 | z | Melandri et al. (2017a); Pian et al. (2017a) |
| REM/ROS2 | 2017 Aug 18 01:24:56 | r | Melandri et al. (2017a); Pian et al. (2017a) |
| Gemini-South/Flamingos-2 | 2017 Aug 18 01:30:00 | Ks | Singer et al. (2017a); Kasliwal et al. (2017) |
| PioftheSky/PioftheSkyNorth | 2017 Aug 18 03:01:39 | visible wide band | Cwiek et al. (2017); Batsch et al. (2017), |
| *Swift*/UVOT | 2017 Aug 18 03:37:00 | uvm2 | Evans et al. (2017a, 2017b) |
| *Swift*/UVOT | 2017 Aug 18 03:50:00 | uvw1 | Evans et al. (2017a, 2017b) |
| *Swift*/UVOT | 2017 Aug 18 03:58:00 | u | Evans et al. (2017a, 2017b) |
| *Swift*/UVOT | 2017 Aug 18 04:02:00 | uvw2 | Evans et al. (2017a, 2017b) |
| Subaru/HyperSuprime-Cam | 2017 Aug 18 05:31:00 | z | Yoshida et al. (2017a, 2017b), Y. Utsumi et al. (2017, in preparation) |
| Pan-STARRS1/GPC1 | 2017 Aug 18 05:33:00 | y | Chambers et al. (2017a); Smartt et al. (2017) |
| Pan-STARRS1/GPC1 | 2017 Aug 18 05:34:00 | z | Chambers et al. (2017a); Smartt et al. (2017) |
| Pan-STARRS1/GPC1 | 2017 Aug 18 05:35:00 | i | Chambers et al. (2017a); Smartt et al. (2017) |
| Pan-STARRS1/GPC1 | 2017 Aug 18 05:36:00 | y | Chambers et al. (2017a); Smartt et al. (2017) |
| Pan-STARRS1/GPC1 | 2017 Aug 18 05:37:00 | z | Chambers et al. (2017a); Smartt et al. (2017) |
| Pan-STARRS1/GPC1 | 2017 Aug 18 05:38:00 | i | Chambers et al. (2017a); Smartt et al. (2017) |
| LasCumbres1-m/Sinistro | 2017 Aug 18 09:10:04 | w | Arcavi et al. (2017b, 2017e) |
| SkyMapper/– | 2017 Aug 18 09:14:00 | i | … |
| SkyMapper/– | 2017 Aug 18 09:35:00 | z | … |
| LasCumbres1-m/Sinistro | 2017 Aug 18 09:37:26 | g | Arcavi et al. (2017e) |
| SkyMapper/– | 2017 Aug 18 09:39:00 | r | … |
| SkyMapper/– | 2017 Aug 18 09:41:00 | g | … |
| LasCumbres1-m/Sinistro | 2017 Aug 18 09:43:11 | r | Arcavi et al. (2017e) |
| T17/– | 2017 Aug 18 09:47:13 | g | Im et al. (2017a, 2017b), Im et al. (2017, in preparation) |
| SkyMapper/– | 2017 Aug 18 09:50:00 | v | … |
| T17/– | 2017 Aug 18 09:56:46 | r | Im et al. (2017a, 2017b), Im et al. (2017, in preparation) |
| SkyMapper/– | 2017 Aug 18 10:01:00 | i | Wolf et al. (2017), |
| SkyMapper/– | 2017 Aug 18 10:03:00 | r | Wolf et al. (2017), |
| SkyMapper/– | 2017 Aug 18 10:05:00 | g | Wolf et al. (2017), |
| T17/– | 2017 Aug 18 10:06:18 | i | Im et al. (2017a, 2017b), Im et al. (2017, in preparation) |
| SkyMapper/– | 2017 Aug 18 10:07:00 | v | Wolf et al. (2017), |
| LSGT/SNUCAM-II | 2017 Aug 18 10:08:01 | m425 | Im et al. (2017a, 2017b), Im et al. (2017, in preparation) |
| SkyMapper/– | 2017 Aug 18 10:09:00 | u | Wolf et al. (2017), |
| LSGT/SNUCAM-II | 2017 Aug 18 10:12:48 | m475 | Im et al. (2017a, 2017b), Im et al. (2017, in preparation) |
| LSGT/SNUCAM-II | 2017 Aug 18 10:15:16 | m525 | Im et al. (2017a, 2017b), Im et al. (2017, in preparation) |
| T17/– | 2017 Aug 18 10:15:49 | z | Im et al. (2017a, 2017b), Im et al. (2017, in preparation) |
| LSGT/SNUCAM-II | 2017 Aug 18 10:21:14 | m575 | Im et al. (2017a, 2017b), Im et al. (2017, in preparation) |
| LSGT/SNUCAM-II | 2017 Aug 18 10:22:33 | m625 | Im et al. (2017a, 2017b), Im et al. (2017, in preparation) |
| AST3-2/wide-fieldcamera | 2017 Aug 18 13:11:49 | g | Hu et al. (2017), |
| *Swift*/UVOT | 2017 Aug 18 13:30:00 | uvm2 | Cenko et al. (2017); Evans et al. (2017b) |
| *Swift*/UVOT | 2017 Aug 18 13:37:00 | uvw1 | Cenko et al. (2017); Evans et al. (2017b) |

**Table 1**
(Continued)

| Telescope/Instrument | UT Date | Band | References |
|---|---|---|---|
| *Swift*/UVOT | 2017 Aug 18 13:41:00 | u | Cenko et al. (2017); Evans et al. (2017b) |
| IRSF/SIRIUS | 2017 Aug 18 16:34:00 | Ks | Utsumi et al. (2017, in press) |
| IRSF/SIRIUS | 2017 Aug 18 16:34:00 | H | Utsumi et al. (2017, in press) |
| IRSF/SIRIUS | 2017 Aug 18 16:48:00 | J | Utsumi et al. (2017, in press) |
| KMTNet-SAAO/wide-fieldcamera | 2017 Aug 18 17:00:36 | B | Im et al. (2017d, 2017c); Troja et al. (2017a) |
| KMTNet-SAAO/wide-fieldcamera | 2017 Aug 18 17:02:55 | V | Im et al. (2017d, 2017c); Troja et al. (2017a) |
| KMTNet-SAAO/wide-fieldcamera | 2017 Aug 18 17:04:54 | R | Im et al. (2017d, 2017c); Troja et al. (2017a) |
| MASTER/– | 2017 Aug 18 17:06:55 | Clear | Lipunov et al. (2017e, 2017a) |
| KMTNet-SAAO/wide-fieldcamera | 2017 Aug 18 17:07:12 | I | Im et al. (2017d, 2017c); Troja et al. (2017a) |
| MASTER/– | 2017 Aug 18 17:17:33 | R | Lipunov et al. (2017c, 2017b, 2017a) |
| MASTER/– | 2017 Aug 18 17:34:02 | B | Lipunov et al. (2017b, 2017a) |
| 1.5 m Boyden/– | 2017 Aug 18 18:12:00 | r | Smartt et al. (2017) |
| MPG2.2 m/GROND | 2017 Aug 18 18:12:00 | g | Smartt et al. (2017) |
| NOT/NOTCam | 2017 Aug 18 20:24:08 | Ks | Malesani et al. (2017a); Tanvir & Levan (2017) |
| NOT/NOTCam | 2017 Aug 18 20:37:46 | J | Malesani et al. (2017a); Tanvir & Levan (2017) |
| PioftheSky/PioftheSkyNorth | 2017 Aug 18 21:44:44 | visible wide band | Cwiek et al. (2017); Batsch et al. (2017), |
| LasCumbres1-m/Sinistro | 2017 Aug 18 23:19:40 | i | Arcavi et al. (2017e) |
| Blanco/DECam/– | 2017 Aug 18 23:25:56 | Y | Cowperthwaite et al. (2017b); Soares-Santos et al. (2017) |
| *Magellan*-Clay/LDSS3-C | 2017 Aug 18 23:26:33 | z | Drout et al. (2017b) |
| Blanco/DECam/– | 2017 Aug 18 23:26:55 | z | Cowperthwaite et al. (2017b); Soares-Santos et al. (2017) |
| Blanco/DECam/– | 2017 Aug 18 23:27:54 | i | Cowperthwaite et al. (2017b); Soares-Santos et al. (2017) |
| KMTNet-CTIO/wide-fieldcamera | 2017 Aug 18 23:28:35 | B | Im et al. (2017d, 2017c); Troja et al. (2017a) |
| Blanco/DECam/– | 2017 Aug 18 23:28:53 | r | Cowperthwaite et al. (2017b); Soares-Santos et al. (2017) |
| Blanco/DECam/– | 2017 Aug 18 23:29:52 | g | Cowperthwaite et al. (2017b); Soares-Santos et al. (2017) |
| KMTNet-CTIO/wide-fieldcamera | 2017 Aug 18 23:30:31 | V | Im et al. (2017d, 2017c); Troja et al. (2017a) |
| Blanco/DECam/– | 2017 Aug 18 23:30:50 | u | Cowperthwaite et al. (2017b); Soares-Santos et al. (2017) |
| *Magellan*-Clay/LDSS3-C | 2017 Aug 18 23:30:55 | i | Drout et al. (2017b) |
| REM/ROS2 | 2017 Aug 18 23:31:02 | z | Melandri et al. (2017c); Pian et al. (2017a) |
| *Magellan*-Clay/LDSS3-C | 2017 Aug 18 23:32:02 | r | Drout et al. (2017b) |
| KMTNet-CTIO/wide-fieldcamera | 2017 Aug 18 23:32:36 | R | Im et al. (2017d, 2017c); Troja et al. (2017a) |
| *Magellan*-Baade/FourStar | 2017 Aug 18 23:32:58 | J | Drout et al. (2017b) |
| KMTNet-CTIO/wide-fieldcamera | 2017 Aug 18 23:34:48 | I | Im et al. (2017d, 2017c); Troja et al. (2017a) |
| *Magellan*-Clay/LDSS3-C | 2017 Aug 18 23:35:20 | B | Drout et al. (2017b) |
| VISTA/VIRCAM | 2017 Aug 18 23:44:00 | J | Tanvir & Levan (2017) |
| *Magellan*-Baade/FourStar | 2017 Aug 18 23:45:49 | H | Drout et al. (2017b) |
| PROMPT5(DLT40)/– | 2017 Aug 18 23:47:00 | r | Yang et al. (2017b), Valenti et al. (submitted) |
| VLT/FORS2 | 2017 Aug 18 23:47:02 | Rspecial | Wiersema et al. (2017); Covino et al. (2017) |
| Swope/DirectCCD | 2017 Aug 18 23:52:29 | V | Kilpatrick et al. (2017); Coulter et al. (2017) |
| VISTA/VIRCAM | 2017 Aug 18 23:53:00 | Y | Tanvir & Levan (2017) |
| TOROS/T80S | 2017 Aug 18 23:53:00 | g | Diaz et al. (2017a, 2017b), Diaz et al. (2017, in preparation) |
| TOROS/T80S | 2017 Aug 18 23:53:00 | r | Diaz et al. (2017a, 2017b), Diaz et al. (2017, in preparation) |
| TOROS/T80S | 2017 Aug 18 23:53:00 | i | Diaz et al. (2017a, 2017b), Diaz et al. (2017, in preparation) |
| MPG2.2 m/GROND | 2017 Aug 18 23:56:00 | i | Smartt et al. (2017) |
| MPG2.2 m/GROND | 2017 Aug 18 23:56:00 | z | Smartt et al. (2017) |
| MPG2.2 m/GROND | 2017 Aug 18 23:56:00 | J | Smartt et al. (2017) |
| MPG2.2 m/GROND | 2017 Aug 18 23:56:00 | r | Smartt et al. (2017) |
| MPG2.2 m/GROND | 2017 Aug 18 23:56:00 | H | Smartt et al. (2017) |
| MPG2.2 m/GROND | 2017 Aug 18 23:56:00 | Ks | Smartt et al. (2017) |
| Gemini-South/Flamingos-2 | 2017 Aug 19 00:00:19 | H | Cowperthwaite et al. (2017b) |
| *Magellan*-Baade/FourStar | 2017 Aug 19 00:02:53 | J1 | Drout et al. (2017b) |
| VLT/X-shooter | 2017 Aug 19 00:08:58 | r | Pian et al. (2017a, 2017a) |
| VLT/X-shooter | 2017 Aug 19 00:10:46 | z | Pian et al. (2017b, 2017b) |
| VLT/X-shooter | 2017 Aug 19 00:14:01 | g | Pian et al. (2017, 2017) |
| *Swift*/UVOT | 2017 Aug 19 00:41:00 | u | Evans et al. (2017b) |
| Swope/DirectCCD | 2017 Aug 19 00:49:15 | B | Kilpatrick et al. (2017); Coulter et al. (2017) |
| Swope/DirectCCD | 2017 Aug 19 01:08:00 | r | Coulter et al. (2017) |
| NTT/– | 2017 Aug 19 01:09:00 | U | Smartt et al. (2017) |
| Swope/DirectCCD | 2017 Aug 19 01:18:57 | g | Coulter et al. (2017) |
| BOOTES-5/JGT/– | 2017 Aug 19 03:08:14 | clear | Castro-Tirado et al. (2017), Zhang et al. (2017, in preparation) |
| Pan-STARRS1/GPC1 | 2017 Aug 19 05:42:00 | y | Chambers et al. (2017b); Smartt et al. (2017) |
| Pan-STARRS1/GPC1 | 2017 Aug 19 05:44:00 | z | Chambers et al. (2017b); Smartt et al. (2017) |

**Table 1**
(Continued)

| Telescope/Instrument | UT Date | Band | References |
|---|---|---|---|
| Pan-STARRS1/GPC1 | 2017 Aug 19 05:46:00 | i | Chambers et al. (2017b); Smartt et al. (2017) |
| MOA-II/MOA-cam3 | 2017 Aug 19 07:26:00 | R | Utsumi et al. (2017, in press) |
| B&C61cm/Tripole5 | 2017 Aug 19 07:26:00 | g | Utsumi et al. (2017, in press) |
| KMTNet-SSO/wide-fieldcamera | 2017 Aug 19 08:32:48 | B | Im et al. (2017d, 2017c); Troja et al. (2017a) |
| KMTNet-SSO/wide-fieldcamera | 2017 Aug 19 08:34:43 | V | Im et al. (2017d, 2017c); Troja et al. (2017a) |
| KMTNet-SSO/wide-fieldcamera | 2017 Aug 19 08:36:39 | R | Im et al. (2017d, 2017c); Troja et al. (2017a) |
| KMTNet-SSO/wide-fieldcamera | 2017 Aug 19 08:38:42 | I | Im et al. (2017d, 2017c); Troja et al. (2017a) |
| T27/– | 2017 Aug 19 09:01:31 | V | Im et al. (2017a, 2017b), Im et al. (2017, in preparation) |
| T30/– | 2017 Aug 19 09:02:27 | V | Im et al. (2017a, 2017b), Im et al. (2017, in preparation) |
| T27/– | 2017 Aug 19 09:02:27 | R | Im et al. (2017a, 2017b), Im et al. (2017, in preparation) |
| T31/– | 2017 Aug 19 09:02:34 | R | Im et al. (2017a, 2017b), Im et al. (2017, in preparation) |
| T27/– | 2017 Aug 19 09:11:30 | I | Im et al. (2017a, 2017b), Im et al. (2017, in preparation) |
| Zadko/CCDimager | 2017 Aug 19 10:57:00 | r | Coward et al. (2017a), |
| MASTER/– | 2017 Aug 19 17:06:57 | Clear | Lipunov et al. (2017b, 2017a) |
| MASTER/– | 2017 Aug 19 17:53:34 | R | Lipunov et al. (2017b, 2017a) |
| LasCumbres1-m/Sinistro | 2017 Aug 19 18:01:26 | V | Arcavi et al. (2017e) |
| LasCumbres1-m/Sinistro | 2017 Aug 19 18:01:26 | z | Arcavi et al. (2017e) |
| MASTER/– | 2017 Aug 19 18:04:32 | B | Lipunov et al. (2017b, 2017a) |
| 1.5 m Boyden/– | 2017 Aug 19 18:16:00 | r | Smartt et al. (2017) |
| REM/ROS2 | 2017 Aug 19 23:12:59 | r | Melandri et al. (2017c); Pian et al. (2017) |
| REM/ROS2 | 2017 Aug 19 23:12:59 | i | Melandri et al. (2017c); Pian et al. (2017) |
| REM/ROS2 | 2017 Aug 19 23:12:59 | g | Melandri et al. (2017c); Pian et al. (2017) |
| MASTER/– | 2017 Aug 19 23:13:20 | Clear | Lipunov et al. (2017b, 2017a) |
| Gemini-South/Flamingos-2 | 2017 Aug 19 23:13:34 | H | Cowperthwaite et al. (2017b) |
| MPG2.2 m/GROND | 2017 Aug 19 23:15:00 | r | Smartt et al. (2017) |
| MPG2.2 m/GROND | 2017 Aug 19 23:15:00 | z | Smartt et al. (2017) |
| MPG2.2 m/GROND | 2017 Aug 19 23:15:00 | H | Smartt et al. (2017) |
| MPG2.2 m/GROND | 2017 Aug 19 23:15:00 | i | Smartt et al. (2017) |
| MPG2.2 m/GROND | 2017 Aug 19 23:15:00 | J | Smartt et al. (2017) |
| TOROS/EABA | 2017 Aug 19 23:18:38 | r | Diaz et al. (2017b), Diaz et al. (2017, in preparation) |
| *Magellan*-Baade/FourStar | 2017 Aug 19 23:18:50 | H | Drout et al. (2017b) |
| Etelman/VIRT/CCDimager | 2017 Aug 19 23:19:00 | R | Gendre et al. (2017), Andreoni et al. (2017, in preparation) |
| Blanco/DECam/– | 2017 Aug 19 23:23:29 | Y | Cowperthwaite et al. (2017b); Soares-Santos et al. (2017) |
| Blanco/DECam/– | 2017 Aug 19 23:26:59 | r | Cowperthwaite et al. (2017b); Soares-Santos et al. (2017) |
| Blanco/DECam/– | 2017 Aug 19 23:27:59 | g | Cowperthwaite et al. (2017b); Soares-Santos et al. (2017) |
| ChilescopeRC-1000/– | 2017 Aug 19 23:30:33 | clear | Pozanenko et al. (2017a, 2017b), Pozanenko et al. (2017, in preparation) |
| *Magellan*-Baade/FourStar | 2017 Aug 19 23:31:06 | J1 | Drout et al. (2017b) |
| Blanco/DECam/– | 2017 Aug 19 23:31:13 | u | Cowperthwaite et al. (2017b); Soares-Santos et al. (2017) |
| *Magellan*-Baade/FourStar | 2017 Aug 19 23:41:59 | Ks | Drout et al. (2017b) |
| *Magellan*-Baade/IMACS | 2017 Aug 20 00:13:32 | r | Drout et al. (2017b) |
| Gemini-South/Flamingos-2 | 2017 Aug 20 00:19:00 | Ks | Kasliwal et al. (2017) |
| LasCumbres1-m/Sinistro | 2017 Aug 20 00:24:28 | g | Arcavi et al. (2017e) |
| Gemini-South/Flamingos-2 | 2017 Aug 20 00:27:00 | J | Kasliwal et al. (2017) |
| NTT/– | 2017 Aug 20 01:19:00 | U | Smartt et al. (2017) |
| Pan-STARRS1/GPC1 | 2017 Aug 20 05:38:00 | y | Chambers et al. (2017c); Smartt et al. (2017) |
| Pan-STARRS1/GPC1 | 2017 Aug 20 05:41:00 | z | Chambers et al. (2017c); Smartt et al. (2017) |
| Pan-STARRS1/GPC1 | 2017 Aug 20 05:45:00 | i | Chambers et al. (2017c); Smartt et al. (2017) |
| T31/– | 2017 Aug 20 09:20:38 | R | Im et al. (2017a, 2017b), Im et al. (2017, in preparation) |
| MASTER/– | 2017 Aug 20 17:04:36 | Clear | Lipunov et al. (2017b, 2017a) |
| MASTER/– | 2017 Aug 20 17:25:56 | R | Lipunov et al. (2017b, 2017a) |
| MASTER/– | 2017 Aug 20 17:36:32 | B | Lipunov et al. (2017b, 2017a) |
| LasCumbres1-m/Sinistro | 2017 Aug 20 17:39:50 | i | Arcavi et al. (2017e) |
| LasCumbres1-m/Sinistro | 2017 Aug 20 17:45:36 | z | Arcavi et al. (2017e) |
| LasCumbres1-m/Sinistro | 2017 Aug 20 17:49:55 | V | Arcavi et al. (2017e) |
| MPG2.2 m/GROND | 2017 Aug 20 23:15:00 | g | Smartt et al. (2017) |
| *Magellan*-Baade/FourStar | 2017 Aug 20 23:20:42 | J | Drout et al. (2017b) |
| ChilescopeRC-1000/– | 2017 Aug 20 23:21:09 | clear | Pozanenko et al. (2017a) |
| VISTA/VIRCAM | 2017 Aug 20 23:24:00 | K | Tanvir & Levan (2017) |
| Blanco/DECam/– | 2017 Aug 20 23:37:06 | u | Cowperthwaite et al. (2017b); Soares-Santos et al. (2017) |
| Swope/DirectCCD | 2017 Aug 20 23:44:36 | V | Coulter et al. (2017) |
| Swope/DirectCCD | 2017 Aug 20 23:53:00 | B | Coulter et al. (2017) |

**Table 1**
(Continued)

| Telescope/Instrument | UT Date | Band | References |
|---|---|---|---|
| MASTER/– | 2017 Aug 21 00:26:31 | Clear | Lipunov et al. (2017b, 2017a) |
| Gemini-South/Flamingos-2 | 2017 Aug 21 00:38:00 | H | Kasliwal et al. (2017); Troja et al. (2017a) |
| Pan-STARRS1/GPC1 | 2017 Aug 21 05:37:00 | y | Chambers et al. (2017d); Smartt et al. (2017) |
| Pan-STARRS1/GPC1 | 2017 Aug 21 05:39:00 | z | Chambers et al. (2017d); Smartt et al. (2017) |
| Pan-STARRS1/GPC1 | 2017 Aug 21 05:42:00 | i | Chambers et al. (2017d); Smartt et al. (2017) |
| AST3-2/wide-fieldcamera | 2017 Aug 21 15:36:50 | g | … |
| MASTER/– | 2017 Aug 21 17:08:14 | Clear | Lipunov et al. (2017b, 2017a) |
| MASTER/– | 2017 Aug 21 18:06:12 | R | Lipunov et al. (2017b, 2017a) |
| MASTER/– | 2017 Aug 21 19:20:23 | B | Lipunov et al. (2017b, 2017a) |
| duPont/RetroCam | 2017 Aug 21 23:17:19 | Y | Drout et al. (2017b) |
| Etelman/VIRT/CCDimager | 2017 Aug 21 23:19:00 | Clear | Gendre et al. (2017); Andreoni et al. (2017, in preparation) |
| MPG2.2 m/GROND | 2017 Aug 21 23:22:00 | Ks | Smartt et al. (2017) |
| VLT/FORS2 | 2017 Aug 21 23:23:11 | R | D'Avanzo et al. (2017); Pian et al. (2017) |
| ChilescopeRC-1000/– | 2017 Aug 21 23:32:09 | clear | Pozanenko et al. (2017c) |
| duPont/RetroCam | 2017 Aug 21 23:34:34 | H | Drout et al. (2017b) |
| LasCumbres1-m/Sinistro | 2017 Aug 21 23:48:28 | w | Arcavi et al. (2017e) |
| Swope/DirectCCD | 2017 Aug 21 23:54:57 | r | Coulter et al. (2017) |
| duPont/RetroCam | 2017 Aug 21 23:57:41 | J | Drout et al. (2017b) |
| Swope/DirectCCD | 2017 Aug 22 00:06:17 | g | Coulter et al. (2017) |
| VLT/FORS2 | 2017 Aug 22 00:09:09 | z | D'Avanzo et al. (2017); Pian et al. (2017) |
| VLT/FORS2 | 2017 Aug 22 00:18:49 | I | D'Avanzo et al. (2017); Pian et al. (2017) |
| *Magellan*-Clay/LDSS3-C | 2017 Aug 22 00:27:40 | g | Drout et al. (2017b) |
| VLT/FORS2 | 2017 Aug 22 00:28:18 | B | D'Avanzo et al. (2017); Pian et al. (2017) |
| VLT/FORS2 | 2017 Aug 22 00:38:20 | V | D'Avanzo et al. (2017); Pian et al. (2017) |
| *HST*/WFC3/IR | 2017 Aug 22 07:34:00 | F110W | Tanvir & Levan (2017); Troja et al. (2017a) |
| LasCumbres1-m/Sinistro | 2017 Aug 22 08:35:31 | r | Arcavi et al. (2017e) |
| *HST*/WFC3/IR | 2017 Aug 22 10:45:00 | F160W | Tanvir & Levan (2017); Troja et al. (2017a) |
| HubbleSpaceTelescope/WFC3 | 2017 Aug 22 20:19:00 | F336W | Adams et al. (2017); Kasliwal et al. (2017) |
| Etelman/VIRT/CCDimager | 2017 Aug 22 23:19:00 | Clear | Gendre et al. (2017); Andreoni et al. (2017, in preparation) |
| VLT/VIMOS | 2017 Aug 22 23:30:00 | z | Tanvir & Levan (2017) |
| duPont/RetroCam | 2017 Aug 22 23:33:54 | Y | Drout et al. (2017b) |
| VLT/VIMOS | 2017 Aug 22 23:42:00 | R | Tanvir & Levan (2017) |
| VLT/VIMOS | 2017 Aug 22 23:53:00 | u | Evans et al. (2017b) |
| VLT/FORS2 | 2017 Aug 22 23:53:31 | Rspecial | Covino et al. (2017) |
| VST/OmegaCam | 2017 Aug 22 23:58:32 | g | Grado et al. (2017a); Pian et al. (2017) |
| VLT/X-shooter | 2017 Aug 23 00:35:20 | r | Pian et al. (2017) |
| VLT/X-shooter | 2017 Aug 23 00:37:08 | z | Pian et al. (2017) |
| VLT/X-shooter | 2017 Aug 23 00:40:24 | g | Pian et al. (2017) |
| Zadko/CCDimager | 2017 Aug 23 11:32:00 | r | Coward et al. (2017a), |
| IRSF/SIRIUS | 2017 Aug 23 17:22:00 | Ks | Kasliwal et al. (2017) |
| IRSF/SIRIUS | 2017 Aug 23 17:22:00 | J | Kasliwal et al. (2017) |
| IRSF/SIRIUS | 2017 Aug 23 17:22:00 | H | Kasliwal et al. (2017) |
| VST/OmegaCam | 2017 Aug 23 23:26:51 | i | Grado et al. (2017a); Pian et al. (2017) |
| VLT/VISIR | 2017 Aug 23 23:35:00 | 8.6um | Kasliwal et al. (2017) |
| VST/OmegaCam | 2017 Aug 23 23:42:49 | r | Grado et al. (2017a); Pian et al. (2017) |
| CTIO1.3 m/ANDICAM | 2017 Aug 24 23:20:00 | Ks | Kasliwal et al. (2017) |
| Swope/DirectCCD | 2017 Aug 24 23:45:07 | i | Coulter et al. (2017) |
| ChilescopeRC-1000/– | 2017 Aug 24 23:53:39 | clear | Pozanenko et al. (2017b), |
| Blanco/DECam/– | 2017 Aug 24 23:56:22 | g | Cowperthwaite et al. (2017b); Soares-Santos et al. (2017) |
| *Magellan*-Clay/LDSS3-C | 2017 Aug 25 00:43:27 | B | Drout et al. (2017b) |
| *HST*/WFC3/UVIS | 2017 Aug 25 13:55:00 | F606W | Tanvir & Levan (2017); Troja et al. (2017a) |
| *HST*/WFC3/UVIS | 2017 Aug 25 15:28:00 | F475W | Tanvir & Levan (2017); Troja et al. (2017a) |
| *HST*/WFC3/UVIS | 2017 Aug 25 15:36:00 | F275W | Levan & Tanvir (2017); Tanvir & Levan (2017), |
| *Magellan*-Clay/LDSS3-C | 2017 Aug 25 23:19:41 | z | Drout et al. (2017b) |
| Blanco/DECam/– | 2017 Aug 25 23:56:05 | r | Cowperthwaite et al. (2017b); Soares-Santos et al. (2017) |
| VLT/FORS2 | 2017 Aug 26 00:13:40 | z | Covino et al. (2017) |
| duPont/RetroCam | 2017 Aug 26 00:14:28 | J | Drout et al. (2017b) |
| VLT/FORS2 | 2017 Aug 26 00:27:16 | B | Pian et al. (2017) |
| IRSF/SIRIUS | 2017 Aug 26 16:57:00 | J | Kasliwal et al. (2017) |
| IRSF/SIRIUS | 2017 Aug 26 16:57:00 | Ks | Kasliwal et al. (2017) |
| IRSF/SIRIUS | 2017 Aug 26 16:57:00 | H | Kasliwal et al. (2017) |

**Table 1**
(Continued)

| Telescope/Instrument | UT Date | Band | References |
|---|---|---|---|
| VISTA/VIRCAM | 2017 Aug 26 23:38:00 | Y | Tanvir & Levan (2017) |
| ApachePointObservatory/NICFPS | 2017 Aug 27 02:15:00 | Ks | Kasliwal et al. (2017) |
| Palomar200inch/WIRC | 2017 Aug 27 02:49:00 | Ks | Kasliwal et al. (2017) |
| *HST*/WFC3/IR | 2017 Aug 27 06:45:56 | F110W | Cowperthwaite et al. (2017b) |
| *HST*/WFC3/IR | 2017 Aug 27 07:06:57 | F160W | Cowperthwaite et al. (2017b) |
| *HST*/WFC3/UVIS | 2017 Aug 27 08:20:49 | F336W | Cowperthwaite et al. (2017b) |
| *HST*/ACS/WFC | 2017 Aug 27 10:24:14 | F475W | Cowperthwaite et al. (2017b) |
| *HST*/ACS/WFC | 2017 Aug 27 11:57:07 | F625W | Cowperthwaite et al. (2017b) |
| *HST*/ACS/WFC | 2017 Aug 27 13:27:15 | F775W | Cowperthwaite et al. (2017b) |
| *HST*/ACS/WFC | 2017 Aug 27 13:45:24 | F850LP | Cowperthwaite et al. (2017b) |
| Gemini-South/Flamingos-2 | 2017 Aug 27 23:16:00 | J | Kasliwal et al. (2017) |
| CTIO1.3 m/ANDICAM | 2017 Aug 27 23:18:00 | Ks | Kasliwal et al. (2017) |
| Blanco/DECam/– | 2017 Aug 27 23:23:33 | Y | Cowperthwaite et al. (2017b); Soares-Santos et al. (2017) |
| MPG2.2 m/GROND | 2017 Aug 27 23:24:00 | J | Smartt et al. (2017) |
| Gemini-South/Flamingos-2 | 2017 Aug 27 23:28:10 | $K_s$ | Cowperthwaite et al. (2017b) |
| Gemini-South/Flamingos-2 | 2017 Aug 27 23:33:07 | H | Cowperthwaite et al. (2017b) |
| duPont/RetroCam | 2017 Aug 27 23:36:25 | H | Drout et al. (2017b) |
| Blanco/DECam/– | 2017 Aug 27 23:40:57 | z | Cowperthwaite et al. (2017b); Soares-Santos et al. (2017) |
| Blanco/DECam/– | 2017 Aug 28 00:00:01 | i | Cowperthwaite et al. (2017b); Soares-Santos et al. (2017) |
| VLT/FORS2 | 2017 Aug 28 00:07:31 | R | Pian et al. (2017a) |
| VLT/FORS2 | 2017 Aug 28 00:15:56 | V | Pian et al. (2017a) |
| MPG2.2 m/GROND | 2017 Aug 28 00:22:00 | H | Smartt et al. (2017) |
| *HST*/WFC3/IR | 2017 Aug 28 01:50:00 | F110W | Tanvir & Levan (2017); Troja et al. (2017a) |
| *HST*/WFC3/IR | 2017 Aug 28 03:25:00 | F160W | Tanvir & Levan (2017); Troja et al. (2017a) |
| *HST*/WFC3/UVIS | 2017 Aug 28 20:56:00 | F275W | Levan & Tanvir (2017); Tanvir & Levan (2017), |
| *HST*/WFC3/UVIS | 2017 Aug 28 22:29:00 | F475W | Tanvir & Levan (2017); Troja et al. (2017a) |
| *HST*/WFC3/UVIS | 2017 Aug 28 23:02:00 | F814W | Tanvir & Levan (2017); Troja et al. (2017a) |
| NTT/– | 2017 Aug 28 23:03:00 | H | Smartt et al. (2017) |
| *HST*/WFC3/UVIS | 2017 Aug 28 23:08:00 | F606W | Tanvir & Levan (2017); Troja et al. (2017a) |
| MPG2.2 m/GROND | 2017 Aug 28 23:22:00 | Ks | Smartt et al. (2017) |
| VISTA/VIRCAM | 2017 Aug 28 23:33:00 | J | Tanvir & Levan (2017) |
| Gemini-South/Flamingos-2 | 2017 Aug 28 23:36:01 | $K_s$ | Cowperthwaite et al. (2017b) |
| VLT/FORS2 | 2017 Aug 29 00:00:13 | I | Pian et al. (2017a) |
| HubbleSpaceTelescope/WFC3/UVIS | 2017 Aug 29 00:36:00 | F275W | Kasliwal et al. (2017) |
| HubbleSpaceTelescope/WFC3/UVIS | 2017 Aug 29 00:36:00 | F225W | Kasliwal et al. (2017) |
| NTT/– | 2017 Aug 29 22:56:00 | Ks | Smartt et al. (2017) |
| VLT/VIMOS | 2017 Aug 29 23:16:00 | R | Tanvir & Levan (2017) |
| SkyMapper/– | 2017 Aug 30 09:26:00 | u | … |
| SkyMapper/– | 2017 Aug 30 09:32:00 | v | … |
| NTT/– | 2017 Aug 30 23:03:00 | Ks | Smartt et al. (2017) |
| VLT/FORS2 | 2017 Aug 31 23:34:46 | z | Pian et al. (2017a) |
| VISTA/VIRCAM | 2017 Aug 31 23:42:00 | K | Tanvir & Levan (2017) |
| Gemini-South/Flamingos-2 | 2017 Aug 31 23:50:00 | H | Singer et al. (2017b); Kasliwal et al. (2017) |
| SkyMapper/– | 2017 Sep 01 09:12:00 | i | … |
| SkyMapper/– | 2017 Sep 01 09:14:00 | z | … |
| SkyMapper/– | 2017 Sep 03 09:21:00 | g | … |
| SkyMapper/– | 2017 Sep 03 09:23:00 | r | … |
| NTT/– | 2017 Sep 04 23:12:00 | Ks | Smartt et al. (2017) |
| Gemini-South/Flamingos-2 | 2017 Sep 04 23:28:45 | $K_s$ | Cowperthwaite et al. (2017b) |
| VLT/VIMOS | 2017 Sep 05 23:23:00 | z | Tanvir & Levan (2017) |
| Gemini-South/Flamingos-2 | 2017 Sep 05 23:48:00 | Ks | Kasliwal et al. (2017) |
| *Magellan*-Baade/FourStar | 2017 Sep 06 23:24:28 | Ks | Drout et al. (2017b) |
| VLT/HAWKI | 2017 Sep 07 23:11:00 | K | Tanvir & Levan (2017) |
| VLT/HAWKI | 2017 Sep 11 23:21:00 | K | Tanvir & Levan (2017) |

**Note.** This is a subset of all the observations made in order to give a sense of the substantial coverage of this event.

time, flux upper limits, and the energy range of the observations, which are summarized here.

At the time of GRB 170817A, three out of six spacecraft of the Inter Planetary Network (Hurley et al. 2013) had a favorable orientation to observe the LIGO-Virgo skymap. However, based on the *Fermi*-GBM (Goldstein et al. 2017b) and *INTEGRAL* analyses, GRB 170817A was too weak to be detected by Konus-*Wind* (Svinkin et al. 2017a). Using the

**Table 2**
Record of Spectroscopic Observations

| Telescope/Instrument | UT Date | Wavelengths (Å) | Resolution (R) | References |
|---|---|---|---|---|
| *Magellan*-Clay/LDSS-3 | 2017 Aug 18 00:26:17 | 3780–10200 | 860 | Drout et al. (2017); Shappee et al. (2017) |
| *Magellan*-Clay/LDSS-3 | 2017 Aug 18 00:40:09 | 3800–6200 | 1900 | Shappee et al. (2017) |
| *Magellan*-Clay/LDSS-3 | 2017 Aug 18 00:52:09 | 6450–10000 | 1810 | Shappee et al. (2017) |
| *Magellan*-Baade/MagE | 2017 Aug 18 01:26:22 | 3650–10100 | 5800 | Shappee et al. (2017) |
| ANU2.3/WiFeS | 2017 Aug 18 09:24:00 | 3200–9800 | B/R 3000 | … |
| SALT/RSS | 2017 Aug 18 17:07:00 | 3600–8000 | 300 | Shara et al. (2017), |
| NTT/EFOSC2Gr#11+16 | 2017 Aug 18 23:19:12 | 3330–9970 | 260/400 | Smartt et al. (2017) |
| VLT/X-shooter | 2017 Aug 18 23:22:25 | 3000–24800 | 4290/8150/5750 | Pian et al. (2017b, 2017b) |
| SOAR/GHTS | 2017 Aug 18 23:22:39 | 4000–8000 | 830 | Nicholl et al. (2017d) |
| *Magellan*-Clay/LDSS-3 | 2017 Aug 18 23:47:37 | 3820–9120 | 860 | Shappee et al. (2017) |
| VLT/MUSE | 2017 Aug 18 23:49:00 | 4650–9300 | 3000 | Levan & Tanvir (2017); Tanvir & Levan (2017) |
| *Magellan*-Clay/MIKE | 2017 Aug 19 00:18:11 | 3900–9400 | 30000 | Shappee et al. (2017) |
| *Magellan*-Baade/MagE | 2017 Aug 19 00:35:25 | 3800–10300 | 4100 | Shappee et al. (2017) |
| Gemini-South/FLAMINGOS2 | 2017 Aug 19 00:42:27 | 9100–18000 | 500 | Chornock et al. (2017a) |
| LCOFaulkesTelescopeSouth/FLOYDS | 2017 Aug 19 08:36:22 | 5500–9250 | 700 | GC21908, McCully et al. (2017b) |
| ANU2.3/WiFeS | 2017 Aug 19 09:26:12 | 3200–9800 | B/R 3000 | … |
| SALT/RSS | 2017 Aug 19 16:58:00 | 3600–8000 | 300 | Shara et al. (2017) |
| SALT/RSS | 2017 Aug 19 16:58:32 | 3600–8000 | 300 | Shara et al. (2017); Shara et al. 2017, McCully et al. (2017b) |
| NTT/EFOSC2Gr#11+16 | 2017 Aug 19 23:25:41 | 3330–9970 | 260/400 | Smartt et al. (2017) |
| SOAR/GHTS | 2017 Aug 19 23:28:32 | 4000–8000 | 830 | Nicholl et al. (2017d) |
| VLT/Xshooterfixed | 2017 Aug 19 23:28:46 | 3700–22790 | 4290/3330/5450 | Smartt et al. (2017) |
| Gemini-South/FLAMINGOS2 | 2017 Aug 19 23:42:56 | 9100–18000 | 500 | Chornock et al. (2017a) |
| *Magellan*-Baade/IMACS | 2017 Aug 20 00:26:28 | 4355–8750 | 1000 | Shappee et al. (2017) |
| GeminiSouth/GMOS | 2017 Aug 20 01:01:54 | 4000–9500 | 400 | McCully et al. (2017a, 2017b) |
| Gemini-South/GMOS | 2017 Aug 20 01:08:00 | 6000–9000 | 1900 | Kasliwal et al. (2017) |
| ANU2.3/WiFeS | 2017 Aug 20 09:21:33 | 3200–9800 | B/R 3000 | … |
| NTT/EFOSC2Gr#11+16 | 2017 Aug 20 23:21:13 | 3330–9970 | 390/600 | Smartt et al. (2017) |
| SOAR/GHTS | 2017 Aug 20 23:23:17 | 5000–9000 | 830 | Nicholl et al. (2017d) |
| VLT/X-shooter | 2017 Aug 20 23:25:28 | 3000–24800 | 4290/8150/5750 | Pian et al. (2017a) |
| *Magellan*-Clay/LDSS-3 | 2017 Aug 20 23:45:53 | 4450–10400 | 860 | Shappee et al. (2017) |
| Gemini-South/GMOS | 2017 Aug 21 00:15:00 | 3800–9200 | 1700 | Troja et al. (2017b); Kasliwal et al. (2017); Troja et al. (2017a) |
| GeminiSouth/GMOS | 2017 Aug 21 00:16:09 | 4000–9500 | 400 | Troja et al. (2017b); McCully et al. (2017b); Troja et al. (2017a) |
| VLT/FORS2 | 2017 Aug 21 00:43:12 | 3500–8600 | 800–1000 | Pian et al. (2017a) |
| ANU2.3/WiFeS | 2017 Aug 21 09:13:00 | 3200–7060 | B 3000 R 7000 | … |
| NTT/SOFIBlueGrism | 2017 Aug 21 23:11:37 | 9380–16460 | 550 | Smartt et al. (2017) |
| SOAR/GHTS | 2017 Aug 21 23:24:49 | 4000–8000 | 830 | Nicholl et al. (2017d) |
| VLT/Xshooterfixed | 2017 Aug 21 23:25:38 | 3700–22790 | 4290/3330/5450 | Smartt et al. (2017) |
| VLT/FORS2 | 2017 Aug 21 23:31:12 | 3500–8600 | 800–1000 | Pian et al. (2017a) |
| Gemini-South/FLAMINGOS2 | 2017 Aug 21 23:40:09 | 9100–18000 | 500 | Chornock et al. (2017a) |
| Gemini-South/Flamingos-2 | 2017 Aug 22 00:21:00 | 12980–25070 | 600 | Kasliwal et al. (2017) |
| Gemini-South/Flamingos-2 | 2017 Aug 22 00:47:00 | 9840–18020 | 600 | Kasliwal et al. (2017) |
| *Magellan*-Clay/LDSS-3 | 2017 Aug 22 00:50:34 | 5010–10200 | 860 | Shappee et al. (2017) |
| *HST*/WFC3/IR-G102 | 2017 Aug 22 09:07:00 | 8000–11150 | 210 | Tanvir & Levan (2017); Troja et al. (2017a) |
| *HST*/WFC3/IR-G141 | 2017 Aug 22 10:53:00 | 10750–17000 | 130 | Tanvir & Levan (2017); Troja et al. (2017a) |
| *Magellan*-Clay/LDSS-3 | 2017 Aug 22 23:34:00 | 5000–10200 | 860 | Shappee et al. (2017) |
| *HST*/STIS | 2017 Aug 23 02:51:54 | 1600–3200 | 700 | Nicholl et al. (2017d) |

**Table 2**
(Continued)

| Telescope/Instrument | UT Date | Wavelengths (Å) | Resolution (R) | References |
|---|---|---|---|---|
| AAT/AAOmega2DF | 2017 Aug 24 08:55:00 | 3750–8900 | 1700 | Andreoni et al. (2017), |
| *HST*/WFC3/IR-G102 | 2017 Aug 24 18:58:00 | 8000–11150 | 210 | Tanvir & Levan (2017); Troja et al. (2017a) |
| *Magellan*-Clay/LDSS-3 | 2017 Aug 24 23:33:51 | 6380–10500 | 1810 | Shappee et al. (2017) |
| SOAR/GHTS | 2017 Aug 24 23:34:31 | 5000–9000 | 830 | Nicholl et al. (2017d) |
| Gemini-South/FLAMINGOS2 | 2017 Aug 24 23:56:32 | 9100–18000 | 500 | Chornock et al. (2017a) |
| KeckI/LRIS | 2017 Aug 25 05:45:00 | 2000–10300 | 1000 | Kasliwal et al. (2017) |
| *Magellan*/Baade/IMACS | 2017 Aug 25 23:37:59 | 4300–9300 | 1100 | Nicholl et al. (2017d) |
| *Magellan*-Clay/LDSS-3 | 2017 Aug 25 23:39:18 | 6380–10500 | 1810 | Shappee et al. (2017) |
| Gemini-South/FLAMINGOS2 | 2017 Aug 26 00:21:24 | 9100–18000 | 500 | Chornock et al. (2017a) |
| *HST*/WFC3/IR-G141 | 2017 Aug 26 22:57:00 | 10750–17000 | 130 | Tanvir & Levan (2017); Troja et al. (2017a) |
| *Magellan*/Baade/IMACS | 2017 Aug 26 23:20:54 | 4300–9300 | 1100 | Nicholl et al. (2017d) |
| Gemini-South/FLAMINGOS2 | 2017 Aug 27 00:12:20 | 9100–18000 | 500 | Chornock et al. (2017a) |
| Gemini-South/FLAMINGOS2 | 2017 Aug 28 00:16:28 | 9100–18000 | 500 | Chornock et al. (2017a) |
| *HST*/WFC3/IR-G102 | 2017 Aug 28 01:58:00 | 8000–11150 | 210 | Tanvir & Levan (2017); Troja et al. (2017a) |
| *HST*/WFC3/IR-G141 | 2017 Aug 28 03:33:00 | 10750–17000 | 130 | Tanvir & Levan (2017); Troja et al. (2017a) |
| Gemini-South/Flamingos-2 | 2017 Aug 29 00:23:00 | 12980–25070 | 600 | Kasliwal et al. (2017) |

**Table 3**
Gamma-Ray Monitoring and Evolution of GW170817

| Observatory | UT Date | Time since GW Trigger | 90% Flux Upper Limit (erg $cm^{-2}$ $s^{-1}$ ) | Energy Band | GCN/Reference |
|---|---|---|---|---|---|
| *Insight*-HXMT/HE | Aug 17 12:34:24 UTC | −400 s | $3.7 \times 10^{-7}$ | 0.2–5 MeV | Li et al. (2017) |
| CALET CGBM | Aug 17 12:41:04 UTC | 0.0 | $1.3 \times 10^{-7}$[a] | 10–1000 keV | Nakahira et al. (2017) |
| Konus-*Wind* | Aug 17 12:41:04.446 UTC | 0.0 | $3.0 \times 10^{-7}$ [erg $cm^{-2}$] | 10 keV–10 MeV | Svinkin et al. (2017a) |
| *Insight*-HXMT/HE | Aug 17 12:41:04.446 UTC | 0.0 | $3.7 \times 10^{-7}$ | 0.2–5 MeV | Li et al. (2017) |
| *Insight*-HXMT/HE | Aug 17 12:41:06.30 UTC | 1.85 s | $6.6 \times 10^{-7}$ | 0.2–5 MeV | Li et al. (2017) |
| *Insight*-HXMT/HE | Aug 17 12:46:04 UTC | 300 s | $1.5 \times 10^{-7}$ | 0.2–5 MeV | Li et al. (2017) |
| AGILE-GRID | Aug 17 12:56:41 UTC | 0.011 days | $3.9 \times 10^{-9}$ | 0.03–3 GeV | V. Verrecchia et al. (2017, in preparation) |
| *Fermi*-LAT | Aug 17 13:00:14 UTC | 0.013 days | $4.0 \times 10^{-10}$ | 0.1–1 GeV | Kocevski et al. (2017) |
| H.E.S.S. | Aug 17 17:59 UTC | 0.22 days | $3.9 \times 10^{-12}$ | 0.28–2.31 TeV | H. Abdalla et al. (H.E.S.S. Collaboration) (2017, in preparation) |
| HAWC | Aug 17 20:53:14—Aug 17 22:55:00 UTC | 0.342 days + 0.425 days | $1.7 \times 10^{-10}$ | 4–100 TeV | Martinez-Castellanos et al. (2017) |
| *Fermi*-GBM | Aug 16 12:41:06—Aug 18 12:41:06 UTC | ±1.0 days | $(8.0–9.9) \times 10^{-10}$ | 20–100 keV | Goldstein et al. (2017a) |
| NTEGRAL IBIS/ISGRI | Aug 18 12:45:10—Aug 23 03:22:34 UTC | 1–5.7 days | $2.0 \times 10^{-11}$ | 20–80 keV | Savchenko et al. (2017) |
| *INTEGRAL* IBIS/ISGRI | Aug 18 12:45:10—Aug 23 03:22:34 UTC | 1–5.7 days | $3.6 \times 10^{-11}$ | 80–300 keV | Savchenko et al. (2017) |
| *INTEGRAL* IBIS/PICsIT | Aug 18 12:45:10—Aug 23 03:22:34 UTC | 1–5.7 days | $0.9 \times 10^{-10}$ | 468–572 keV | Savchenko et al. (2017) |
| *INTEGRAL* IBIS/PICsIT | Aug 18 12:45:10—Aug 23 03:22:34 UTC | 1–5.7 days | $4.4 \times 10^{-10}$ | 572–1196 keV | Savchenko et al. (2017) |
| *INTEGRAL* SPI | Aug 18 12:45:10—Aug 23 03:22:34 UTC | 1–5.7 days | $2.4 \times 10^{-10}$ | 300–500 keV | Savchenko et al. (2017) |
| *INTEGRAL* SPI | Aug 18 12:45:10—Aug 23 03:22:34 UTC | 1–5.7 days | $7.0 \times 10^{-10}$ | 500–1000 keV | Savchenko et al. (2017) |
| *INTEGRAL* SPI | Aug 18 12:45:10—Aug 23 03:22:34 UTC | 1–5.7 days | $1.5 \times 10^{-9}$ | 1000–2000 keV | Savchenko et al. (2017) |
| *INTEGRAL* SPI | Aug 18 12:45:10—Aug 23 03:22:34 UTC | 1–5.7 days | $2.9 \times 10^{-9}$ | 2000–4000 keV | Savchenko et al. (2017) |
| H.E.S.S. | Aug 18 17:55 UTC | 1.22 days | $3.3 \times 10^{-12}$ | 0.27–3.27 TeV | H. Abdalla et al. (H.E.S.S. Collaboration) (2017, in preparation) |
| H.E.S.S. | Aug 19 17:56 UTC | 2.22 days | $1.0 \times 10^{-12}$ | 0.31–2.88 TeV | H. Abdalla et al. (H.E.S.S. Collaboration) (2017, in preparation) |
| H.E.S.S. | Aug 21 + Aug 22 18:15 UTC | 4.23 days + 5.23 days | $2.9 \times 10^{-12}$ | 0.50–5.96 TeV | H. Abdalla et al. (H.E.S.S. Collaboration) (2017, in preparation) |

**Note.**
[a] Assuming no shielding by the structures of ISS.

Earth Occultation technique (Wilson-Hodge et al. 2012), *Fermi*-GBM placed limits on persistent emission for the 48 hr period centered at the *Fermi*-GBM trigger time over the 90% credible region of the GW170817 localization. Using the offline targeted search for transient signals (Blackburn et al. 2015), *Fermi*-GBM also set constraining upper limits on precursor and extended emission associated with GRB 170817A (Goldstein et al. 2017b). *INTEGRAL* (Winkler et al. 2003) continued uninterrupted observations after GRB 170817A for 10 hr. Using the PiCSIT (Labanti et al. 2003) and SPI-ACS detectors, the presence of a steady source 10 times weaker than the prompt emission was excluded (Savchenko et al. 2017).

The High Energy telescope on board *Insight*-HXMT monitored the entire GW170817 skymap from T0 − 650 s to T0 + 450 s but, due to the weak and soft nature of GRB 170817A, did not detect any significant excess at T0 (Liao et al. 2017). Upper limits from 0.2–5 MeV for GRB 170817A and other emission episodes are reported in Li et al. (2017).

The Calorimetric Electron Telescope (CALET) Gamma-ray Burst Monitor (CGBM) found no significant excess around T0. Upper limits may be affected due to the location of SSS17a/AT 2017gfo being covered by the large structure of the International Space Station at the time of GRB 170817A (Nakahira et al. 2017). *AstroSat* CZTI (Singh et al. 2014; Bhalerao et al. 2017) reported upper limits for the 100 s interval centered on T0 (Balasubramanian et al. 2017); the position of SSS17a/AT 2017gfo was occulted by the Earth, however, at the time of the trigger.

For the AstroRivelatore Gamma a Immagini Leggero (AGILE) satellite (Tavani et al. 2009) the first exposure of the GW170817 localization region by the Gamma Ray Imaging Detector (GRID), which was occulted by the Earth at the time of GRB 170817A, started at T0 + 935 s. The GRID observed the field before and after T0, typically with 150 s exposures. No gamma-ray source was detected above $3\sigma$ in the energy range 30 MeV–30 GeV (V. Verrecchia et al. 2017, in preparation).

At the time of the trigger, *Fermi* was entering the South Atlantic Anomaly (SAA) and the Large Area Telescope (LAT) was not collecting science data (*Fermi*-GBM uses different SAA boundaries and was still observing). *Fermi*-LAT resumed data taking at roughly T0 + 1153 s, when 100% of the low-latency GW170817 skymap (LIGO Scientific Collaboration & Virgo Collaboration et al. 2017b) was in the field of view for ∼1000 s. No significant source of high-energy emission was detected. Additional searches over different timescales were performed for the entire time span of LAT data, and no significant excess was detected at the position of SSS17a/AT 2017gfo (Kocevski et al. 2017).

The High Energy Stereoscopic System (H.E.S.S.) array of imaging atmospheric Cherenkov telescopes observed from August 17 18:00 UTC with three pointing positions. The first, at T0 + 5.3 hr, covered SSS17a/AT 2017gfo. Observations repeated the following nights until the location moved outside the visibility window, with the last pointing performed on August 22 18:15 UTC. A preliminary analysis with an energy threshold of ∼500 GeV revealed no significant gamma-ray emission (de Naurois et al. 2017), confirmed by the final, offline analysis (see H. Abdalla et al. (H.E.S.S. Collaboration) 2017, in preparation, for more results).

For the High-Altitude Water Cherenkov (HAWC) Observatory (Abeysekara et al. 2017) the LIGO-Virgo localization region first became visible on August 17 between 19:57 and 23:25 UTC. SSS17a/AT 2017gfo was observed for 2.03 hr starting at 20:53 UTC. Upper limits from HAWC for energies >40 TeV assuming an $E^{-2.5}$ spectrum are reported in Martinez-Castellanos et al. (2017).

*INTEGRAL* (3 keV–8 MeV) carried out follow-up observations of the LIGO-Virgo localization region, centered on the optical counterpart, starting 24 hr after the event and spanning 4.7 days. Hard X-ray emission is mostly constrained by IBIS (Ubertini et al. 2003), while above 500 keV SPI (Vedrenne et al. 2003) is more sensitive. Besides the steady flux limits reported in Table 3, these observations exclude delayed bursting activity at the level of giant magnetar flares. No gamma-ray lines from a kilonova or $e^{+/-}$ pair plasma annihilation were detected (see Savchenko et al. 2017).

### 3.3. Discovery of the X-Ray Counterpart

While the UV, optical, and IR observations mapped the emission from the sub-relativistic ejecta, X-ray observations probed a different physical regime. X-ray observations of GRB afterglows are important to constrain the geometry of the outflow, its energy output, and the orientation of the system with respect to the observers' line of sight.

The earliest limits at X-ray wavelengths were provided by the Gas Slit Camera (GSC) of the *Monitor of All-Sky X-ray Image* (*MAXI*; Matsuoka et al. 2009). Due to an unfavorable sky position, the location of GW170817 was not observed by *MAXI* until August 17 17:21 UTC (T0 + 0.19 days). No X-ray emission was detected at this time to a limiting flux of $8.6 \times 10^{-9}$ erg cm$^{-2}$ s$^{-1}$ (2–10 keV; Sugita et al. 2017; S. Sugita 2017, in preparation). *MAXI* obtained three more scans over the location with no detections before the more sensitive pointed observations began.

In addition, the Super-*AGILE* detector (Feroci et al. 2007) on board the *AGILE* mission (Tavani et al. 2009) observed the location of GW170817 starting at August 18 01:16:34.84 UTC (T0 + 0.53 days). No X-ray source was detected at the location of GW170817, with a $3\sigma$ upper limit of $3.0 \times 10^{-9}$ erg cm$^{-2}$ s$^{-1}$ (18–60 keV; V. Verrecchia et al. 2017, in preparation).

The first pointed X-ray observations of GW170817 were obtained by the X-Ray Telescope (Burrows et al. 2005) on the *Swift* satellite (Gehrels 2004) and the *NUclear Spectroscopic Telescope ARray* (*NuSTAR*; Harrison et al. 2013), beginning at T0 + 0.62 days and T0 + 0.70 days, respectively. No X-ray emission was detected at the location of GW170817 to limiting fluxes of $2.7 \times 10^{-13}$ erg cm$^{-2}$ s$^{-1}$ (0.3–10.0 keV; Evans et al. 2017a, 2017b) and $2.6 \times 10^{-14}$ erg cm$^{-2}$ s$^{-1}$ (3.0–10.0 keV; Evans et al. 2017a, 2017b). *Swift* continued to monitor the field, and after stacking several epochs of observations, a weak X-ray source was detected near the location of GW170817 at a flux of $2.6 \times 10^{-14}$ erg cm$^{-2}$ s$^{-1}$ (Evans et al. 2017c).

*INTEGRAL* (see Section 3.2) performed pointed follow-up observations from one to about six days after the trigger. The X-ray monitor JEM-X (Lund et al. 2003) constrained the average X-ray luminosity at the location of the optical transient to be $<2 \times 10^{-11}$ erg cm$^{-2}$ s$^{-1}$ (3–10.0 keV) and $<7 \times 10^{-12}$ erg cm$^{-2}$ s$^{-1}$ (10–25 keV; Savchenko et al. 2017).

*Chandra* obtained a series of observations of GW170817 beginning at August 19 17:10 UTC (T0 + 2.2 days) and continuing until the emission from NGC 4993 became unobservable because of SSS17a/AT 2017gfo's proximity to the Sun (Fong et al. 2017; Haggard et al. 2017b; Margutti et al. 2017a; Troja et al. 2017c, 2017e). Two days post-trigger, Margutti et al. (2017a) reported an X-ray non-detection for SSS17a/AT 2017gfo in a $\simeq$25 ks *Chandra* exposure,[964] along with the detection of an extended X-ray source whose position was consistent with the host NGC 4993 (Margutti et al. 2017b). Refined astrometry from subsequent *Swift* observations confirmed that the previously reported candidate was indeed associated with the host nucleus (Evans et al. 2017a, 2017b).

Nine days post-trigger, Troja et al. (2017c) reported the discovery of the X-ray counterpart with *Chandra*. In a 50 ks exposure observation, they detected significant X-ray emission at the same position of the optical/IR counterpart (Troja et al. 2017a; top right panel in Figure 2)[965]. Fifteen days post-trigger, two additional 50 ks *Chandra* observations were made, which confirmed the continued presence of X-ray emission. Based on the first of these two observations[966,967]: Fong et al. (2017) reported the detection of the X-ray counterpart and the presence of an additional X-ray point source in the near vicinity (Margutti et al. 2017b), and Troja et al. (2017e) reported a flux of $4.5 \times 10^{-15}$ erg cm$^{-2}$ s$^{-1}$ for the X-ray counterpart. One day later, Haggard et al. (2017b) reported another deep observation showing continued distinct X-ray emission coincident with SSS17a/AT 2017gfo, NGC 4993, and the additional point source (Haggard et al. 2017a, 2017b).[10]

Neither *Swift* nor *Chandra* can currently observe GW170817 because it is too close to the Sun ($<47°$ for *Swift*, $<46°$ for *Chandra*). Hence, until early 2017 December, *NuSTAR* is the only sensitive X-ray observatory that can continue to observe the location of GW170817.

All X-ray observations of GW170817 are summarized in Table 4.

### 3.4. Discovery of the Radio Counterpart

Radio emission traces fast-moving ejecta from a neutron star coalescence, providing information on the energetics of the explosion, the geometry of the ejecta, as well as the environment of the merger. The spectral and temporal evolution of such emission, coupled with X-ray observations, are likely to constrain several proposed models (see, e.g., Nakar & Piran 2011; Piran et al. 2013; Hotokezaka & Piran 2015; Hotokezaka et al. 2016; Gottlieb et al. 2017).

Prior to detection of SSS17a/AT 2017gfo, a blind radio survey of cataloged galaxies in the gravitational-wave localization volume commenced with the Australia Telescope Compact Array (ATCA; Wilson et al. 2011), and observed the merger events' location on 2017 August 18 at 01:46 UTC (Kaplan et al. 2017a). In addition, the Long Wavelength Array 1 (LWA1; Ellingson et al. 2013) followed up the gravitational-wave localization with observations at $t_c + 6.5$ hr, then on 2017 August 23 and 30 (Callister et al. 2017a; Callister et al. 2017b) using four beams (one centered on NGC 4993, one off-center, and two off NGC 4993). These observations set 3$\sigma$ upper limits for the appearance of a radio source in the beam centered on NGC 4993, about 8 hours after the GW event, as $\sim$200 Jy at 25 MHz and $\sim$100 Jy at 45 MHz.

The first reported radio observations of the optical transient SSS17a/AT 2017gfo's location occurred on August 18 at 02:09:00 UTC (T0+13.5 hr) with the Karl G. Jansky Very Large Array (VLA) by Alexander et al. (2017d).[968] Initially attributed to the optical transient, this radio source was later established to be an AGN in the nucleus of the host galaxy, NGC 4993 (Alexander et al. 2017e, 2017c). Subsequent observations with several radio facilities spanning a wide range of radio and millimeter frequencies continued to detect the AGN, but did not reveal radio emission at the position of the transient (Alexander et al. 2017f; Bannister et al. 2017b; Corsi et al. 2017a, 2017b, 2017c; De et al. 2017a, 2017b; Kaplan et al. 2017a; Lynch et al. 2017a, 2017b, 2017c; Mooley et al. 2017a; Resmi et al. 2017).

The first radio counterpart detection consistent with the *HST* position (refined by *Gaia* astrometry) of SSS17a/AT 2017gfo (Adams et al. 2017) was obtained with the VLA on 2017 September 2 and 3 at two different frequencies ($\approx$3 GHz and $\approx$6 GHz) via two independent observations: the Jansky VLA mapping of Gravitational Wave bursts as Afterglows in Radio (JAGWAR[969]; Mooley et al. 2017b) and VLA/16A-206[970] (Corsi et al. 2017d). Marginal evidence for radio excess emission at the location of SSS17a/AT 2017gfo was also confirmed in ATCA images taken on September 5 at similar radio frequencies ($\approx$7.25 GHz; Murphy et al. 2017). Subsequent repeated detections spanning multiple frequencies have confirmed an evolving transient (Hallinan et al. 2017a, 2017b; Corsi et al. 2017d; Mooley et al. 2017b). Independent observations carried out on 2017 September 5 with the same frequency and exposure time used by Corsi et al. (2017d) did not detect any emission to a 5$\sigma$ limit[971] (Alexander et al. 2017a), but this group also subsequently detected the radio counterpart on 2017 September 25 (Alexander et al. 2017b, 2017c).

SSS17a/AT 2017gfo, as well as other parts of the initial gravitational-wave localization area, were and are also being continuously monitored at a multitude of different frequencies with the Atacama Large Millimeter/submillimeter Array (ALMA; Wootten & Thompson 2009; Schulze et al. 2017; Kim et al. 2017, in preparation; Alexander et al. 2017c; Williams et al. 2017a), the Australian Square Kilometre Array Pathfinder (ASKAP; Johnston et al. 2007), ASKAP-Fast Radio Burst (Bannister et al. 2017a, 2017c), ATCA, Effelsberg-100 m (Barr et al. 2013), the Giant Metrewave Radio Telescope (GMRT; Swarup et al. 1991), the Low-Frequency Array (LOFAR; van Haarlem et al. 2013), the Long Wavelength Array (LWA1), MeerKAT (Goedhart et al. 2017a), the Murchison Widefield Array (MWA; Tingay et al. 2013), Parkes-64 m (SUPERB; Bailes et al. 2017a; Keane et al. 2017), Sardinia Radio Telescope (SRT; Prandoni et al. 2017), VLA, VLA Low Band Ionosphere and Transient Experiment (VLITE; Clarke & Kassim 2016), and also using the very long baseline interferometry (VLBI) technique with e-MERLIN (Moldon et al. 2017a, 2017b), the European VLBI Network (Paragi et al. 2017a, 2017b), and the Very Long Baseline Array (VLBA; Deller et al. 2017a, 2017b). The latter have the potential to resolve (mildly) relativistic ejecta on a timescale of months.

Table 5 summarizes the radio observations of GW170817.

[964] *Chandra* OBSID-18955, PI: Fong.
[965] *Chandra* OBSID-19294, PI: Troja.
[966] *Chandra* OBSID-20728, PI: Troja (Director's Discretionary Time observation distributed also to Haggard, Fong, and Margutti).
[967] *Chandra* OBSID-18988, PI: Haggard.
[968] VLA/17A-218, PI: Fong.
[969] VLA/17A-374, PI: Mooley.
[970] VLA/16A-206, PI: Corsi.
[971] VLA/17A-231, PI: Alexander.

**Table 4**
X-Ray Monitoring and Evolution of GW170817

| Observatory | UT Date (Start) | Time since GW trigger (days) | $f_x$ ( erg cm$^{-2}$ s$^{-1}$ ) | $L_x$ (erg s$^{-1}$) | Energy (keV) | GCN/Reference |
|---|---|---|---|---|---|---|
| *MAXI* | Aug 17 17:21:54 UTC | 0.19 | $<8.6 \times 10^{-9}$ | $<1.65 \times 10^{45}$ | 2–10 | S. Sugita et al. (2017, in preparation) |
| *MAXI* | Aug 17 18:54:27 UTC | 0.26 | $<7.7 \times 10^{-8}$ | $<1.47 \times 10^{46}$ | 2–10 | S. Sugita et al. (2017, in preparation) |
| *MAXI* | Aug 18 00:44:59 UTC | 0.50 | $<4.2 \times 10^{-9}$ | $<8.0 \times 10^{44}$ | 2–10 | S. Sugita et al. (2017, in preparation) |
| Super-*AGILE* | Aug 18 01:16:34 UTC | 0.53 | $<3.0 \times 10^{-9}$ | $<5.4 \times 10^{44}$ | 18–60 | V. Verrecchia et al. (2017, in preparation) |
| *MAXI* | Aug 18 02:18:08 UTC | 0.57 | $<2.2 \times 10^{-9}$ | $<4.2 \times 10^{44}$ | 2–10 | S. Sugita et al. (2017, in preparation) |
| *Swift*-XRT | Aug 18 03:34:33 UTC | 0.62 | $<2.74 \times 10^{-13}$ | $<5.25 \times 10^{40}$ | 0.3–10 | Evans et al. (2017b) |
| *NuSTAR* | Aug 18 05:25 UTC | 0.7 | $<2.62 \times 10^{-14}$ | $<5.01 \times 10^{39}$ | 3–10 | Evans et al. (2017b) |
| *Swift*-XRT | Aug 18 12:11:49 UTC | 0.98 | $<2.62 \times 10^{-12}$ | $<5.01 \times 10^{41}$ | 0.3–10 | Evans et al. (2017b) |
| *INTEGRAL* JEM-X | Aug 18 12:45:10 UTC | 1–5.7 | $<1.9 \times 10^{-11}$ | $<3.6 \times 10^{42}$ | 3–10 | Savchenko et al. (2017) |
| *INTEGRAL* JEM-X | Aug 18 12:45:10 UTC | 1–5.7 | $<7.0 \times 10^{-12}$ | $<1.3 \times 10^{42}$ | 10–25 | Savchenko et al. (2017) |
| *Swift*-XRT | Aug 18 13:29:43 UTC | 1.03 | $<1.77 \times 10^{-13}$ | $<3.39 \times 10^{40}$ | 0.3–10 | Evans et al. (2017b) |
| *Swift*-XRT | Aug 19 00:18:22 UTC | 1.48 | $<1.31 \times 10^{-13}$ | $<2.51 \times 10^{40}$ | 0.3–10 | Evans et al. (2017b) |
| *Chandra* | Aug 19 17:10:09 UTC | 2.20 | non-detection | … | 0.3–10 | Margutti et al. (2017a) |
| *Swift*-XRT | Aug 19 13:24:05 UTC | 2.03 | $<1.02 \times 10^{-13}$ | $<1.95 \times 10^{40}$ | 0.3–10 | Evans et al. (2017b) |
| *Swift*-XRT | Aug 19 18:30:52 UTC | 2.24 | $<1.34 \times 10^{-13}$ | $<2.57 \times 10^{40}$ | 0.3–10 | Evans et al. (2017b) |
| *Swift*-XRT | Aug 20 03:24:44 UTC | 2.61 | $<1.41 \times 10^{-13}$ | $<2.69 \times 10^{40}$ | 0.3–10 | Evans et al. (2017b) |
| *Swift*-XRT | Aug 20 08:28:05 UTC | 2.82 | $<3.87 \times 10^{-14}$ | $<7.41 \times 10^{39}$ | 0.3–10 | Evans et al. (2017b) |
| *Swift*-XRT | Aug 21 01:43:44 UTC | 3.54 | $<6.73 \times 10^{-14}$ | $<1.29 \times 10^{40}$ | 0.3–10 | Evans et al. (2017b) |
| *NuSTAR* | Aug 21 20:45:00 UTC | 4.3 | $<2.08 \times 10^{-14}$ | $<3.98 \times 10^{39}$ | 3–10 | Evans et al. (2017b) |
| *Swift*-XRT | Aug 22 00:05:57 UTC | 4.48 | $<6.28 \times 10^{-14}$ | $<1.20 \times 10^{40}$ | 0.3–10 | Evans et al. (2017b) |
| *Swift*-XRT | Aug 23 06:22:57 UTC | 5.74 | $<6.89 \times 10^{-14}$ | $<1.32 \times 10^{40}$ | 0.3–10 | Evans et al. (2017b) |
| *Swift*-XRT | Aug 23 23:59:57 UTC | 6.47 | $<7.21 \times 10^{-14}$ | $<1.38 \times 10^{40}$ | 0.3–10 | Evans et al. (2017b) |
| *Chandra* | Aug 26 10:33:50 UTC | 8.9 | Detection | … | 0.5–8.0 | Troja et al. (2017c, 2017a) |
| *Swift*-XRT | Aug 26 23:59:57 UTC | 9.47 | $<8.67 \times 10^{-14}$ | $<1.66 \times 10^{40}$ | 0.3–10 | Evans et al. (2017b) |
| *Swift*-XRT | Aug 28 10:46:17 UTC | 10.92 | $<1.41 \times 10^{-13}$ | $<2.69 \times 10^{40}$ | 0.3–10 | Evans et al. (2017b) |
| *Swift*-XRT | Aug 29 01:04:57 UTC | 11.52 | $<6.00 \times 10^{-14}$ | $<1.15 \times 10^{40}$ | 0.3–10 | Evans et al. (2017b) |
| *Swift*-XRT | Aug 30 01:00:57 UTC | 12.51 | $<5.47 \times 10^{-14}$ | $<1.05 \times 10^{40}$ | 0.3–10 | Evans et al. (2017b) |
| *Swift*-XRT | Aug 31 02:27:52 UTC | 13.57 | $<3.87 \times 10^{-14}$ | $<7.41 \times 10^{39}$ | 0.3–10 | Evans et al. (2017b) |
| *Swift*-XRT | Sep 01 05:53:04 UTC | 14.72 | $<4.45 \times 10^{-14}$ | $<8.51 \times 10^{39}$ | 0.3–10 | Evans et al. (2017b) |
| *Chandra* | Sep 01 15:22:22 UTC | 15.1 | | … | … | Fong et al. (2017); Margutti et al. (2017b) |
| *Chandra* | Sep 01 15:22:22 UTC | 15.1 | $4.5 \times 10^{-15}$ | $9 \times 10^{38}$ | 0.5–8.0 | Troja et al. (2017e, 2017a) |
| *Chandra* | Sep 02 15:22:22 UTC | 15.1 | $3.5 \times 10^{-15}$ | $2.7 \times 10^{38}$ | 0.3–10 | Haggard et al. (2017b, 2017a) |
| *Chandra* | Sep 02 00:00:00 UTC | 16.1 | $3.8 \times 10^{-15}$ | $3.0 \times 10^{38}$ | 0.3–10 | Haggard et al. (2017b, 2017a) |
| *Swift*-XRT | Sep 02 08:40:56 UTC | 15.83 | $<1.51 \times 10^{-13}$ | $<2.88 \times 10^{40}$ | 0.3–10 | Evans et al. (2017b) |
| *NuSTAR* | Sep 04 17:56 UTC | 18.2 | $<6.58 \times 10^{-14}$ | $<1.26 \times 10^{40}$ | 3–10 | Evans et al. (2017b) |
| *NuSTAR* | Sep 05 14:51 UTC | 19.1 | $<4.15 \times 10^{-14}$ | $<7.94 \times 10^{39}$ | 3–10 | Evans et al. (2017b) |
| *NuSTAR* | Sep 06 17:56 UTC | 20.1 | $<3.30 \times 10^{-14}$ | $<6.31 \times 10^{39}$ | 3–10 | Evans et al. (2017b) |
| *NuSTAR* | Sep 21 11:10 UTC | 34.9 | $<1.65 \times 10^{-14}$ | $<3.16 \times 10^{39}$ | 3–10 | Evans et al. (2017b) |

**Table 5**
Radio Monitoring and Evolution of GW170817

| Telescope | UT Date | Time since GW Trigger (days) | Central Frequency (GHz) | Bandwidth (GHz) | Flux ($\mu$ Jy), $3\sigma$ | GCN/Reference |
|---|---|---|---|---|---|---|
| LWA1 | Aug 17 13:09:51 UTC | 0.02 | 0.02585 | 0.020 | ··· | Callister et al. (2017a) |
| LWA1 | Aug 17 13:09:51 UTC | 0.02 | 0.04545 | 0.020 | ··· | Callister et al. (2017a) |
| LWA1 | Aug 17 19:15:00 UTC | 0.27 | 0.02585 | 0.020 | $<2 \times 10^8$ | Callister et al. (2017a) |
| LWA1 | Aug 17 19:15:00 UTC | 0.27 | 0.04545 | 0.020 | $<1 \times 10^8$ | Callister et al. (2017a) |
| VLBA | Aug 17 19:58:00 UTC | 0.30 | 8.7 | 0.26 | | Deller et al. (2017a) |
| VLA | Aug 18 02:18:00 UTC | 0.57 | 10.0 | ··· | | Alexander et al. (2017d, 2017e) |
| ATCA | Aug 18 01:00:00 UTC | 1 | 8.5 | 2.049 | <120 | Bannister et al. (2017d) Kaplan et al. (2017a) Hallinan et al. (2017a) |
| ATCA | Aug 18 01:00:00 UTC | 1 | 10.5 | 2.049 | <150 | Bannister et al. (2017d) Kaplan et al. (2017a) Hallinan et al. (2017a) |
| ATCA | Aug 18 01:00:00 UTC | 1 | 16.7 | 2.049 | <130 | Kaplan et al. (2017a) Hallinan et al. (2017a) |
| ATCA | Aug 18 01:00:00 UTC | 1 | 21.2 | 2.049 | <140 | Kaplan et al. (2017a) Hallinan et al. (2017a) |
| VLITE | Aug 18 22:23:31 UTC | 1.44 | 0.3387 | 0.034 | <34800 | Hallinan et al. (2017a) |
| ASKAP | Aug 18 04:05:35 UTC | 0.67 | 1.34 | 0.19 | | Bannister et al. (2017e, 2017c) |
| MWA | Aug 18 07:07:50 UTC | 1 | 0. 185 | 0.03 | <51 000 | Kaplan et al. (2017b) |
| ASKAP | Aug 18 08:57:33 UTC | 0.86 | 1.34 | 0.19 | | Bannister et al. (2017e, 2017c) |
| VLA | Aug 18 22:04:57 UTC | 1 | 10.0 | 3.8 | <17.0 | Alexander et al. (2017f) |
| ALMA | Aug 18 22:50:40 UTC | 1.4 | 338.5 | 7.5 | ··· | Schulze et al. (2017) |
| GMRT | Aug 18 11:00:00 UTC | 1 | 10.0 | 0.032 | <195 | De et al. (2017a) Hallinan et al. (2017a) |
| Parkes | Aug 18 00:00:00 UTC | 1.38 | 1.34 | 0.34 | $<1.4 \times 10^6$ | Bailes et al. (2017a) |
| Parkes | Aug 18 00:00:00 UTC | 1.46 | 1.34 | 0.34 | $<1.4 \times 10^6$ | Bailes et al. (2017a) |
| ASKAP | Aug 19 02:08:00 UTC | 1.58 | 1.34 | 0.19 | | Bannister et al. (2017e) |
| ASKAP | Aug 19 05:34:33 UTC | 2 | 1.345 | ··· | <900 | Dobie et al. (2017a) |
| VLA | Aug 19 22:01:48 UTC | 2 | 6.0 | 4 | <22 | Corsi et al. (2017a) |
| VLA | Aug 19 22:01:48 UTC | 2 | 6.0 | 4 | <22 | Corsi et al. (2017a) |
| VLITE | Aug 19 22:29:29 UTC | 2.44 | 0.3387 | 0.034 | <28800 | Hallinan et al. (2017a) |
| VLA | Aug 19 22:30:10 UTC | 2.42 | 15.0 | 6 | <22 | Corsi et al. (2017e) Hallinan et al. (2017a) |
| VLA | Aug 19 23:04:06 UTC | 2.44 | 10.0 | 4 | <17 | Corsi et al. (2017b) Hallinan et al. (2017a) |
| VLA | Aug 19 23:33:30 UTC | 2.46 | 6.0 | ··· | <20 | Corsi et al. (2017a) Hallinan et al. (2017a) |
| ALMA | Aug 19 22:31:43 UTC | 2 | 97.5 | ··· | <50 | Williams et al. (2017a) |
| Parkes | Aug 20 00:00:00 UTC | 3.17 | 1.34 | 0.34 | $<1.4 \times 10^6$ | Bailes et al. (2017a) |
| Parkes | Aug 20 00:00:00 UTC | 3.21 | 1.34 | 0.34 | $<1.4 \times 10^6$ | Bailes et al. (2017a) |
| VLITE | Aug 20 20.49:36 UTC | 3.34 | 0.3387 | 0.034 | <44700 | Hallinan et al. (2017a) |
| VLA | Aug 20 00:01:24 UTC | 3 | 9.7 | 4 | <18 | Corsi et al. (2017b) |
| GMRT | Aug 20 08:00:00 UTC | 3 | 0.4 | 0.2 | <780 | De et al. (2017b) |
| GMRT | Aug 20 08:00:00 UTC | 3 | 1.2 | 0.4 | <98 | De et al. (2017b) |
| VLA | Aug 20 21:07:00 UTC | 3 | 6.2 | 4 | <19 | Corsi et al. (2017c) |
| VLA/JAGWAR | Aug 20 22:20:00 UTC | 3 | 3.0 | ··· | <32 | Mooley et al. (2017a) |
| ATCA | Aug 20 23:31:03 UTC | 3 | 8.5 | 2.049 | <20 | Lynch et al. (2017a) |

**Table 5**
(Continued)

| Telescope | UT Date | Time since GW Trigger (days) | Central Frequency (GHz) | Bandwidth (GHz) | Flux ($\mu$ Jy), $3\sigma$ | GCN/Reference |
|---|---|---|---|---|---|---|
| ATCA | Aug 20 23:31:03 UTC | 3 | 10.5 | 2.049 | <135 | Lynch et al. (2017a) |
| ALMA | Aug 20 22:40:16 UTC | 3 | 338.5 | 7.5 | ⋯ | Schulze et al. (2017) |
| VLBA | Aug 20 21:36:00 UTC | 3 | 8.7 | ⋯ | <48 | Deller et al. (2017b) |
| ALMA | Aug 21 20:58:51 UTC | 4.3 | 338.5 | 7.5 | ⋯ | Schulze et al. (2017) |
| VLA | Aug 22 23:50:18 UTC | 5.48 | 10.0 | ⋯ | | Alexander et al. (2017c) |
| e-MERLIN | Aug 23 12:00:00 UTC | 6 | 5.0 | 0.512 | <108 | Moldon et al. (2017a) |
| e-MERLIN | Aug 24 12:00:00 UTC | 7 | 5.0 | 0.512 | <96 | Moldon et al. (2017a) |
| LWA1 | Aug 24 19:50:00 UTC | 7 | 0.02585 | 0.016 | | Callister et al. (2017b) |
| LWA1 | Aug 24 19:50:00 UTC | 7 | 0.04545 | 0.016 | | Callister et al. (2017b) |
| e-MERLIN | Aug 25 12:00:00 UTC | 8 | 5.0 | 512 | <96 | Moldon et al. (2017a) |
| VLITE | Aug 25 20:38:22 UTC | 8.37 | 0.3387 | 0.034 | <37500 | Hallinan et al. (2017a) |
| GMRT | Aug 25 09:30:00 UTC | 7.9 | 1.39 | 0.032 | <130 | Resmi et al. (2017) |
| VLA | Aug 25 19:15:12 UTC | 8.29 | 10.0 | ⋯ | | Alexander et al. (2017c) |
| ALMA | Aug 25 22:35:17 UTC | 8.4 | 338.5 | 7.5 | ⋯ | Schulze et al. (2017) |
| MeerKAT | Aug 26 08:43:00 UTC | 10 | 1.48 | 0.22 | <70 | Goedhart et al. (2017a) |
| ALMA | Aug 26 22:49:25 UTC | 9.43 | 97.5 | ⋯ | | Williams et al. (2017a) |
| ALMA | Aug 26 22:58:41 UTC | 9.4 | 338.5 | 7.5 | ⋯ | Schulze et al. (2017); S. Kim et al. (2017, in preparation) |
| EVN | Aug 26 12:15:00 UTC | 9 | 5.0 | 0.256 | <96 | Paragi et al. (2017a) |
| e-MERLIN | Aug 26 12:00:00 UTC | 9 | 5.0 | 0.512 | <114 | Moldon et al. (2017a) |
| e-MERLIN | Aug 27 12:00:00 UTC | 10 | 5.0 | 0.512 | <90 | Moldon et al. (2017a) |
| ATCA | Aug 27 23:26:25 UTC | 10 | 8.5 | 2. 049 | <54 | Lynch et al. (2017b) |
| ATCA | Aug 27 23:26:25 UTC | 10 | 10.5 | 2.049 | <39 | Lynch et al. (2017b) |
| e-MERLIN | Aug 28 12:00:00 UTC | 11 | 5.0 | 0.512 | <90 | Moldon et al. (2017a) |
| VLITE | Aug 30 23:10:28 UTC | 13.45 | 0.3387 | 0.034 | <20400 | Hallinan et al. (2017a) |
| LWA1 | Aug 30 19:50:00 UTC | 13 | 0.02585 | 0.016 | | Callister et al. (2017) |
| LWA1 | Aug 30 19:50:00 UTC | 13 | 0.04545 | 0.016 | | Callister et al. (2017) |
| VLA | Aug 30 22:09:24 UTC | 13.41 | 10.0 | ⋯ | | Alexander et al. (2017c) |
| e-MERLIN | Aug 31 13:00:00 UTC | 14 | 5.0 | 0.512 | <109 | Moldon et al. (2017b) |
| VLITE | Sep 1 20:44:59 UTC | 15.37 | 0.3387 | 0.034 | <11400 | Hallinan et al. (2017a) |
| ATCA | Sep 1 12:00:00 UTC | 15 | 16.7 | ⋯ | <50 | Troja et al. (2017f) |
| ATCA | Sep 1 12:00:00 UTC | 15 | 21.2 | ⋯ | <50 | Troja et al. (2017f) |
| ATCA | Sep 1 12:00:00 UTC | 15 | 43.0 | ⋯ | <90 | Troja et al. (2017f) |
| ATCA | Sep 1 12:00:00 UTC | 15 | 45.0 | ⋯ | <90 | Troja et al. (2017f) |
| e-MERLIN | Sep 1 13:00:00 UTC | 15 | 5.0 | 0.512 | <114 | Moldon et al. (2017b) |
| ALMA | Sep 120:22:05 UTC | 15.33 | 97.5 | ⋯ | | Alexander et al. (2017c) |
| VLA/JAGWAR | Sep 2 00:00:00 UTC | 16 | 3.0 | | Detection | Mooley et al. (2017b); Hallinan et al. (2017a) |
| e-MERLIN | Sep 2 13:00:00 UTC | 16 | 5.0 | 0.512 | <144 | Moldon et al. (2017b) |
| VLITE | Sep 2 18:51:34 UTC | 16.36 | 0.3387 | 0.034 | <11700 | Hallinan et al. (2017a) |
| e-MERLIN | Sep 3 13:00:00 UTC | 17 | 5.0 | 0.512 | <166 | Moldon et al. (2017b) |
| VLA | Sep 3 23:30:00 UTC | 17 | 6.0 | | Detection | Corsi et al. (2017d); Hallinan et al. (2017a) |
| VLITE | Sep 3 20:08:05 UTC | 17.40 | 0.3387 | 0.034 | <6900 | Hallinan et al. (2017a) |
| e-MERLIN | Sep 4 13:00:00 UTC | 18 | 5.0 | 0.512 | <147 | Moldon et al. (2017b) |
| ATCA | Sep 5 10:03:04 UTC | 19 | 7.25 | | Detection | Murphy et al. (2017) |
| e-MERLIN | Sep 5 13:00:00 UTC | 19 | 5.0 | 0.512 | <162 | Moldon et al. (2017b) |
| VLA | Sep 5 22:12:00 UTC | 19.47 | 6.0 | ⋯ | | Alexander et al. (2017a) |
| VLA | Sep 5 23:26:06 UTC | 19.43 | 10.0 | ⋯ | | Alexander et al. (2017c) |
| MeerKAT | Sep 6 03:22:00 UTC | 20 | 1.48 | 0.22 | <75 | Goedhart et al. (2017a) |

**Table 5**
(Continued)

| Telescope | UT Date | Time since GW Trigger (days) | Central Frequency (GHz) | Bandwidth (GHz) | Flux ($\mu$ Jy), $3\sigma$ | GCN/Reference |
|---|---|---|---|---|---|---|
| VLITE | Sep 7 19:09:43 UTC | 21.36 | 0.3387 | 0.034 | <8100 | Hallinan et al. (2017a) |
| SRT | Sep 7 10:41:00 UTC | 20.92 | 7.2 | 0.68 | <1200 | Aresu et al. (2017) |
| ATCA | Sep 8 12:00:00 UTC | 22 | 17.0 | ⋯ | <35 | Wieringa et al. (2017) |
| ATCA | Sep 8 12:00:00 UTC | 22 | 21.0 | ⋯ | <35 | Wieringa et al. (2017) |
| SRT | Sep 8 11:00:00 UTC | 21.93 | 7.2 | 0.68 | <1500 | Aresu et al. (2017) |
| VLITE | Sep 8 19:05:35 UTC | 22.37 | 0.3387 | 0.034 | <6300 | Hallinan et al. (2017a) |
| SRT | Sep 9 10:37:00 UTC | 22.92 | 7.2 | 0.68 | <1800 | Aresu et al. (2017) |
| VLITE | Sep 9 18:52:45 UTC | 23.36 | 0.3387 | 0.034 | <4800 | Hallinan et al. (2017a) |
| GMRT | Sep 9 11:30:00 UTC | 23.0 | 1.39 | 0.032 | ⋯ | Resmi et al. (2017), S. Kim et al. (2017, in preparation) |
| e-MERLIN | Sep 10 13:00:00 UTC | 24 | 5.0 | 0.512 | <126 | Moldon et al. (2017b) |
| Effelsberg | Sep 10 13:10 UTC | 24 | 5 | 2 | <30000 | Kramer et al. (2017) |
| Effelsberg | Sep 10 13:35 UTC | 24 | 32 | 2 | <90000 | Kramer et al. (2017) |
| VLITE | Sep 10 18:36:48 UTC | 24.35 | 0.3387 | 0.034 | <6600 | Hallinan et al. (2017a) |
| e-MERLIN | Sep 11 13:00:00 UTC | 25 | 5.0 | 0.512 | <151 | Moldon et al. (2017b) |
| e-MERLIN | Sep 12 13:00:00 UTC | 26 | 5.0 | 0.512 | <113 | Moldon et al. (2017b) |
| e-MERLIN | Sep 14 13:00:00 UTC | 28 | 5.0 | 0.512 | <147 | Moldon et al. 2017b |
| e-MERLIN | Sep 15 13:00:00 UTC | 29 | 5.0 | 0.512 | <106 | Moldon et al. 2017b |
| GMRT | Sep 16 07:30:00 UTC | 29.8 | 1.39 | 0.032 | ⋯ | Resmi et al. (2017); S. Kim et al. (2017, in preparation) |
| e-MERLIN | Sep 16 13:00:00 UTC | 30 | 5.0 | 0.512 | <118 | Moldon et al. 2017b |
| ALMA | Sep 16 20:36:21 UTC | 30.34 | 97.5 | ⋯ | | Alexander et al. (2017c) |
| MeerKAT | Sep 17 07:16:00 UTC | 31 | 1.48 | 0.22 | <60 | Goedhart et al. (2017a) |
| e-MERLIN | Sep 17 13:00:00 UTC | 31 | 5.0 | 0.512 | <111 | Moldon et al. (2017b) |
| e-MERLIN | Sep 18 13:00:00 UTC | 32 | 5.0 | 0.512 | 111 | Moldon et al. (2017b) |
| SRT | Sep 19 11:38:00 UTC | 32.96 | 7.2 | 0.68 | <1200 | Aresu et al. (2017) |
| EVN | Sep 20 10:00:00 UTC | 34 | 5.0 | 0.256 | <84 | Paragi et al. (2017b) |
| e-MERLIN | Sep 21 13:00:00 UTC | 35 | 5.0 | 0.512 | <132 | Moldon et al. (2017b) |
| e-MERLIN | Sep 22 13:00:00 UTC | 36 | 5.0 | 0.512 | <121 | Paragi et al. (2017b) |
| VLA | Sep 25 16:51:45 UTC | 39.2 | 6.0 GHz | | Detection | Alexander et al. (2017b) |

**Table 6**
Gamma-ray Coordinates Network (GCN) Notices and Circulars related to GW170817 until 2017 October 1 UTC

| Telescope | UT Date | $\Delta t$ (days) | Obs. Wavelength | References |
|---|---|---|---|---|
| Fermi/GBM | 2017 Aug 17 12:41:20 | 0.0 | gamma-ray | GCN Notice 524666471, Fermi-GBM (2017) |
| LIGO-Virgo/– | 2017 Aug 17 13:21:42 | 0.03 | gw | GCN 21505, LIGO Scientific Collaboration & Virgo Collaboration et al. (2017a) |
| Fermi/GBM | 2017 Aug 17 13:47:37 | 0.05 | gamma-ray | GCN 21506, Connaughton et al. (2017) |
| *INTEGRAL*/SPI-ACS | 2017 Aug 17 13:57:47 | 0.05 | gamma-ray | GCN 21507, Savchenko et al. (2017a) |
| IceCube/– | 2017 Aug 17 14:05:11 | 0.06 | neutrino | GCN 21508, Bartos et al. (2017a) |
| LIGO-Virgo/– | 2017 Aug 17 14:09:25 | 0.06 | gw | GCN 21509, LIGO Scientific Collaboration & Virgo Collaboration et al. (2017d) |
| LIGO-Virgo/– | 2017 Aug 17 14:38:46 | 0.08 | gw | GCN 21510, LIGO Scientific Collaboration & Virgo Collaboration et al. (2017e) |
| IceCube/– | 2017 Aug 17 14:54:58 | 0.09 | neutrino | GCN 21511, Bartos et al. (2017c) |
| LIGO-Virgo/– | 2017 Aug 17 17:54:51 | 0.22 | gw | GCN 21513, LIGO Scientific Collaboration & Virgo Collaboration et al. (2017b) |
| Astrosat/CZTI | 2017 Aug 17 18:16:42 | 0.23 | gamma-ray | GCN 21514, Balasubramanian et al. (2017) |
| IPN/– | 2017 Aug 17 18:35:12 | 0.25 | gamma-ray | GCN 21515, Svinkin et al. (2017b) |
| –/– | 2017 Aug 17 18:55:12 | 0.26 | | GCN 21516, Dalya et al. (2016) |
| Insight-HXMT/HE | 2017 Aug 17 19:35:28 | 0.29 | gamma-ray | GCN 21518, Liao et al. (2017) |
| –/– | 2017 Aug 17 20:00:07 | 0.3 | | GCN 21519, Cook et al. (2017a) |
| Fermi/GBM | 2017 Aug 17 20:00:07 | 0.3 | gamma-ray | GCN 21520, von Kienlin et al. (2017) |
| –/– | 2017 Aug 17 20:12:41 | 0.31 | | GCN 21521, Cook et al. (2017b) |
| ANTARES/– | 2017 Aug 17 20:35:31 | 0.33 | neutrino | GCN 21522, Ageron et al. (2017a) |
| *Swift*/BAT | 2017 Aug 17 21:34:36 | 0.37 | gamma-ray | GCN 21524, Barthelmy et al. (2017) |
| AGILE/MCAL | 2017 Aug 17 22:01:26 | 0.39 | gamma-ray | GCN 21525, Pilia et al. (2017) |
| AGILE/GRID | 2017 Aug 17 22:22:43 | 0.4 | gamma-ray | GCN 21526, Piano et al. (2017) |
| LIGO-Virgo/– | 2017 Aug 17 23:54:40 | 0.47 | gw | GCN 21527, LIGO Scientific Collaboration & Virgo Collaboration et al. (2017c) |
| Fermi/GBM | 2017 Aug 18 00:36:12 | 0.5 | gamma-ray | GCN 21528, Goldstein et al. (2017b) |
| Swope/– | 2017 Aug 18 01:05:23 | 0.52 | optical | GCN 21529, Coulter et al. (2017a) |
| DECam/– | 2017 Aug 18 01:15:01 | 0.52 | optical | GCN 21530, Allam et al. (2017) |
| DLT40/– | 2017 Aug 18 01:41:13 | 0.54 | optical | GCN 21531, Yang et al. (2017a) |
| REM-ROS2/– | 2017 Aug 18 02:00:40 | 0.56 | optical, IR | GCN 21532, Melandri et al. (2017a) |
| ASAS-SN/– | 2017 Aug 18 02:06:30 | 0.56 | optical | GCN 21533, Cowperthwaite et al. (2017a) |
| Fermi/LAT | 2017 Aug 18 02:09:53 | 0.56 | gamma-ray | GCN 21534, Kocevski et al. (2017) |
| –/– | 2017 Aug 18 02:48:50 | 0.59 | | GCN 21535, Cook et al. (2017c) |
| *HST*/– | 2017 Aug 18 03:01:20 | 0.6 | optical | GCN 21536, Foley et al. (2017a) |
| ATCA/– | 2017 Aug 18 04:04:00 | 0.64 | radio | GCN 21537, Bannister et al. (2017d) |
| LasCumbres/– | 2017 Aug 18 04:06:31 | 0.64 | optical | GCN 21538, Arcavi et al. (2017a) |
| DLT40/– | 2017 Aug 18 04:11:35 | 0.65 | optical | GCN 21539, Yang et al. (2017c) |
| DECam/– | 2017 Aug 18 04:44:32 | 0.67 | optical | GCN 21541, Nicholl et al. (2017a) |
| SkyMapper/– | 2017 Aug 18 04:46:27 | 0.67 | optical | GCN 21542, Moller et al. (2017) |
| LasCumbres/– | 2017 Aug 18 04:54:23 | 0.68 | optical | GCN 21543, Arcavi et al. (2017d) |
| VISTA/VIRCAM | 2017 Aug 18 05:03:48 | 0.68 | optical, IR | GCN 21544, Tanvir et al. (2017a) |
| VLA/– | 2017 Aug 18 05:07:58 | 0.69 | radio | GCN 21545, Alexander et al. (2017d) |
| MASTER/– | 2017 Aug 18 05:37:59 | 0.71 | optical | GCN 21546, Lipunov et al. (2017d) |
| *Magellan*/– | 2017 Aug 18 05:46:33 | 0.71 | optical | GCN 21547, Drout et al. (2017) |
| VLA/– | 2017 Aug 18 06:56:44 | 0.76 | radio | GCN 21548, Alexander et al. (2017e) |
| Subaru/HSC | 2017 Aug 18 07:07:07 | 0.77 | optical | GCN 21549, Yoshida et al. (2017a) |
| *Swift*/UVOT,XRT | 2017 Aug 18 07:24:04 | 0.78 | x-ray, uv | GCN 21550, Evans et al. (2017a) |
| *Magellan*/LDSS-3 | 2017 Aug 18 07:54:23 | 0.8 | optical | GCN 21551, Simon et al. (2017) |
| Gemini-South/Flamingos-2 | 2017 Aug 18 08:00:58 | 0.81 | IR | GCN 21552, Singer et al. (2017a) |
| Pan-STARRS/– | 2017 Aug 18 08:37:20 | 0.83 | optical | GCN 21553, Chambers et al. (2017a) |

**Table 6**
(Continued)

| Telescope | UT Date | Δ*t* (days) | Obs. Wavelength | References |
|---|---|---|---|---|
| HCT/HFOSC | 2017 Aug 18 09:54:21 | 0.88 | optical | GCN 21554, Pavana et al. (2017) |
| *MAXI*/GSC/– | 2017 Aug 18 10:43:45 | 0.92 | x-ray | GCN 21555, Sugita et al. (2017) |
| REM-ROS2/– | 2017 Aug 18 10:54:42 | 0.93 | optical | GCN 21556, Melandri et al. (2017b) |
| –/– | 2017 Aug 18 12:15:23 | 0.98 | | GCN 21557, Foley et al. (2017b) |
| TZAC/TAROT-Reunion | 2017 Aug 18 13:04:25 | 1.02 | optical | GCN 21558, Klotz et al. (2017) |
| ATCA/– | 2017 Aug 18 13:27:25 | 1.03 | radio | GCN 21559, Bannister et al. (2017b) |
| SkyMapper/– | 2017 Aug 18 13:54:11 | 1.05 | optical | GCN 21560, Wolf et al. (2017) |
| Subaru/HSC | 2017 Aug 18 14:27:26 | 1.07 | optical | GCN 21561, Yoshida et al. (2017b) |
| ASKAP/– | 2017 Aug 18 14:36:00 | 1.08 | radio | GCN 21562, Bannister et al. (2017e) |
| LSGT,T17/SNUCAM-II | 2017 Aug 18 14:45:33 | 1.09 | optical | GCN 21563, Im et al. (2017a) |
| AGILE/GRID | 2017 Aug 18 15:22:43 | 1.11 | gamma-ray | GCN 21564, Bulgarelli et al. (2017) |
| LasCumbres/– | 2017 Aug 18 15:58:41 | 1.14 | optical | GCN 21565, Arcavi et al. (2017b) |
| LSGT,T17/SNUCAM-II | 2017 Aug 18 17:15:43 | 1.19 | optical | GCN 21566, Im et al. (2017b) |
| Swope/– | 2017 Aug 18 17:19:22 | 1.19 | optical | GCN 21567, Coulter et al. (2017b) |
| IceCube/– | 2017 Aug 18 17:27:25 | 1.2 | neutrino | GCN 21568, Bartos et al. (2017b) |
| Gemini-South/– | 2017 Aug 18 17:44:26 | 1.21 | optical, IR | GCN 21569, Singer et al. (2017c) |
| MASTER/– | 2017 Aug 18 18:06:51 | 1.23 | optical | GCN 21570, Lipunov et al. (2017e) |
| VLA/– | 2017 Aug 18 18:16:30 | 1.23 | radio | GCN 21571, Williams et al. (2017b) |
| *Swift*/UVOT,XRT | 2017 Aug 18 18:32:37 | 1.24 | x-ray, uv | GCN 21572, Cenko et al. (2017) |
| ATCA/– | 2017 Aug 18 20:19:00 | 1.32 | radio | GCN 21574, Kaplan et al. (2017a) |
| 2MASS,Spitzer/– | 2017 Aug 18 20:23:05 | 1.32 | IR | GCN 21575, Eikenberry et al. (2017) |
| VISTA/VIRCam | 2017 Aug 18 21:16:32 | 1.36 | IR | GCN 21576, Tanvir et al. (2017b) |
| –/– | 2017 Aug 18 23:00:31 | 1.43 | | GCN 21577, Malesani et al. (2017b) |
| –/– | 2017 Aug 18 23:11:30 | 1.44 | | GCN 21578, Cowperthwaite et al. (2017c) |
| PROMPT5/– | 2017 Aug 19 00:18:04 | 1.48 | optical | GCN 21579, Yang et al. (2017b) |
| DECam/– | 2017 Aug 19 00:22:23 | 1.49 | optical | GCN 21580, Nicholl et al. (2017b) |
| LasCumbres/– | 2017 Aug 19 01:26:07 | 1.53 | optical | GCN 21581, Arcavi et al. (2017c) |
| NTT/– | 2017 Aug 19 01:46:26 | 1.55 | optical, IR | GCN 21582, Lyman et al. (2017) |
| Swope/– | 2017 Aug 19 01:54:36 | 1.55 | optical | GCN 21583, Kilpatrick et al. (2017) |
| GROND/– | 2017 Aug 19 01:58:14 | 1.55 | optical, IR | GCN 21584, Wiseman et al. (2017) |
| SOAR/GoodmanSpectrograph | 2017 Aug 19 03:10:19 | 1.6 | IR, optical | GCN 21585, Nicholl et al. (2017c) |
| Subaru/HSC | 2017 Aug 19 06:52:33 | 1.76 | optical | GCN 21586, Yoshida et al. (2017c) |
| MASTER/– | 2017 Aug 19 08:10:30 | 1.81 | optical | GCN 21587, Lipunov et al. (2017c) |
| VLBA/– | 2017 Aug 19 09:36:26 | 1.87 | radio | GCN 21588, Deller et al. (2017a) |
| VLA/– | 2017 Aug 19 09:51:33 | 1.88 | radio | GCN 21589, Alexander et al. (2017f) |
| Pan-STARRS/– | 2017 Aug 19 10:14:53 | 1.9 | optical | GCN 21590, Chambers et al. (2017b) |
| NOT/NOTCam | 2017 Aug 19 12:00:05 | 1.97 | IR | GCN 21591, Malesani et al. (2017a) |
| ESO-VLT/X-shooter | 2017 Aug 19 12:16:37 | 1.98 | IR, optical | GCN 21592, Pian et al. (2017b) |
| ESO-VLT/FORS2 | 2017 Aug 19 14:13:15 | 2.06 | optical | GCN 21594, Wiersema et al. (2017) |
| Subaru/HSC | 2017 Aug 19 14:46:41 | 2.09 | optical | GCN 21595, Tominaga et al. (2017) |
| REM-ROS2/– | 2017 Aug 19 16:38:19 | 2.16 | optical | GCN 21596, Melandri et al. (2017c) |
| KMTNet/wide-fieldcamera | 2017 Aug 19 16:55:08 | 2.18 | optical | GCN 21597, Im et al. (2017d) |
| ESO-VST/OmegaCam | 2017 Aug 19 17:37:19 | 2.21 | optical | GCN 21598, Grado et al. (2017c) |
| LaSilla-QUEST/– | 2017 Aug 19 18:04:05 | 2.22 | optical | GCN 21599, Rabinowitz et al. (2017) |
| GMRT/– | 2017 Aug 19 21:18:21 | 2.36 | radio | GCN 21603, De et al. (2017a) |
| PROMPT5/– | 2017 Aug 19 23:31:25 | 2.45 | optical | GCN 21606, Valenti et al. (2017) |

**Table 6**
(Continued)

| Telescope | UT Date | $\Delta t$ (days) | Obs. Wavelength | References |
|---|---|---|---|---|
| GROND/– | 2017 Aug 20 04:49:21 | 2.67 | optical, IR | GCN 21608, Chen et al. (2017) |
| VIRT/– | 2017 Aug 20 05:27:49 | 2.7 | optical | GCN 21609, Gendre et al. (2017) |
| SALT/– | 2017 Aug 20 06:14:37 | 2.73 | optical | GCN 21610, Shara et al. (2017) |
| *Swift*/XRT | 2017 Aug 20 08:42:40 | 2.83 | x-ray | GCN 21612, Evans et al. (2017c) |
| VLA/– | 2017 Aug 20 09:17:57 | 2.86 | radio | GCN 21613, Corsi et al. (2017b) |
| VLA/– | 2017 Aug 20 10:26:01 | 2.91 | radio | GCN 21614, Corsi et al. (2017a) |
| Pan-STARRS/– | 2017 Aug 20 13:59:50 | 3.05 | optical | GCN 21617, Chambers et al. (2017c) |
| ChilescopeRC-1000/– | 2017 Aug 20 14:24:47 | 3.07 | optical | GCN 21618, Pozanenko et al. (2017d) |
| TOROS/– | 2017 Aug 20 14:48:49 | 3.09 | optical | GCN 21619, Diaz et al. (2017a) |
| TOROS/– | 2017 Aug 20 15:03:42 | 3.1 | optical | GCN 21620, Diaz et al. (2017c) |
| –/– | 2017 Aug 20 15:40:35 | 3.12 | … | GCN 21621, Lipunov (2017) |
| Kanata/HONIR | 2017 Aug 20 16:37:38 | 3.16 | IR | GCN 21623, Nakaoka et al. (2017) |
| BOOTES-5/– | 2017 Aug 20 21:59:59 | 3.39 | optical | GCN 21624, Castro-Tirado et al. (2017) |
| ASKAP/– | 2017 Aug 21 00:58:33 | 3.51 | radio | GCN 21625, Dobie et al. (2017b) |
| *NuSTAR*/– | 2017 Aug 21 04:33:27 | 3.66 | x-ray | GCN 21626, Harrison et al. (2017) |
| Zadko/– | 2017 Aug 21 05:57:23 | 3.72 | optical | GCN 21627, Coward et al. (2017b) |
| ATCA/– | 2017 Aug 21 07:45:30 | 3.79 | radio | GCN 21628, Lynch et al. (2017c) |
| ATCA/– | 2017 Aug 21 09:02:12 | 3.85 | radio | GCN 21629, Lynch et al. (2017d) |
| ANTARES/– | 2017 Aug 21 15:08:00 | 4.1 | neutrino | GCN 21631, Ageron et al. (2017b) |
| KMTNet,iTelescope.NET/– | 2017 Aug 21 15:49:41 | 4.13 | optical | GCN 21632, Im et al. (2017c) |
| Pan-STARRS/– | 2017 Aug 21 16:03:52 | 4.14 | optical | GCN 21633, Chambers et al. (2017d) |
| TOROS/CASLEO | 2017 Aug 21 16:05:22 | 4.14 | optical | GCN 21634, Diaz et al. (2017d) |
| ChilescopeRC-1000/– | 2017 Aug 21 16:11:53 | 4.15 | optical | GCN 21635, Pozanenko et al. (2017a) |
| VLA/– | 2017 Aug 21 18:40:08 | 4.25 | radio | GCN 21636, Corsi et al. (2017e) |
| MWA/– | 2017 Aug 22 00:59:36 | 4.51 | radio | GCN 21637, Kaplan et al. (2017c) |
| Gemini-South/Flamingos-2 | 2017 Aug 22 05:20:11 | 4.69 | IR | GCN 21638, Chornock et al. (2017c) |
| ASKAP/– | 2017 Aug 22 07:23:04 | 4.78 | radio | GCN 21639, Dobie et al. (2017a) |
| CALET/CGBM | 2017 Aug 22 09:36:51 | 4.87 | gamma-ray | GCN 21641, Nakahira et al. (2017) |
| ChilescopeRC-1000/– | 2017 Aug 22 15:23:04 | 5.11 | optical | GCN 21644, Pozanenko et al. (2017c) |
| 6dFGS/– | 2017 Aug 22 16:55:17 | 5.18 | optical | GCN 21645, Sadler et al. (2017) |
| *Chandra*/*CXO* | 2017 Aug 22 18:06:23 | 5.23 | x-ray | GCN 21648, Margutti et al. (2017b) |
| VLA/JAGWAR | 2017 Aug 22 19:13:38 | 5.27 | radio | GCN 21650, Mooley et al. (2017a) |
| ESO-VLT/FORS2 | 2017 Aug 23 07:52:38 | 5.8 | optical | GCN 21653, D'Avanzo et al. (2017) |
| VLA/– | 2017 Aug 23 18:25:07 | 6.24 | radio | GCN 21664, Corsi et al. (2017c) |
| *HST*/Pan-STARRS1/GPC1 | 2017 Aug 24 01:39:20 | 6.54 | optical | GCN 21669, Yu et al. (2017) |
| ATCA/– | 2017 Aug 24 04:30:05 | 6.66 | radio | GCN 21670, Lynch et al. (2017a) |
| ASKAP/– | 2017 Aug 24 06:10:24 | 6.73 | radio | GCN 21671, Bannister et al. (2017c) |
| *INTEGRAL*/SPI,IBIS,JEM-X,OMC | 2017 Aug 24 09:03:02 | 6.85 | gamma-ray, x-ray, optical | GCN 21672, Savchenko et al. (2017b) |
| H.E.S.S./– | 2017 Aug 24 10:35:02 | 6.91 | gamma-ray | GCN 21674, de Naurois et al. (2017) |
| LOFAR/ILT | 2017 Aug 24 13:35:06 | 7.04 | radio | GCN 21676, Broderick et al. (2017) |
| AAT/AAO | 2017 Aug 24 15:31:25 | 7.12 | optical | GCN 21677, Andreoni et al. (2017) |
| LWA/LWA1 | 2017 Aug 24 16:08:17 | 7.14 | radio | GCN 21680, Callister et al. (2017a) |
| ESO-VLT/MUSEIntegralFieldUnit | 2017 Aug 24 19:28:30 | 7.28 | optical | GCN 21681, Levan et al. (2017b) |
| Gemini-South/Flamingos-2,GMOS | 2017 Aug 24 19:31:19 | 7.28 | optical, IR | GCN 21682, Troja et al. (2017b) |
| HAWC/– | 2017 Aug 24 19:35:19 | 7.29 | gamma-ray | GCN 21683, Martinez-Castellanos et al. (2017) |
| Gemini-South/Flamingos-2 | 2017 Aug 25 04:04:17 | 7.64 | IR | GCN 21684, Chornock et al. (2017b) |

**Table 6**
(Continued)

| Telescope | UT Date | $\Delta t$ (days) | Obs. Wavelength | References |
|---|---|---|---|---|
| Subaru/HSC | 2017 Aug 25 07:38:17 | 7.79 | optical | GCN 21685, Yoshida et al. (2017d) |
| Auger/SurfaceDetector | 2017 Aug 25 08:13:23 | 7.81 | neutrino | GCN 21686, Alvarez-Muniz et al. (2017) |
| MASTER/MASTER-II | 2017 Aug 25 08:48:24 | 7.84 | optical | GCN 21687, Lipunov et al. (2017b) |
| ESO-VST/OmegaCAM | 2017 Aug 25 22:15:33 | 8.4 | optical | GCN 21703, Grado et al. (2017a) |
| GMRT/– | 2017 Aug 26 01:23:58 | 8.53 | radio | GCN 21708, De et al. (2017b) |
| ATCA/– | 2017 Aug 29 03:49:22 | 11.63 | radio | GCN 21740, Lynch et al. (2017b) |
| Zadko/– | 2017 Aug 29 08:29:39 | 11.83 | optical | GCN 21744, Coward et al. (2017a) |
| Konus-Wind/– | 2017 Aug 29 10:55:08 | 11.93 | gamma-ray | GCN 21746, Svinkin et al. (2017a) |
| ALMA/– | 2017 Aug 29 12:37:56 | 12.0 | radio | GCN 21747, Schulze et al. (2017) |
| ALMA/– | 2017 Aug 29 14:55:15 | 12.09 | radio | GCN 21750, Williams et al. (2017a) |
| OVRO/– | 2017 Aug 30 03:23:28 | 12.61 | radio | GCN 21760, Pearson et al. (2017) |
| EVN/VLBI | 2017 Aug 30 09:48:26 | 12.88 | radio | GCN 21763, Paragi et al. (2017a) |
| *Chandra*/*CXO* | 2017 Aug 30 12:07:12 | 12.98 | x ray | GCN 21765, Troja et al. (2017c) |
| GMRT/– | 2017 Aug 30 16:06:24 | 13.14 | radio | GCN 21768, Resmi et al. (2017) |
| Gemini-South/– | 2017 Aug 31 18:28:50 | 14.24 | IR | GCN 21778, Troja et al. (2017d) |
| Gemini-South/Flamingos-2 | 2017 Aug 31 18:32:01 | 14.24 | IR | GCN 21779, Singer et al. (2017b) |
| *HST*/– | 2017 Aug 31 20:33:24 | 14.33 | optical, IR | GCN 21781, Levan et al. (2017a) |
| PioftheSky/PioftheSkyNorth | 2017 Sep 01 21:54:25 | 15.38 | optical | GCN 21783, Cwiek et al. (2017) |
| AGILE/GRID | 2017 Sep 02 16:54:59 | 16.18 | gamma-ray | GCN 21785, Verrecchia et al. (2017) |
| *Chandra*/*CXO* | 2017 Sep 02 16:57:54 | 16.18 | x ray | GCN 21786, Fong et al. (2017) |
| *Chandra*/*CXO* | 2017 Sep 02 17:06:21 | 16.18 | x ray | GCN 21787, Troja et al. (2017e) |
| *Chandra*/*CXO* | 2017 Sep 03 20:24:16 | 17.32 | x ray | GCN 21798, Haggard et al. (2017b) |
| ATCA/– | 2017 Sep 04 02:26:14 | 17.57 | radio | GCN 21803, Troja et al. (2017f) |
| e-MERLIN/– | 2017 Sep 04 07:48:43 | 17.8 | radio | GCN 21804, Moldon et al. (2017a) |
| VLA/– | 2017 Sep 04 22:14:55 | 18.4 | radio | GCN 21814, Mooley et al. (2017b) |
| VLA/– | 2017 Sep 04 22:14:59 | 18.4 | radio | GCN 21815, Corsi et al. (2017d) |
| *HST*/*HST*,*Gaia* | 2017 Sep 05 00:30:09 | 18.49 | optical, IR, uv | GCN 21816, Adams et al. (2017) |
| ESO-VST/OMEGACam | 2017 Sep 06 15:07:27 | 20.1 | optical | GCN 21833, Grado et al. (2017b) |
| ATCA/– | 2017 Sep 07 02:31:55 | 20.58 | radio | GCN 21842, Murphy et al. (2017) |
| LWA/LWA1 | 2017 Sep 08 02:47:01 | 21.59 | radio | GCN 21848, Callister et al. (2017b) |
| VLBA/– | 2017 Sep 08 11:16:27 | 21.94 | radio | GCN 21850, Deller et al. (2017b) |
| VLA/– | 2017 Sep 08 13:23:16 | 22.03 | radio | GCN 21851, Alexander et al. (2017a) |
| ATCA/– | 2017 Sep 14 05:25:42 | 27.7 | radio | GCN 21882, Wieringa et al. (2017) |
| AST3-2/– | 2017 Sep 15 03:45:21 | 28.63 | optical | GCN 21883, Hu et al. (2017) |
| ATLAS/– | 2017 Sep 15 11:24:15 | 28.95 | optical | GCN 21886, Tonry et al. (2017) |
| DanishTel/– | 2017 Sep 15 16:40:07 | 29.17 | optical | GCN 21889, Cano et al. (2017) |
| MeerKAT/– | 2017 Sep 15 20:16:29 | 29.32 | radio | GCN 21891, Goedhart et al. (2017b) |
| DFN/– | 2017 Sep 18 13:45:29 | 32.04 | optical | GCN 21894, Hancock et al. (2017) |
| T80S,EABA/– | 2017 Sep 18 16:22:27 | 32.15 | optical | GCN 21895, Diaz et al. (2017b) |
| VLBA/– | 2017 Sep 19 07:51:22 | 32.8 | radio | GCN 21897, Deller et al. (2017c) |
| ChilescopeRC-1000/– | 2017 Sep 19 18:09:03 | 33.23 | optical | GCN 21898, Pozanenko et al. (2017b) |
| Parkes/– | 2017 Sep 21 02:38:29 | 34.58 | radio | GCN 21899, Bailes et al. (2017a) |
| ATCA/– | 2017 Sep 21 06:42:36 | 34.75 | radio | GCN 21900, Ricci et al. (2017) |
| LasCumbres/FLOYDS,Gemini | 2017 Sep 22 03:24:44 | 35.61 | optical | GCN 21908, McCully et al. (2017a) |
| SRT/– | 2017 Sep 22 19:06:44 | 36.27 | radio | GCN 21914, Aresu et al. (2017) |
| Effelsberg/– | 2017 Sep 23 20:34:41 | 37.33 | radio | GCN 21920, Kramer et al. (2017) |

**Table 6**
(Continued)

| Telescope | UT Date | $\Delta t$ (days) | Obs. Wavelength | References |
|---|---|---|---|---|
| MWA/– | 2017 Sep 25 22:30:34 | 39.41 | radio | GCN 21927, Kaplan et al. (2017b) |
| Parkes/– | 2017 Sep 26 02:00:59 | 39.56 | radio | GCN 21928, Bailes et al. (2017b) |
| VLA/– | 2017 Sep 26 05:14:16 | 39.69 | radio | GCN 21929, Hallinan et al. (2017b) |
| PioftheSky/PioftheSkyNorth | 2017 Sep 26 21:17:49 | 40.36 | optical | GCN 21931, Batsch et al. (2017) |
| MeerKAT/– | 2017 Sep 27 13:19:14 | 41.03 | radio | GCN 21933, Goedhart et al. (2017a) |
| VLA/– | 2017 Sep 27 19:03:46 | 41.27 | radio | GCN 21935, Alexander et al. (2017b) |
| EVN/– | 2017 Sep 28 10:35:27 | 41.91 | radio | GCN 21939, Paragi et al. (2017b) |
| e-MERLIN/– | 2017 Sep 28 11:12:37 | 41.94 | radio | GCN 21940, Moldon et al. (2017b) |

### 3.5. Neutrinos

The detection of GW170817 was rapidly followed up by the IceCube (Aartsen et al. 2017) and ANTARES (Ageron et al. 2011) neutrino observatories and the Pierre Auger Observatory (Aab et al. 2015a) to search for coincident, high-energy (GeV–EeV) neutrinos emitted in the relativistic outflow produced by the BNS merger. The results from these observations, described briefly below, can be used to constrain the properties of relativistic outflows driven by the merger (A. Albert et al. 2017, in preparation).

In a search for muon–neutrino track candidates (Aartsen et al. 2016), and contained neutrino events of any flavor (Aartsen et al. 2015), IceCube identified no neutrinos that were directionally coincident with the final localization of GW170817 at 90% credible level, within ±500 s of the merger (Bartos et al. 2017a, 2017b). Additionally, no MeV supernova neutrino burst signal was detected coincident with the merger. Following the identification via electromagnetic observations of the host galaxy of the event, IceCube also carried out an extended search in the direction of NGC 4993 for neutrinos within the 14 day period following the merger, but found no significant neutrino emission (A. Albert et al. 2017, in preparation).

A neutrino search for upgoing high-energy muon neutrinos was carried out using the online ANTARES data stream (Ageron et al. 2017a). No upgoing neutrino candidates were found over a $t_c \pm 500$ s time window. The final localization of GW170817 (LIGO Scientific Collaboration & Virgo Collaboration et al. 2017c) was above the ANTARES horizon at the time of the GW event. A search for downgoing muon neutrinos was thus performed, and no neutrinos were found over $t_c$ 500 s (Ageron et al. 2017b). A search for neutrinos originating from below the ANTARES horizon, over an extended period of 14 days after the merger, was also performed, without yielding significant detection (A. Albert et al. 2017, in preparation).

The Pierre Auger Observatory carried out a search for ultra-high-energy (UHE) neutrinos above $\sim 10^{17}$ eV using its Surface Detector (Aab et al. 2015a). UHE neutrino-induced extensive air showers produced either by interactions of downward-going neutrinos in the atmosphere or by decays of tau leptons originating from tau neutrino interactions in the Earth's crust can be efficiently identified above the background of the more numerous ultra-high-energy cosmic rays (Aab et al. 2015b). Remarkably, the position of the transient in NGC 4993 was just between $0^\circ\!.3$ and $3^\circ\!.2$ below the horizon during $t_c \pm 500$ s. This region corresponds to the most efficient geometry for Earth-skimming tau neutrino detection at $10^{18}$ eV energies. No neutrino candidates were found in $t_c \pm 500$ s (Alvarez-Muniz et al. 2017) nor in the 14 day period after it (A. Albert et al. 2017, in preparation).

## 4. Conclusion

For the first time, gravitational and electromagnetic waves from a single source have been observed. The gravitational-wave observation of a binary neutron star merger is the first of its kind. The electromagnetic observations further support the interpretation of the nature of the binary, and comprise three components at different wavelengths: (i) a prompt sGRB that demonstrates that BNS mergers are the progenitor of at least a fraction of such bursts; (ii) an ultraviolet, optical, and infrared transient (kilonova), which allows for the identification of the host galaxy and is associated with the aftermath of the BNS merger; and (iii) delayed X-ray and radio counterparts that provide information on the environment of the binary. These observations, described in detail in the companion articles cited above, offer a comprehensive, sequential description of the physical processes related to the merger of a binary neutron star. Table 6 collects all of the Gamma-ray Coordinates Network (GCN) notices and circulars related to GW170817 through 2017 October 1 UTC. The results of this campaign demonstrate the importance of collaborative gravitational-wave, electromagnetic, and neutrino observations and mark a new era in multi-messenger, time-domain astronomy.

(*1M2H*) We thank J. McIver for alerting us to the LVC circular. We thank J. Mulchaey (Carnegie Observatories director), L. Infante (Las Campanas Observatory director), and the entire Las Campanas staff for their extreme dedication, professionalism, and excitement, all of which were critical in the discovery of the first gravitational-wave optical counterpart and its host galaxy as well as the observations used in this study. We thank I. Thompson and the Carnegie Observatory Time Allocation Committee for approving the Swope Supernova Survey and scheduling our program. We thank the University of Copenhagen, DARK Cosmology Centre, and the Niels Bohr International Academy for hosting D.A.C., R.J.F., A.M.B., E.R., and M.R.S. during the discovery of GW170817/SSS17a. R.J.F., A.M.B., and E.R. were participating in the Kavli Summer Program in Astrophysics, "Astrophysics with gravitational wave detections." This program was supported by the the Kavli Foundation, Danish National Research Foundation, the Niels Bohr International Academy, and the DARK Cosmology Centre. The UCSC group is supported in part by NSF grant AST–1518052, the Gordon & Betty Moore Foundation, the Heising-Simons Foundation, generous donations from many individuals through a UCSC Giving Day grant, and from fellowships from the Alfred P. Sloan Foundation (R.J.F.), the David and Lucile Packard Foundation (R.J.F. and E.R.) and the Niels Bohr Professorship from the DNRF (E.R.). AMB acknowledges support from a UCMEXUS-CONACYT Doctoral Fellowship. Support for this work was provided by NASA through Hubble Fellowship grants HST–HF–51348.001 (B.J.S.) and HST–HF–51373.001 (M.R.D.) awarded by the Space Telescope Science Institute, which is operated by the Association of Universities for Research in Astronomy, Inc., for NASA, under contract NAS5–26555. This paper includes data gathered with the 1 meter Swope and 6.5 meter Magellan Telescopes located at Las Campanas Observatory, Chile.

*(AGILE)* The AGILE Team thanks the ASI management, the technical staff at the ASI Malindi ground station, the technical support team at the ASI Space Science Data Center, and the Fucino AGILE Mission Operation Center. AGILE is an ASI space mission developed with programmatic support by INAF and INFN. We acknowledge partial support through the ASI grant No. I/028/12/2. We also thank INAF, Italian Institute of Astrophysics, and ASI, Italian Space Agency.

*(ANTARES)* The ANTARES Collaboration acknowledges the financial support of: Centre National de la Recherche Scientifique (CNRS), Commissariat à l'énergie atomique et aux énergies alternatives (CEA), Commission Européenne (FEDER fund and Marie Curie Program), Institut Universitaire de France (IUF), IdEx program and UnivEarthS Labex program at Sorbonne Paris Cité (ANR-10-LABX-0023 and ANR-11-IDEX-0005-02), Labex OCEVU (ANR-11-LABX-0060) and the A*MIDEX project (ANR-11-IDEX-0001-02), Région

Île-de-France (DIM-ACAV), Région Alsace (contrat CPER), Région Provence-Alpes-Côte d'Azur, Département du Var and Ville de La Seyne-sur-Mer, France; Bundesministerium für Bildung und Forschung (BMBF), Germany; Istituto Nazionale di Fisica Nucleare (INFN), Italy; Nederlandse organisatie voor Wetenschappelijk Onderzoek (NWO), the Netherlands; Council of the President of the Russian Federation for young scientists and leading scientific schools supporting grants, Russia; National Authority for Scientific Research (ANCS), Romania; Ministerio de Economía y Competitividad (MINECO): Plan Estatal de Investigación (refs. FPA2015-65150-C3-1-P, -2-P and -3-P; MINECO/FEDER), Severo Ochoa Centre of Excellence and MultiDark Consolider (MINECO), and Prometeo and Grisolía programs (Generalitat Valenciana), Spain; Ministry of Higher Education, Scientific Research and Professional Training, Morocco. We also acknowledge the technical support of Ifremer, AIM and Foselev Marine for the sea operation and the CC-IN2P3 for the computing facilities.

*(AST3)* The AST3 project is supported by the National Basic Research Program (973 Program) of China (Grant Nos. 2013CB834901, 2013CB834900, 2013CB834903), and the Chinese Polar Environment Comprehensive Investigation & Assessment Program (grant No. CHINARE2016-02-03-05). The construction of the AST3 telescopes has received fundings from Tsinghua University, Nanjing University, Beijing Normal University, University of New South Wales, and Texas A&M University, the Australian Antarctic Division, and the National Collaborative Research Infrastructure Strategy (NCRIS) of Australia. It has also received funding from Chinese Academy of Sciences through the Center for Astronomical Mega-Science and National Astronomical Observatory of China (NAOC).

*(Auger)* The successful installation, commissioning, and operation of the Pierre Auger Observatory would not have been possible without the strong commitment and effort from the technical and administrative staff in Malargüe. We are very grateful to the following agencies and organizations for financial support: Argentina—Comisión Nacional de Energía Atómica; Agencia Nacional de Promoción Científica y Tecnológica (ANPCyT); Consejo Nacional de Investigaciones Científicas y Técnicas (CONICET); Gobierno de la Provincia de Mendoza; Municipalidad de Malargüe; NDM Holdings and Valle Las Leñas; in gratitude for their continuing cooperation over land access; Australia—the Australian Research Council; Brazil—Conselho Nacional de Desenvolvimento Científico e Tecnológico (CNPq); Financiadora de Estudos e Projetos (FINEP); Fundação de Amparo à Pesquisa do Estado de Rio de Janeiro (FAPERJ); São Paulo Research Foundation (FAPESP) grant Nos. 2010/07359-6 and 1999/05404-3; Ministério da Ciência, Tecnologia, Inovações e Comunicações (MCTIC); Czech Republic—grant Nos. MSMT CR LG15014, LO1305, LM2015038 and CZ.02.1.01/0.0/0.0/16_013/0001402; France—Centre de Calcul IN2P3/CNRS; Centre National de la Recherche Scientifique (CNRS); Conseil Régional Ile-de-France; Département Physique Nucléaire et Corpusculaire (PNC-IN2P3/CNRS); Département Sciences de l'Univers (SDU-INSU/CNRS); Institut Lagrange de Paris (ILP) grant No. LABEX ANR-10-LABX-63 within the Investissements d'Avenir Programme Grant No. ANR-11-IDEX-0004-02; Germany—Bundesministerium für Bildung und Forschung (BMBF); Deutsche Forschungsgemeinschaft (DFG); Finanzministerium Baden-Württemberg; Helmholtz Alliance for Astroparticle Physics (HAP); Helmholtz-Gemeinschaft Deutscher Forschungszentren (HGF); Ministerium für Innovation, Wissenschaft und Forschung des Landes Nordrhein-Westfalen; Ministerium für Wissenschaft, Forschung und Kunst des Landes Baden-Württemberg; Italy—Istituto Nazionale di Fisica Nucleare (INFN); Istituto Nazionale di Astrofisica (INAF); Ministero dell'Istruzione, dell'Universitá e della Ricerca (MIUR); CETEMPS Center of Excellence; Ministero degli Affari Esteri (MAE); Mexico—Consejo Nacional de Ciencia y Tecnología (CONACYT) No. 167733; Universidad Nacional Autónoma de México (UNAM); PAPIIT DGAPA-UNAM; The Netherlands – Ministerie van Onderwijs, Cultuur en Wetenschap; Nederlandse Organisatie voor Wetenschappelijk Onderzoek (NWO); Stichting voor Fundamenteel Onderzoek der Materie (FOM); Poland—National Centre for Research and Development, grant Nos. ERA-NET-ASPERA/01/11 and ERA-NET-ASPERA/02/11; National Science Centre, grant Nos. 2013/08/M/ST9/00322, 2013/08/M/ST9/00728, and HARMONIA 5–2013/10/M/ST9/00062, UMO-2016/22/M/ST9/00198; Portugal—Portuguese national funds and FEDER funds within Programa Operacional Factores de Competitividade through Fundação para a Ciência e a Tecnologia (COMPETE); Romania—Romanian Authority for Scientific Research ANCS; CNDI-UEFISCDI partnership projects grant Nos. 20/2012 and 194/2012 and PN 16 42 01 02; Slovenia—Slovenian Research Agency; Spain—Comunidad de Madrid; Fondo Europeo de Desarrollo Regional (FEDER) funds; Ministerio de Economía y Competitividad; Xunta de Galicia; European Community 7th Framework Program grant No. FP7-PEOPLE-2012-IEF-328826; USA—Department of Energy, Contract Nos. DE-AC02-07CH11359, DE-FR02-04ER41300, DE-FG02-99ER41107, and DE-SC0011689; National Science Foundation, grant No. 0450696; The Grainger Foundation; Marie Curie-IRSES/EPLANET; European Particle Physics Latin American Network; European Union 7th Framework Program, grant No. PIRSES-2009-GA-246806; European Union's Horizon 2020 research and innovation programme (grant No. 646623); and UNESCO.

*(Australian Radio)* T.M. acknowledges the support of the Australian Research Council through grant FT150100099. S.O. acknowledges the Australian Research Council grant Laureate Fellowship FL15010014. D.L.K. and I.S.B. are additionally supported by NSF grant AST-141242. P.A.B. and the DFN team acknowledge the Australian Research Council for support under their Australian Laureate Fellowship scheme. The Australia Telescope Compact Array is part of the Australia Telescope National Facility, which is funded by the Australian Government for operation as a National Facility managed by CSIRO. This scientific work makes use of the Murchison Radio-astronomy Observatory, operated by CSIRO. We acknowledge the Wajarri Yamatji people as the traditional owners of the Observatory site. Support for the operation of the MWA is provided by the Australian Government (NCRIS), under a contract to Curtin University administered by Astronomy Australia Limited. We acknowledge the Pawsey Supercomputing Centre, which is supported by the Western Australian and Australian Governments. The Australian SKA Pathfinder is part of the Australia Telescope National Facility, which is managed by CSIRO. Operation of ASKAP is funded by the Australian Government with support from the National Collaborative Research Infrastructure Strategy. ASKAP uses the resources of the Pawsey Supercomputing Centre. Establishment of ASKAP, the Murchison Radio-astronomy Observatory and the Pawsey Supercomputing Centre are initiatives of the Australian Government, with support from the Government of Western Australia and the Science and Industry Endowment Fund. Parts of this research were conducted by the Australian Research Council Centre of Excellence for All-sky Astrophysics in 3D (ASTRO 3D) through project number CE170100013.

*(Berger Time-Domain Group)* The Berger Time-Domain Group at Harvard is supported in part by the NSF through grants AST-1411763 and AST-1714498, and by NASA through grants NNX15AE50G and NNX16AC22G.

*(Bootes)* A.J.C.T. acknowledges support from the Spanish Ministry Project AYA 2015-71718-R (including FEDER funds) and Junta de Andalucia Proyecto de Excelencia TIC-2839. I.H.P. acknowledges the support of the National Research Foundation (NRF-2015R1A2A1A01006870). S.J. acknowledges the support of Korea Basic Science Research Program (NRF2014R1A6A3A03057484 and NRF-2015R1D1A4A01020961). The BOOTES-5/JGT observations were carried out at Observatorio Astronómico Nacional in San Pedro Mártir (OAN-SPM, México), operated by Instituto de Astronomía, UNAM and with support from Consejo Nacional de Ciencia y Tecnología (México) through the Laboratorios Nacionales Program (México), Instituto de Astrofísica de Andalucía (IAA-CSIC, Spain) and Sungkyunkwan University (SKKU, South Korea). We also thank the staff of OAN-SPM for their support in carrying out the observations.

*(CAASTRO)* Parts of this research were conducted by the Australian Research Council Centre of Excellence for All-sky Astrophysics (CAASTRO), through project number CE110001020. The national facility capability for SkyMapper has been funded through ARC LIEF grant LE130100104 from the Australian Research Council, awarded to the University of Sydney, the Australian National University, Swinburne University of Technology, the University of Queensland, the University of Western Australia, the University of Melbourne, Curtin University of Technology, Monash University, and the Australian Astronomical Observatory. SkyMapper is owned and operated by The Australian National University's Research School of Astronomy and Astrophysics.

*(CALET)* The CALET team gratefully acknowledges support from NASA, ASI, JAXA, and MEXT KAKENHI grant numbers JP 17H06362, JP26220708, and JP17H02901.

*(Chandra/McGill)* This work was supported in part by Chandra Award Number GO7-18033X, issued by the Chandra X-ray Observatory Center, which is operated by the Smithsonian Astrophysical Observatory for and on behalf of the National Aeronautics Space Administration (NASA) under contract NAS8-03060. D.H., M.N., and J.J.R. acknowledge support from a Natural Sciences and Engineering Research Council of Canada (NSERC) Discovery Grant and a Fonds de recherche du Québec–Nature et Technologies (FRQNT) Nouveaux Chercheurs Grant. P.A.E. acknowledges UKSA support. J.A.K. acknowledges the support of NASA grant NAS5-00136. D.H. also acknowledges support from the Canadian Institute for Advanced Research (CIFAR).

*(CZTI/AstroSat)* CZTI is built by a TIFR-led consortium of institutes across India, including VSSC, ISAC, IUCAA, SAC, and PRL. The Indian Space Research Organisation funded, managed, and facilitated the project.

*(DLT40)* D.J.S. acknowledges support for the DLT40 program from NSF grant AST-1517649.

*(EuroVLBI)* The European VLBI Network is a joint facility of independent European, African, Asian, and North American radio astronomy institutes. Scientific results from data presented in this publication are derived from the following EVN project code: RP029. e-MERLIN is a National Facility operated by the University of Manchester at Jodrell Bank Observatory on behalf of STFC. The collaboration between LIGO/Virgo and EVN/e-MERLIN is part of a project that has received funding from the European Unions Horizon 2020 research and innovation programme under grant agreement No. 653477.

*(ePESSTO)* We acknowledge ESO programs 199.D-0143 and 099.D-0376. PS1 and ATLAS are supported by NASA grants NNX08AR22G, NNX12AR65G, NNX14AM74G, and NNX12AR55G. We acknowledge the Leibniz-Prize to Prof. G. Hasinger (DFG grant HA 1850/28-1), EU/FP7-ERC grants 291222, 615929, 647208, 725161, STFC grants ST/P000312/1 and ERF ST/M005348/1, ST/P000495/1. Marie Sklodowska-Curie grant No 702538. Polish NCN grant OPUS 2015/17/B/ST9/03167, Knut and Alice Wallenberg Foundation. PRIN-INAF 2014. David and Ellen Lee Prize Postdoctoral Fellowship at the California Institute of Technology. Alexander von Humboldt Sofja Kovalevskaja Award. Royal Society—Science Foundation Ireland Vilho, Yrjö and Kalle Väisälä Foundation. FONDECYT grant number 3160504. US NSF grant AST-1311862. Swedish Research Council and the Swedish Space Board. The Quantum Universe I-Core program, the ISF, BSF, and Kimmel award. IRC grant GOIPG/2017/1525. Australian Research Council CAASTRO CE110001020 and grant FT160100028. We acknowledge Millennium Science Initiative grant IC120009.

*(Fermi-GBM)* B.C., V.C., A.G., and W.S.P. gratefully acknowledge NASA funding through contract NNM13AA43C. M.S.B., R.H., P.J., C.A.M., S.P., R.D.P., M.S., and P.V. gratefully acknowledge NASA funding from cooperative agreement NNM11AA01A. E.B. is supported by an appointment to the NASA Postdoctoral Program at the Goddard Space Flight Center, administered by Universities Space Research Association under contract with NASA. D.K., C.A.W.H., C.M.H., and J.R. gratefully acknowledge NASA funding through the *Fermi*-GBM project. Support for the German contribution to GBM was provided by the Bundesministerium für Bildung und Forschung (BMBF) via the Deutsches Zentrum für Luft und Raumfahrt (DLR) under contract number 50 QV 0301. A.v.K. was supported by the Bundesministeriums für Wirtschaft und Technologie (BMWi) through DLR grant 50 OG 1101. S.M.B. acknowledges support from Science Foundation Ireland under grant 12/IP/1288.

*(Fermi-LAT)* The *Fermi*-LAT Collaboration acknowledges support for LAT development, operation, and data analysis from NASA and DOE (United States), CEA/Irfu and IN2P3/CNRS (France), ASI and INFN (Italy), MEXT, KEK, and JAXA (Japan), and the K.A. Wallenberg Foundation, the Swedish Research Council and the National Space Board (Sweden). Science analysis support in the operations phase from INAF (Italy) and CNES (France) is also gratefully acknowledged. This work performed in part under DOE Contract DE-AC02-76SF00515.

*(FRBSG)* S.L.L. is supported by NSF grant PHY-1607291 (LIU). Construction of the LWA has been supported by the Office of Naval Research under Contract N00014-07-C-0147. Support for operations and continuing development of the LWA1 is provided by the National Science Foundation under grants AST-1139963 and AST-1139974 of the University Radio Observatory program.

*(GRAWITA)* We acknowledge INAF for supporting the project "Gravitational Wave Astronomy with the first detections of adLIGO and adVIRGO experiments—GRAWITA" PI: E. Brocato. Observations are made with ESO Telescopes at the Paranal Observatory under programmes ID 099.D-0382 (PI: E. Pian), 099.D-0622 (PI: P. D'Avanzo), 099.D-0191 (PI: A. Grado), 099.D-0116 (PI: S. Covino) and with the REM telescope at the ESO La Silla Observatory under program ID 35020 (PI: S. Campana). We thank the ESO operation staff for excellent support of this program. The Sardinia Radio Telescope (SRT) is funded by the Department of University and Research (MIUR), the Italian Space Agency (ASI), and the Autonomous Region of Sardinia (RAS) and is operated as National Facility by the National Institute for Astrophysics (INAF). Z.J. is supported by the External Cooperation Program of BIC (number 114332KYSB20160007). J.M. is supported by the Hundred Talent Program, the Major Program of the Chinese Academy of Sciences (KJZD-EW-M06), the National Natural Science Foundation of China 11673062, and the Oversea Talent Program of Yunnan Province. R.L.C. Starling, K.W., A.B.H., N.R.T., and C.G.M. are supported by the STFC (Science and Technology Facilities Council). D.K., acknowledges the financial support from the Slovenian Research Agency (P1-0188). S.K. and A.N.G. acknowledge support by grant DFG Kl 766/16-3. D.G. acknowledges the financial support of the UnivEarthS Labex program at Sorbonne Paris Cité (ANR-10-LABX-0023 and ANR-11-IDEX-0005-02). K.T. was supported by JSPS grant 15H05437 and by a JST Consortia grant.

*(GROND)* Part of the funding for GROND was generously granted from the Leibniz-Prize to Prof. G. Hasinger (DFG grant HA 1850/28-1). "We acknowledge the excellent help in obtaining GROND data from Angela Hempel, Markus Rabus and Régis Lachaume on La Silla."

*(GROWTH, JAGWAR, Caltech-NRAO, TTU-NRAO, and NuSTAR)* This work was supported by the GROWTH (Global Relay of Observatories Watching Transients Happen) project funded by the National Science Foundation under PIRE grant No. 1545949. GROWTH is a collaborative project among California Institute of Technology (USA), University of Maryland College Park (USA), University of Wisconsin–Milwaukee (USA), Texas Tech University (USA), San Diego State University (USA), Los Alamos National Laboratory (USA), Tokyo Institute of Technology (Japan), National Central University (Taiwan), Indian Institute of Astrophysics (India), Inter-University Center for Astronomy and Astrophysics (India), Weizmann Institute of Science (Israel), The Oskar Klein Centre at Stockholm University (Sweden), Humboldt University (Germany), Liverpool John Moores University (UK). A.H. acknowledges support by the I-Core Program of the Planning and Budgeting Committee and the Israel Science Foundation. T.M. acknowledges the support of the Australian Research Council through grant FT150100099. Parts of this research were conducted by the Australian Research Council Centre of Excellence for All-sky Astrophysics (CAASTRO), through project number CE110001020. The Australia Telescope Compact Array is part of the Australia Telescope National Facility which is funded by the Australian Government for operation as a National Facility managed by CSIRO. D.L.K. is additionally supported by NSF grant AST-1412421. A.A.M. is funded by the Large Synoptic Survey Telescope Corporation in support of the Data Science Fellowship Program. P.C.Y., C.C.N., and W.H.I. thank the support from grants MOST104-2923-M-008-004-MY5 and MOST106-2112-M-008-007. A.C. acknowledges support from the National Science Foundation CAREER award 1455090, "CAREER: Radio and gravitational-wave emission from the largest explosions since the Big Bang." T.P. acknowledges the support of Advanced ERC grant TReX. B.E.C. thanks SMARTS 1.3 m Queue Manager Bryndis Cruz for prompt scheduling of the SMARTS observations. Basic research in radio astronomy at the Naval Research Laboratory (NRL) is funded by 6.1 Base funding. Construction and installation of VLITE was supported by NRL Sustainment Restoration and Maintenance funding. K.P.M.'s research is supported by the Oxford Centre for Astrophysical Surveys, which is funded through the Hintze Family Charitable Foundation. J.S. and A. G. are grateful for support from the Knut and Alice Wallenberg Foundation. GREAT is funded by the Swedish Research Council (V.R.). E.O.O. is grateful for the support by grants from the Israel Science Foundation, Minerva, Israeli ministry of Science, the US-Israel Binational Science Foundation, and the I-CORE Program of the Planning and Budgeting Committee and The Israel Science Foundation. We thank the staff of the GMRT that made these observations possible. The GMRT is run by the National Centre for Radio Astrophysics of the Tata Institute of Fundamental Research. AYQH was supported by a National Science Foundation Graduate Research Fellowship under grant No. DGE-1144469. S.R. has been supported by the Swedish Research Council (VR) under grant number 2016 03657 3, by the Swedish National Space Board under grant number Dnr. 107/16 and by the research environment grant "Gravitational Radiation and Electromagnetic Astrophysical Transients (GREAT)" funded by the Swedish Research council (V.R.) under Dnr. 2016-06012. We acknowledge the support of the Science and Engineering Research Board, Department of Science and Technology, India and the Indo-US Science and Technology Foundation for the GROWTH-India project.

*(HAWC)* We acknowledge the support from: the US National Science Foundation (NSF); the US Department of Energy Office of High-Energy Physics; the Laboratory Directed Research and Development (LDRD) program of Los Alamos National Laboratory; Consejo Nacional de Ciencia y Tecnología (CONACyT), Mexico (grants 271051, 232656, 167281, 260378, 179588, 239762, 254964, 271737, 258865, 243290); Red HAWC, Mexico; DGAPA-UNAM (grants RG100414, IN111315, IN111716-3, IA102715, 109916); VIEP-BUAP; the University of Wisconsin Alumni Research Foundation; the Institute of Geophysics, Planetary Physics, and Signatures at Los Alamos National Laboratory; Polish Science Centre grant DEC-2014/13/B/ST9/945. We acknowledge the support of the Science and Engineering Research Board, Department of Science and Technology, India and the Indo-US Science and Technology Foundation for the GROWTH-India project.

*(H.E.S.S.)* The support of the Namibian authorities and of the University of Namibia in facilitating the construction and operation of H.E.S.S. is gratefully acknowledged, as is the support by the German Ministry for Education and Research (BMBF), the Max Planck Society, the German Research Foundation (DFG), the Alexander von Humboldt Foundation, the Deutsche Forschungsgemeinschaft, the French Ministry for Research, the CNRS-IN2P3 and the Astroparticle

Interdisciplinary Programme of the CNRS, the U.K. Science and Technology Facilities Council (STFC), the IPNP of the Charles University, the Czech Science Foundation, the Polish National Science Centre, the South African Department of Science and Technology and National Research Foundation, the University of Namibia, the National Commission on Research, Science and Technology of Namibia (NCRST), the Innsbruck University, the Austrian Science Fund (FWF), and the Austrian Federal Ministry for Science, Research and Economy, the University of Adelaide and the Australian Research Council, the Japan Society for the Promotion of Science and by the University of Amsterdam. We appreciate the excellent work of the technical support staff in Berlin, Durham, Hamburg, Heidelberg, Palaiseau, Paris, Saclay, and in Namibia in the construction and operation of the equipment. This work benefited from services provided by the H.E.S.S. Virtual Organisation, supported by the national resource providers of the EGI Federation.

*(Insight-HXMT)* The *Insight*-HXMT team acknowledges the support from the China National Space Administration (CNSA), the Chinese Academy of Sciences (CAS; grant No. XDB23040400), and the Ministry of Science and Technology of China (MOST; grant No. 2016YFA0400800).

*(IceCube)* We acknowledge the support from the following agencies: U.S. National Science Foundation-Office of Polar Programs, U.S. National Science Foundation-Physics Division, University of Wisconsin Alumni Research Foundation, the Grid Laboratory of Wisconsin (GLOW) grid infrastructure at the University of Wisconsin—Madison, the Open Science Grid (OSG) grid infrastructure; U.S. Department of Energy, and National Energy Research Scientific Computing Center, the Louisiana Optical Network Initiative (LONI) grid computing resources; Natural Sciences and Engineering Research Council of Canada, WestGrid and Compute/Calcul Canada; Swedish Research Council, Swedish Polar Research Secretariat, Swedish National Infrastructure for Computing (SNIC), and Knut and Alice Wallenberg Foundation, Sweden; German Ministry for Education and Research (BMBF), Deutsche Forschungsgemeinschaft (DFG), Helmholtz Alliance for Astroparticle Physics (HAP), Initiative and Networking Fund of the Helmholtz Association, Germany; Fund for Scientific Research (FNRS-FWO), FWO Odysseus programme, Flanders Institute to encourage scientific and technological research in industry (IWT), Belgian Federal Science Policy Office (Belspo); Marsden Fund, New Zealand; Australian Research Council; Japan Society for Promotion of Science (JSPS); the Swiss National Science Foundation (SNSF), Switzerland; National Research Foundation of Korea (NRF); Villum Fonden, Danish National Research Foundation (DNRF), Denmark.

*(IKI-GW)* A.S.P., A.A.V., E.D.M., and P.Y.u.M. acknowledge the support from the Russian Science Foundation (grant 15-12-30015). V.A.K., A.V.K., and I.V.R. acknowledge the Science and Education Ministry of Kazakhstan (grant No. 0075/GF4). R.I. is grateful to the grant RUSTAVELI FR/379/6-300/14 for partial support. We acknowledge the excellent help in obtaining Chilescope data from Sergei Pogrebsskiy and Ivan Rubzov.

*(INTEGRAL)* This work is based on observations with INTEGRAL, an ESA project with instruments and science data center funded by ESA member states (especially the PI countries: Denmark, France, Germany, Italy, Switzerland, Spain), and with the participation of Russia and the USA. The INTEGRAL SPI project has been completed under the responsibility and leadership of CNES. The SPI-ACS detector system has been provided by MPE Garching/Germany. The SPI team is grateful to ASI, CEA, CNES, DLR, ESA, INTA, NASA, and OSTC for their support. The Italian INTEGRAL team acknowledges the support of ASI/INAF agreement No. 2013-025-R.1. R.D. and A.v.K. acknowledge the German INTEGRAL support through DLR grant 50 OG 1101. A.L. and R.S. acknowledge the support from the Russian Science Foundation (grant 14-22-00271). A.D. is funded by Spanish MINECO/FEDER grant ESP2015-65712-C5-1-R.

*(IPN)* K.H. is grateful for support under NASA grant NNX15AE60G. R.L.A. and D.D.F. are grateful for support under RFBR grant 16-29-13009-ofi-m.

*(J-GEM)* MEXT KAKENHI (JP17H06363, JP15H00788, JP24103003, JP10147214, JP10147207), JSPS KAKENHI (JP16H02183, JP15H02075, JP15H02069, JP26800103, JP25800103), Inter-University Cooperation Program of the MEXT, the NINS program for cross-disciplinary science study, the Toyota Foundation (D11-R-0830), the Mitsubishi Foundation, the Yamada Science Foundation, Inoue Foundation for Science, the National Research Foundation of South Africa.

*(KU)* The Korea-Uzbekistan Consortium team acknowledges the support from the NRF grant No. 2017R1A3A3001362, and the KASI grant 2017-1-830-03. This research has made use of the KMTNet system operated by KASI.

*(Las Cumbres)* Support for I.A. and J.B. was provided by NASA through the Einstein Fellowship Program, grants PF6-170148 and PF7-180162, respectively. D.A.H., C.M., and G.H. are supported by NSF grant AST-1313484. D.P. and D..M acknowledge support by Israel Science Foundation grant 541/17. This work makes use of observations from the LCO network.

*(LIGO and Virgo)* The authors gratefully acknowledge the support of the United States National Science Foundation (NSF) for the construction and operation of the LIGO Laboratory and Advanced LIGO as well as the Science and Technology Facilities Council (STFC) of the United Kingdom, the Max-Planck-Society (MPS), and the State of Niedersachsen/Germany for support of the construction of Advanced LIGO and construction and operation of the GEO600 detector. Additional support for advanced LIGO was provided by the Australian Research Council. The authors gratefully acknowledge the Italian Istituto Nazionale di Fisica Nucleare (INFN), the French Centre National de la Recherche Scientifique (CNRS) and the Foundation for Fundamental Research on Matter supported by the Netherlands Organisation for Scientific Research, for the construction and operation of the Virgo detector and the creation and support of the EGO consortium. The authors also gratefully acknowledge research support from these agencies as well as by the Council of Scientific and Industrial Research of India, the Department of Science and Technology, India, the Science & Engineering Research Board (SERB), India, the Ministry of Human Resource Development, India, the Spanish Agencia Estatal de Investigación, the Vicepresidència i Conselleria d'Innovació Recerca i Turisme and the Conselleria d'Educació i Universitat del Govern de les Illes Balears, the Conselleria d'Educació Investigació Cultura i Esport de la Generalitat Valenciana, the National Science Centre of Poland, the Swiss National Science Foundation (SNSF), the Russian Foundation for Basic Research, the Russian Science Foundation, the European Commission, the European Regional Development Funds (ERDF), the Royal Society, the Scottish Funding Council, the Scottish Universities Physics Alliance, the Hungarian Scientific Research Fund (OTKA), the Lyon Institute of Origins (LIO), the National Research, Development and Innovation Office Hungary (NKFI), the National Research Foundation of Korea, Industry Canada and the Province of Ontario through the Ministry of Economic Development and Innovation, the Natural Science and Engineering Research Council Canada, the Canadian Institute for Advanced Research, the Brazilian Ministry of Science, Technology, Innovations, and Communications, the International Center for Theoretical Physics South American Institute for Fundamental Research (ICTP-SAIFR), the Research Grants Council of Hong Kong, the National Natural Science Foundation of China (NSFC), the China National Space Administration (CNSA) and the Chinese Academy of Sciences (CAS), the Ministry of Science and Technology of China (MOST), the Leverhulme Trust, the Research Corporation, the Ministry of Science and Technology (MOST), Taiwan and the Kavli Foundation. The authors gratefully acknowledge the support of the NSF, STFC, MPS, INFN, CNRS, and the State of Niedersachsen/Germany for provision of computational resources. The *MAXI* team acknowledges the support by JAXA, RIKEN, and MEXT KAKENHI grant number JP 17H06362. The National Radio Astronomy Observatory is a facility of the National Science Foundation operated under cooperative agreement by Associated Universities, Inc. The European VLBI Network is a joint facility of independent European, African, Asian, and North American radio astronomy institutes. Scientific results from data presented in this publication are derived from the following EVN project code: RP029. e-MERLIN is a National Facility operated by the University of Manchester at Jodrell Bank Observatory on behalf of STFC. The collaboration between LIGO/Virgo and EVN/e-MERLIN is part of a project that has received funding from the European Union's Horizon 2020 research and innovation programme under grant agreement No. 653477. We thank Britt Griswold (NASA/GSFC) for graphic arts. P.G.J. acknowledges ERC–Consolidator grant No. 647208. We thank the GMRT staff for prompt scheduling of these observations. The GMRT is run by the National Center for Radio Astrophysics of the Tata Institute of Fundamental Research. INAF, Italian Institute of Astrophysics ASI, Italian Space Agency. This work is part of the research program Innovational Research Incentives Scheme (Vernieuwingsimpuls), which is financed by the Netherlands Organization for Scientific Research through the NWO VIDI grant No. 639.042.612-Nissanke and NWO TOP grant No. 62002444–Nissanke. We thank ESO for granting full access to all the LVC MoU partners of the observations of GW170817 obtained with NACO and VISIR under the Observatory program 60.A-9392.

*(LOFAR)* LOFAR, the Low-Frequency Array designed and constructed by ASTRON, has facilities in several countries that are owned by various parties (each with their own funding sources) and that are collectively operated by the International LOFAR Telescope (ILT) foundation under a joint scientific policy. P.G.J. acknowledges support from ERC grant number 647208. R.F. was partially funded by ERC Advanced Investigator Grant 267607 "4 PI SKY."

*(MASTER)* Development Programme of Lomonosov Moscow State University, Sergey Bodrov of Moscow Union OPTICA, Russian Scientific Foundation 16-12-00085, National Research Foundation of South Africa, Russian Federation Ministry of Education and Science (14.B25.31.0010, 14.593.21.0005, 3.10131.2017/NM), RFBR 17-52-80133

*(MAXI)* The *MAXI* team acknowledges support by JAXA, RIKEN, and MEXT KAKENHI grant number JP 17H06362.

*(Nordic Optical Telescope)* J.P.U.F. acknowledges the Carlsberg foundation for funding for the NTE project. D.X. acknowledges the support by the One-Hundred-Talent Program of the Chinese Academy of Sciences (CAS) and by the Strategic Priority Research Program "Multi-wavelength Gravitational Wave Universe" of the CAS (No. XDB23000000). Based on observations made with the Nordic Optical Telescope (program 55-013), operated by the Nordic Optical Telescope Scientific Association.

*(OzGrav)* Part of this research was funded by the Australian Research Council Centre of Excellence for Gravitational Wave Discovery (OzGrav), CE170100004 and the Australian Research Council Centre of Excellence for All-sky Astrophysics (CAASTRO), CE110001020. J.C. acknowledges the Australian Research Council Future Fellowship grant FT130101219. Research support to I.A. is provided by the Australian Astronomical Observatory (AAO). A.T.D. acknowledges the support of an Australian Research Council Future Fellowship (FT150100415). Based in part on data acquired through the Australian Astronomical Observatory. We acknowledge the traditional owners of the land on which the AAT stands, the Gamilaraay people, and pay our respects to elders past and present. The Etelman/VIRT team acknowledge NASA grant NNX13AD28A.

*(Pan-STARRS)* The Pan-STARRS1 observations were supported in part by NASA grant No. NNX14AM74G issued through the SSO Near Earth Object Observations Program and the Queen's University Belfast. The Pan-STARRS1 Surveys were made possible through contributions by the Institute for Astronomy, the University of Hawaii, the Pan-STARRS Project Office, the Max-Planck Society and its participating institutes, the Max Planck Institute for Astronomy, Heidelberg and the Max Planck Institute for Extraterrestrial Physics, Garching, The Johns Hopkins University, Durham University, the University of Edinburgh, the Queen's University Belfast, the Harvard-Smithsonian Center for Astrophysics, the LCO Global Telescope Network Incorporated, the National Central University of Taiwan, the Space Telescope Science Institute, and the National Aeronautics and Space Administration under grant No. NNX08AR22G issued through the Planetary Science Division of the NASA Science Mission Directorate, the National Science Foundation grant No. AST-1238877, the University of Maryland, Eotvos Lorand University (ELTE), and the Los Alamos National Laboratory. The Pan-STARRS1 Surveys are archived at the Space Telescope Science Institute (STScI) and can be accessed through MAST, the Mikulski Archive for Space Telescopes. Additional support for the Pan-STARRS1 public science archive is provided by the Gordon and Betty Moore Foundation.

*(Pi of the Sky)* The Pi of the Sky team is grateful for the support of the ESAt/INTA-CEDEA personnel in Mazagón, Huelva (Spain). Analysis of the Pi of the Sky data was based on the LUIZA software developed within the GLORIA project, funded from the European Union Seventh Framework Programme (FP7/2007-2013) under grant 283783.

*(SALT)* D.B., S.M.C., E.R.C., S.B.P., P.V., and T.W. acknowledge support from the South African National Research Foundation. M.M.S. gratefully acknowledges the support of the late Paul Newman and the Newmans Own Foundation. We are most grateful for the DDT allocation for the SALT observations.

*(SKA)* R.F. was partially funded by ERC Advanced Investigator Grant 267607 "4 PI SKY."

*(Swift)* Funding for the *Swift* mission in the UK is provided by the UK Space Agency. The *Swift* team at the MOC at Penn State acknowledges support from NASA contract NAS5-00136. The Italian *Swift* team acknowledge support from ASI-INAF grant I/004/11/3.

*(TOROS)* We thank support from the USA Air Force Office of International Scientific Research (AFOSR/IO), the Dirección de Investigación de la Universidad de La Serena, the Consejo Nacional de Investigaciones Científicas y Técnicas of Argentina, the FAPESP, and the Observatorio Nacional-MCT of Brasil.

*(TTU Group)* A.C. and N.T.P. acknowledge support from the NSF CAREER Award 1455090: "CAREER: Radio and gravitational-wave emission from the largest explosions since the Big Bang." The National Radio Astronomy Observatory is a facility of the National Science Foundation operated under cooperative agreement by Associated Universities, Inc.

*(VINROUGE)* Based on observations made with ESO telescopes at the La Silla Paranal Observatory under programmes ID 099.D-0668, 099.D-0116, 099.D-0622, 179.A-2010, and 198. D-2010; and with the NASA/ESA *Hubble Space Telescope* observations under programs GO 14771, GO 14804, GO 14850. The VISTA observations were processed by C.G.F. at the Cambridge Astronomy Survey Unit (CASU), which is funded by the UK Science and Technology Research Council under grant ST/N005805/1. This research used resources provided by the Los Alamos National Laboratory Institutional Computing Program, which is supported by the U.S. Department of Energy National Nuclear Security Administration under Contract No. DE-AC52-06NA25396. We acknowledge support to the following bodies: the ERC (grant No. 725246); STFC via grant ST/P000495/1; VILLUM FONDEN (investigator grant project number 16599); the Spanish project AYA 2014-58381-P; the Juan de la Cierva Incorporación fellowship IJCI-2014-21669; the Juan de la Cierva Incorporación fellowship IJCI-2015-26153; the NRFK grant No. 2017R1A3A3001362; grants GO718062A and HSTG014850001A; the Swedish Research Council (VR) under grant number 2016-03657-3; the Swedish National Space Board under grant number Dnr. 107/16; the research environment grant "Gravitational Radiation and Electromagnetic Astrophysical Transients (GREAT)" under Dnr 2016-06012; UKSA.

*(Zadko)* The Zadko Telescope was made possible by a philanthropic donation by James Zadko to the University of Western Australia (UWA). Zadko Telescope operations are supported by UWA and the Australian Research Council Centre of Excellence OzGrav CE170100004. The TAROT network of telescopes is supported by the French Centre National de la Recherche Scientifique (CNRS), the Observatoire de la Côte d'Azur (OCA), and we thank the expertise and support of the Observatoire des Sciences de l'Univers, Institut Pythéas, Aix-Marseille University. The FIGARONet network is supported under the Agence Nationale de la Recherche (ANR) grant 14-CE33. The paper-writing team would like to thank Britt Griswold (NASA/GSFC) and Aaron Geller (Northwestern/NUIT/CIERA) for assistance with graphics.

B. P. Abbott[1], R. Abbott[1], T. D. Abbott[2], F. Acernese[3,4], K. Ackley[5,6], C. Adams[7], T. Adams[8], P. Addesso[9], R. X. Adhikari[1], V. B. Adya[10], C. Affeldt[10], M. Afrough[11], B. Agarwal[12], M. Agathos[13], K. Agatsuma[14], N. Aggarwal[15], O. D. Aguiar[16], L. Aiello[17,18], A. Ain[19], P. Ajith[20], B. Allen[10,21,22], G. Allen[12], A. Allocca[23,24], P. A. Altin[25], A. Amato[26], A. Ananyeva[1], S. B. Anderson[1], W. G. Anderson[21], S. V. Angelova[27], S. Antier[28], S. Appert[1], K. Arai[1], M. C. Araya[1], J. S. Areeda[29], N. Arnaud[28,30], K. G. Arun[31], S. Ascenzi[32,33], G. Ashton[10], M. Ast[34], S. M. Aston[7], P. Astone[35], D. V. Atallah[36], P. Aufmuth[22], C. Aulbert[10], K. AultONeal[37], C. Austin[2], A. Avila-Alvarez[29], S. Babak[38], P. Bacon[39], M. K. M. Bader[14], S. Bae[40], P. T. Baker[41], F. Baldaccini[42,43], G. Ballardin[30], S. W. Ballmer[44], S. Banagiri[45], J. C. Barayoga[1], S. E. Barclay[46], B. C. Barish[1], D. Barker[47], K. Barkett[48], F. Barone[3,4], B. Barr[46], L. Barsotti[15], M. Barsuglia[39], D. Barta[49], S. D. Barthelmy[50], J. Bartlett[47], I. Bartos[51,5], R. Bassiri[52], A. Basti[23,24], J. C. Batch[47], M. Bawaj[53,43], J. C. Bayley[46], M. Bazzan[54,55], B. Bécsy[56], C. Beer[10], M. Bejger[57], I. Belahcene[28], A. S. Bell[46], B. K. Berger[1], G. Bergmann[10], J. J. Bero[58], C. P. L. Berry[59], D. Bersanetti[60], A. Bertolini[14], J. Betzwieser[7], S. Bhagwat[44], R. Bhandare[61], I. A. Bilenko[62], G. Billingsley[1], C. R. Billman[5], J. Birch[7], R. Birney[63], O. Birnholtz[10], S. Biscans[1,15], S. Biscoveanu[64,6], A. Bisht[22], M. Bitossi[30,24], C. Biwer[44], M. A. Bizouard[28], J. K. Blackburn[1], J. Blackman[48], C. D. Blair[1,65], D. G. Blair[65], R. M. Blair[47], S. Bloemen[66], O. Bock[10], N. Bode[10], M. Boer[67], G. Bogaert[67], A. Bohe[38], F. Bondu[68], E. Bonilla[52], R. Bonnand[8], B. A. Boom[14], R. Bork[1], V. Boschi[30,24], S. Bose[69,19], K. Bossie[7], Y. Bouffanais[39], A. Bozzi[30], C. Bradaschia[24], P. R. Brady[21], M. Branchesi[17,18], J. E. Brau[70], T. Briant[71], A. Brillet[67], M. Brinkmann[10], V. Brisson[28], P. Brockill[21], J. E. Broida[72], A. F. Brooks[1], D. A. Brown[44], D. D. Brown[73], S. Brunett[1], C. C. Buchanan[2], A. Buikema[15], T. Bulik[74], H. J. Bulten[75,14], A. Buonanno[38,76], D. Buskulic[8], C. Buy[39], R. L. Byer[52], M. Cabero[10], L. Cadonati[77], G. Cagnoli[26,78], C. Cahillane[1], J. Calderón Bustillo[77], T. A. Callister[1], E. Calloni[79,4], J. B. Camp[50], M. Canepa[60,80], P. Canizares[66], K. C. Cannon[81], H. Cao[73], J. Cao[82], C. D. Capano[10], E. Capocasa[39], F. Carbognani[30], S. Caride[83], M. F. Carney[84], J. Casanueva Diaz[28], C. Casentini[32,33], S. Caudill[14,21], M. Cavaglià[11], F. Cavalier[28], R. Cavalieri[30], G. Cella[24], C. B. Cepeda[1], P. Cerdá-Durán[85], G. Cerretani[23,24], E. Cesarini[33,86], S. J. Chamberlin[64], M. Chan[46], S. Chao[87], P. Charlton[88], E. Chase[89], E. Chassande-Mottin[39], D. Chatterjee[21], K. Chatziioannou[90], B. D. Cheeseboro[41], H. Y. Chen[91], X. Chen[65], Y. Chen[48], H.-P. Cheng[5], H. Chia[5], A. Chincarini[60], A. Chiummo[30], T. Chmiel[84], H. S. Cho[92], M. Cho[76], J. H. Chow[25], N. Christensen[72,67], Q. Chu[65], A. J. K. Chua[13], S. Chua[71], A. K. W. Chung[93], S. Chung[65], G. Ciani[5,54,55], R. Ciolfi[94,95], C. E. Cirelli[52], A. Cirone[60,80], F. Clara[47], J. A. Clark[77], P. Clearwater[96], F. Cleva[67], C. Cocchieri[11], E. Coccia[17,18], P.-F. Cohadon[71], D. Cohen[28], A. Colla[97,35], C. G. Collette[98],

L. R. Cominsky[99], M. Constancio Jr.[16], L. Conti[55], S. J. Cooper[59], P. Corban[7], T. R. Corbitt[2], I. Cordero-Carrión[100], K. R. Corley[51], N. Cornish[101], A. Corsi[83], S. Cortese[30], C. A. Costa[16], M. W. Coughlin[72,1], S. B. Coughlin[89], J.-P. Coulon[67], S. T. Countryman[51], P. Couvares[1], P. B. Covas[102], E. E. Cowan[77], D. M. Coward[65], M. J. Cowart[7], D. C. Coyne[1], R. Coyne[83], J. D. E. Creighton[21], T. D. Creighton[103], J. Cripe[2], S. G. Crowder[104], T. J. Cullen[29,2], A. Cumming[46], L. Cunningham[46], E. Cuoco[30], T. Dal Canton[50], G. Dálya[56], S. L. Danilishin[22,10], S. D'Antonio[33], K. Danzmann[22,10], A. Dasgupta[105], C. F. Da Silva Costa[5], V. Dattilo[30], I. Dave[61], M. Davier[28], D. Davis[44], E. J. Daw[106], B. Day[77], S. De[44], D. DeBra[52], J. Degallaix[26], M. De Laurentis[17,4], S. Deléglise[71], W. Del Pozzo[59,23,24], N. Demos[15], T. Denker[10], T. Dent[10], R. De Pietri[107,108], V. Dergachev[38], R. De Rosa[79,4], R. T. DeRosa[7], C. De Rossi[26,30], R. DeSalvo[109], O. de Varona[10], J. Devenson[27], S. Dhurandhar[19], M. C. Díaz[103], L. Di Fiore[4], M. Di Giovanni[110,95], T. Di Girolamo[51,79,4], A. Di Lieto[23,24], S. Di Pace[97,35], I. Di Palma[97,35], F. Di Renzo[23,24], Z. Doctor[91], V. Dolique[26], F. Donovan[15], K. L. Dooley[11], S. Doravari[10], I. Dorrington[36], R. Douglas[46], M. Dovale Álvarez[59], T. P. Downes[21], M. Drago[10], C. Dreissigacker[10], J. C. Driggers[47], Z. Du[82], M. Ducrot[8], P. Dupej[46], S. E. Dwyer[47], T. B. Edo[106], M. C. Edwards[72], A. Effler[7], H.-B. Eggenstein[38,10], P. Ehrens[1], J. Eichholz[1], S. S. Eikenberry[5], R. A. Eisenstein[15], R. C. Essick[15], D. Estevez[8], Z. B. Etienne[41], T. Etzel[1], M. Evans[15], T. M. Evans[7], M. Factourovich[51], V. Fafone[32,33,17], H. Fair[44], S. Fairhurst[36], X. Fan[82], S. Farinon[60], B. Farr[91], W. M. Farr[59], E. J. Fauchon-Jones[36], M. Favata[111], M. Fays[36], C. Fee[84], H. Fehrmann[10], J. Feicht[1], M. M. Fejer[52], A. Fernandez-Galiana[15], I. Ferrante[23,24], E. C. Ferreira[16], F. Ferrini[30], F. Fidecaro[23,24], D. Finstad[44], I. Fiori[30], D. Fiorucci[39], M. Fishbach[91], R. P. Fisher[44], M. Fitz-Axen[45], R. Flaminio[26,112], M. Fletcher[46], H. Fong[90], J. A. Font[85,113], P. W. F. Forsyth[25], S. S. Forsyth[77], J.-D. Fournier[67], S. Frasca[97,35], F. Frasconi[24], Z. Frei[56], A. Freise[59], R. Frey[70], V. Frey[28], E. M. Fries[1], P. Fritschel[15], V. V. Frolov[7], P. Fulda[5], M. Fyffe[7], H. Gabbard[46], B. U. Gadre[19], S. M. Gaebel[59], J. R. Gair[114], L. Gammaitoni[42], M. R. Ganija[73], S. G. Gaonkar[19], C. Garcia-Quiros[102], F. Garufi[79,4], B. Gateley[47], S. Gaudio[37], G. Gaur[115], V. Gayathri[116], N. Gehrels[50,954], G. Gemme[60], E. Genin[30], A. Gennai[24], D. George[12], J. George[61], L. Gergely[117], V. Germain[8], S. Ghonge[77], Abhirup Ghosh[20], Archisman Ghosh[20,14], S. Ghosh[66,14,21], J. A. Giaime[2,7], K. D. Giardina[7], A. Giazotto[24], K. Gill[37], L. Glover[109], E. Goetz[118], R. Goetz[5], S. Gomes[36], B. Goncharov[6], G. González[2], J. M. Gonzalez Castro[23,24], A. Gopakumar[119], M. L. Gorodetsky[62], S. E. Gossan[1], M. Gosselin[30], R. Gouaty[8], A. Grado[120,4], C. Graef[46], M. Granata[26], A. Grant[46], S. Gras[15], C. Gray[47], G. Greco[121,122], A. C. Green[59], E. M. Gretarsson[37], B. Griswold[70], P. Groot[66], H. Grote[10], S. Grunewald[38], P. Gruning[28], G. M. Guidi[121,122], X. Guo[82], A. Gupta[64], M. K. Gupta[105], K. E. Gushwa[1], E. K. Gustafson[1], R. Gustafson[118], O. Halim[18,17], B. R. Hall[69], E. D. Hall[15], E. Z. Hamilton[36], G. Hammond[46], M. Haney[123], M. M. Hanke[10], J. Hanks[47], C. Hanna[64], M. D. Hannam[36], O. A. Hannuksela[93], J. Hanson[7], T. Hardwick[2], J. Harms[17,18], G. M. Harry[124], I. W. Harry[38], M. J. Hart[46], C.-J. Haster[90], K. Haughian[46], J. Healy[58], A. Heidmann[71], M. C. Heintze[7], H. Heitmann[67], P. Hello[28], G. Hemming[30], M. Hendry[46], I. S. Heng[46], J. Hennig[46], A. W. Heptonstall[1], M. Heurs[10,22], S. Hild[46], T. Hinderer[66], D. Hoak[30], D. Hofman[26], K. Holt[7], D. E. Holz[91], P. Hopkins[36], C. Horst[21], J. Hough[46], E. A. Houston[46], E. J. Howell[65], A. Hreibi[67], Y. M. Hu[10], E. A. Huerta[12], D. Huet[28], B. Hughey[37], S. Husa[102], S. H. Huttner[46], T. Huynh-Dinh[7], N. Indik[10], R. Inta[83], G. Intini[97,35], H. N. Isa[46], J.-M. Isac[71], M. Isi[1], B. R. Iyer[20], K. Izumi[47], T. Jacqmin[71], K. Jani[77], P. Jaranowski[125], S. Jawahar[63], F. Jiménez-Forteza[102], W. W. Johnson[2], D. I. Jones[126], R. Jones[46], R. J. G. Jonker[14], L. Ju[65], J. Junker[10], C. V. Kalaghatgi[36], V. Kalogera[89], B. Kamai[1], S. Kandhasamy[7], G. Kang[40], J. B. Kanner[1], S. J. Kapadia[21], S. Karki[70], K. S. Karvinen[10], M. Kasprzack[2], M. Katolik[12], E. Katsavounidis[15], W. Katzman[7], S. Kaufer[22], K. Kawabe[47], F. Kéfélian[67], D. Keitel[46], A. J. Kemball[12], R. Kennedy[106], C. Kent[36], J. S. Key[127], F. Y. Khalili[62], I. Khan[17,33], S. Khan[10], Z. Khan[105], E. A. Khazanov[128], N. Kijbunchoo[25], Chunglee Kim[129], J. C. Kim[130], K. Kim[93], W. Kim[73], W. S. Kim[131], Y.-M. Kim[92], S. J. Kimbrell[77], E. J. King[73], P. J. King[47], M. Kinley-Hanlon[124], R. Kirchhoff[10], J. S. Kissel[47], L. Kleybolte[34], S. Klimenko[5], T. D. Knowles[41], P. Koch[10], S. M. Koehlenbeck[10], S. Koley[14], V. Kondrashov[1], A. Kontos[15], M. Korobko[34], W. Z. Korth[1], I. Kowalska[74], D. B. Kozak[1], C. Krämer[10], V. Kringel[10], B. Krishnan[10], A. Królak[132,133], G. Kuehn[10], P. Kumar[90], R. Kumar[105], S. Kumar[20], L. Kuo[87], A. Kutynia[132], S. Kwang[21], B. D. Lackey[38], K. H. Lai[93], M. Landry[47], R. N. Lang[134], J. Lange[58], B. Lantz[52], R. K. Lanza[15], S. L. Larson[89], A. Lartaux-Vollard[28], P. D. Lasky[6], M. Laxen[7], A. Lazzarini[1], C. Lazzaro[55], P. Leaci[97,35], S. Leavey[46], C. H. Lee[92], H. K. Lee[135], H. M. Lee[136], H. W. Lee[130], K. Lee[46], J. Lehmann[10], A. Lenon[41], M. Leonardi[110,95], N. Leroy[28], N. Letendre[8], Y. Levin[6], T. G. F. Li[93], S. D. Linker[109], T. B. Littenberg[137], J. Liu[65], R. K. L. Lo[93], N. A. Lockerbie[63], L. T. London[36], J. E. Lord[44], M. Lorenzini[17,18], V. Loriette[138], M. Lormand[7], G. Losurdo[24], J. D. Lough[10], C. O. Lousto[58], G. Lovelace[29], H. Lück[22,10], D. Lumaca[32,33], A. P. Lundgren[10], R. Lynch[15], Y. Ma[48], R. Macas[36], S. Macfoy[27], B. Machenschalk[10], M. MacInnis[15], D. M. Macleod[36], I. Magaña Hernandez[21], F. Magaña-Sandoval[44], L. Magaña Zertuche[44], R. M. Magee[64], E. Majorana[35], I. Maksimovic[138], N. Man[67], V. Mandic[45], V. Mangano[46], G. L. Mansell[25], M. Manske[21,25], M. Mantovani[30], F. Marchesoni[53,43], F. Marion[8], S. Márka[51], Z. Márka[51], C. Markakis[12], A. S. Markosyan[52], A. Markowitz[1], E. Maros[1], A. Marquina[100], P. Marsh[127], F. Martelli[121,122], L. Martellini[67], I. W. Martin[46], R. M. Martin[111], D. V. Martynov[15], K. Mason[15], E. Massera[106], A. Masserot[8], T. J. Massinger[1], M. Masso-Reid[46], S. Mastrogiovanni[97,35], A. Matas[45], F. Matichard[1,15], L. Matone[51], N. Mavalvala[15], N. Mazumder[69], R. McCarthy[47], D. E. McClelland[25], S. McCormick[7], L. McCuller[15], S. C. McGuire[139], G. McIntyre[1], J. McIver[1], D. J. McManus[25], L. McNeill[6], T. McRae[25], S. T. McWilliams[41], D. Meacher[64], G. D. Meadors[38,10], M. Mehmet[10],

J. Meidam[14], E. Mejuto-Villa[9], A. Melatos[96], G. Mendell[47], R. A. Mercer[21], E. L. Merilh[47], M. Merzougui[67], S. Meshkov[1], C. Messenger[46], C. Messick[64], R. Metzdorff[71], P. M. Meyers[45], H. Miao[59], C. Michel[26], H. Middleton[59], E. E. Mikhailov[140], L. Milano[79,4], A. L. Miller[5,97,35], B. B. Miller[89], J. Miller[15], M. Millhouse[101], M. C. Milovich-Goff[109], O. Minazzoli[67,141], Y. Minenkov[33], J. Ming[38], C. Mishra[142], S. Mitra[19], V. P. Mitrofanov[62], G. Mitselmakher[5], R. Mittleman[15], D. Moffa[84], A. Moggi[24], K. Mogushi[11], M. Mohan[30], S. R. P. Mohapatra[15], M. Montani[121,122], C. J. Moore[13], D. Moraru[47], G. Moreno[47], S. R. Morriss[103], B. Mours[8], C. M. Mow-Lowry[59], G. Mueller[5], A. W. Muir[36], Arunava Mukherjee[10], D. Mukherjee[21], S. Mukherjee[103], N. Mukund[19], A. Mullavey[7], J. Munch[73], E. A. Muñiz[44], M. Muratore[37], P. G. Murray[46], K. Napier[77], I. Nardecchia[32,33], L. Naticchioni[97,35], R. K. Nayak[143], J. Neilson[109], G. Nelemans[66,14], T. J. N. Nelson[7], M. Nery[10], A. Neunzert[118], L. Nevin[1], J. M. Newport[124], G. Newton[46,955], K. K. Y. Ng[93], P. Nguyen[70], T. T. Nguyen[25], D. Nichols[66], A. B. Nielsen[10], S. Nissanke[66,14], A. Nitz[10], A. Noack[10], F. Nocera[30], D. Nolting[7], C. North[36], L. K. Nuttall[36], J. Oberling[47], G. D. O'Dea[109], G. H. Ogin[144], J. J. Oh[131], S. H. Oh[131], F. Ohme[10], M. A. Okada[16], M. Oliver[102], P. Oppermann[10], Richard J. Oram[7], B. O'Reilly[7], R. Ormiston[45], L. F. Ortega[5], R. O'Shaughnessy[58], S. Ossokine[38], D. J. Ottaway[73], H. Overmier[7], B. J. Owen[83], A. E. Pace[64], J. Page[137], M. A. Page[65], A. Pai[116,145], S. A. Pai[61], J. R. Palamos[70], O. Palashov[128], C. Palomba[35], A. Pal-Singh[34], Howard Pan[87], Huang-Wei Pan[87], B. Pang[48], P. T. H. Pang[93], C. Pankow[89], F. Pannarale[36], B. C. Pant[61], F. Paoletti[24], A. Paoli[30], M. A. Papa[38,21,10], A. Parida[19], W. Parker[7], D. Pascucci[46], A. Pasqualetti[30], R. Passaquieti[23,24], D. Passuello[24], M. Patil[133], B. Patricelli[146,24], B. L. Pearlstone[46], M. Pedraza[1], R. Pedurand[26,147], L. Pekowsky[44], A. Pele[7], S. Penn[148], C. J. Perez[47], A. Perreca[1,110,95], L. M. Perri[89], H. P. Pfeiffer[90,38], M. Phelps[46], O. J. Piccinni[97,35], M. Pichot[67], F. Piergiovanni[121,122], V. Pierro[9], G. Pillant[30], L. Pinard[26], I. M. Pinto[9], M. Pirello[47], M. Pitkin[46], M. Poe[21], R. Poggiani[23,24], P. Popolizio[30], E. K. Porter[39], A. Post[10], J. Powell[46,149], J. Prasad[19], J. W. W. Pratt[37], G. Pratten[102], V. Predoi[36], T. Prestegard[21], L. R. Price[1], M. Prijatelj[10], M. Principe[9], S. Privitera[38], G. A. Prodi[110,95], L. G. Prokhorov[62], O. Puncken[10], M. Punturo[43], P. Puppo[35], M. Pürrer[38], H. Qi[21], V. Quetschke[103], E. A. Quintero[1], R. Quitzow-James[70], F. J. Raab[47], D. S. Rabeling[25], H. Radkins[47], P. Raffai[56], S. Raja[61], C. Rajan[61], B. Rajbhandari[83], M. Rakhmanov[103], K. E. Ramirez[103], A. Ramos-Buades[102], P. Rapagnani[97,35], V. Raymond[38], M. Razzano[23,24], J. Read[29], T. Regimbau[67], L. Rei[60], S. Reid[63], D. H. Reitze[1,5], W. Ren[12], S. D. Reyes[44], F. Ricci[97,35], P. M. Ricker[12], S. Rieger[10], K. Riles[118], M. Rizzo[58], N. A. Robertson[1,46], R. Robie[46], F. Robinet[28], A. Rocchi[33], L. Rolland[8], J. G. Rollins[1], V. J. Roma[70], R. Romano[3,4], C. L. Romel[47], J. H. Romie[7], D. Rosińska[150,57], M. P. Ross[151], S. Rowan[46], A. Rüdiger[10], P. Ruggi[30], G. Rutins[27], K. Ryan[47], S. Sachdev[1], T. Sadecki[47], L. Sadeghian[21], M. Sakellariadou[152], L. Salconi[30], M. Saleem[116], F. Salemi[10], A. Samajdar[143], L. Sammut[6], L. M. Sampson[89], E. J. Sanchez[1], L. E. Sanchez[1], N. Sanchis-Gual[85], V. Sandberg[47], J. R. Sanders[44], B. Sassolas[26], B. S. Sathyaprakash[64,36], P. R. Saulson[44], O. Sauter[118], R. L. Savage[47], A. Sawadsky[34], P. Schale[70], M. Scheel[48], J. Scheuer[89], J. Schmidt[205], P. Schmidt[1,66], R. Schnabel[34], R. M. S. Schofield[70], A. Schönbeck[34], E. Schreiber[10], D. Schuette[10,22], B. W. Schulte[10], B. F. Schutz[36,10], S. G. Schwalbe[37], J. Scott[46], S. M. Scott[25], E. Seidel[12], D. Sellers[7], A. S. Sengupta[153], D. Sentenac[30], V. Sequino[32,33,17], A. Sergeev[128], D. A. Shaddock[25], T. J. Shaffer[47], A. A. Shah[137], M. S. Shahriar[89], M. B. Shaner[109], L. Shao[38], B. Shapiro[52], P. Shawhan[76], A. Sheperd[21], D. H. Shoemaker[15], D. M. Shoemaker[77], K. Siellez[77], X. Siemens[21], M. Sieniawska[57], D. Sigg[47], A. D. Silva[16], L. P. Singer[50], A. Singh[38,10,22], A. Singhal[17,35], A. M. Sintes[102], B. J. J. Slagmolen[25], B. Smith[7], J. R. Smith[29], R. J. E. Smith[1,6], S. Somala[154], E. J. Son[131], J. A. Sonnenberg[21], B. Sorazu[46], F. Sorrentino[60], T. Souradeep[19], A. P. Spencer[46], A. K. Srivastava[105], K. Staats[37], A. Staley[51], M. Steinke[10], J. Steinlechner[34,46], S. Steinlechner[34], D. Steinmeyer[10], S. P. Stevenson[59,149], R. Stone[103], D. J. Stops[59], K. A. Strain[46], G. Stratta[121,122], S. E. Strigin[62], A. Strunk[47], R. Sturani[155], A. L. Stuver[7], T. Z. Summerscales[156], L. Sun[96], S. Sunil[105], J. Suresh[19], P. J. Sutton[36], B. L. Swinkels[30], M. J. Szczepańczyk[37], M. Tacca[14], S. C. Tait[46], C. Talbot[6], D. Talukder[70], D. B. Tanner[5], M. Tápai[117], A. Taracchini[38], J. D. Tasson[72], J. A. Taylor[137], R. Taylor[1], S. V. Tewari[148], T. Theeg[10], F. Thies[10], E. G. Thomas[59], M. Thomas[7], P. Thomas[47], K. A. Thorne[7], K. S. Thorne[48], E. Thrane[6], S. Tiwari[17,95], V. Tiwari[36], K. V. Tokmakov[63], K. Toland[46], M. Tonelli[23,24], Z. Tornasi[46], A. Torres-Forné[85], C. I. Torrie[1], D. Töyrä[59], F. Travasso[30,43], G. Traylor[7], J. Trinastic[5], M. C. Tringali[110,95], L. Trozzo[157,24], K. W. Tsang[14], M. Tse[15], R. Tso[1], L. Tsukada[81], D. Tsuna[81], D. Tuyenbayev[103], K. Ueno[21], D. Ugolini[158], C. S. Unnikrishnan[119], A. L. Urban[1], S. A. Usman[36], H. Vahlbruch[22], G. Vajente[1], G. Valdes[2], N. van Bakel[14], M. van Beuzekom[14], J. F. J. van den Brand[75,14], C. Van Den Broeck[14], D. C. Vander-Hyde[44], L. van der Schaaf[14], J. V. van Heijningen[14], A. A. van Veggel[46], M. Vardaro[54,55], V. Varma[48], S. Vass[1], M. Vasúth[49], A. Vecchio[59], G. Vedovato[55], J. Veitch[46], P. J. Veitch[73], K. Venkateswara[151], G. Venugopalan[1], D. Verkindt[8], F. Vetrano[121,122], A. Viceré[121,122], A. D. Viets[21], S. Vinciguerra[59], D. J. Vine[27], J.-Y. Vinet[67], S. Vitale[15], T. Vo[44], H. Vocca[42,43], C. Vorvick[47], S. P. Vyatchanin[62], A. R. Wade[1], L. E. Wade[84], M. Wade[84], R. Walet[14], M. Walker[29], L. Wallace[1], S. Walsh[38,10,21], G. Wang[17,122], H. Wang[59], J. Z. Wang[64], W. H. Wang[103], Y. F. Wang[93], R. L. Ward[25], J. Warner[47], M. Was[8], J. Watchi[98], B. Weaver[47], L.-W. Wei[10,22], M. Weinert[10], A. J. Weinstein[1], R. Weiss[15], L. Wen[65], E. K. Wessel[12], P. Wessels[10], J. Westerweck[10], T. Westphal[10], K. Wette[25], J. T. Whelan[58], S. E. Whitcomb[1], B. F. Whiting[5], C. Whittle[6], D. Wilken[10], D. Williams[46], R. D. Williams[1], A. R. Williamson[66], J. L. Willis[1,159], B. Willke[22,10], M. H. Wimmer[10], W. Winkler[10], C. C. Wipf[1], H. Wittel[10,22], G. Woan[46], J. Woehler[10], J. Wofford[58], K. W. K. Wong[93], J. Worden[47], J. L. Wright[46], D. S. Wu[10], D. M. Wysocki[58], S. Xiao[1], H. Yamamoto[1], C. C. Yancey[76], L. Yang[160], M. J. Yap[25],

M. Yazback[5], Hang Yu[15], Haocun Yu[15], M. Yvert[8], A. Zadrożny[132], M. Zanolin[37], T. Zelenova[30], J.-P. Zendri[55], M. Zevin[89], L. Zhang[1], M. Zhang[140], T. Zhang[46], Y.-H. Zhang[58], C. Zhao[65], M. Zhou[89], Z. Zhou[89], S. J. Zhu[38,10], X. J. Zhu[6], A. B. Zimmerman[90], M. E. Zucker[1,15], J. Zweizig[1],
(LIGO Scientific Collaboration and Virgo Collaboration),
C. A. Wilson-Hodge[137], E. Bissaldi[161,162], L. Blackburn[163,15], M. S. Briggs[164], E. Burns[50], W. H. Cleveland[165], V. Connaughton[165], M. H. Gibby[166], M. M Giles[166], A. Goldstein[165], R. Hamburg[164], P. Jenke[164], C. M. Hui[137], R. M. Kippen[167], D. Kocevski[137], S. McBreen[168], C. A. Meegan[164], W. S. Paciesas[165], S. Poolakkil[164], R. D. Preece[164], J. Racusin[50], O. J. Roberts[165], M. Stanbro[164], P. Veres[164], A. von Kienlin[169],
(Fermi GBM),
V. Savchenko[170], C. Ferrigno[170], E. Kuulkers[171], A. Bazzano[172], E. Bozzo[170], S. Brandt[173], J. Chenevez[173], T. J.-L. Courvoisier[170], R. Diehl[169], A. Domingo[174], L. Hanlon[168], E. Jourdain[175], P. Laurent[176,177], F. Lebrun[176], A. Lutovinov[178,179], A. Martin-Carrillo[168], S. Mereghetti[180], L. Natalucci[172], J. Rodi[172], J.-P. Roques[175], R. Sunyaev[178,181], P. Ubertini[172],
(INTEGRAL),
M. G. Aartsen[182], M. Ackermann[183], J. Adams[184], J. A. Aguilar[185], M. Ahlers[186], M. Ahrens[187], I. Al Samarai[188], D. Altmann[189], K. Andeen[190], T. Anderson[191], I. Ansseau[185], G. Anton[189], C. Argüelles[192], J. Auffenberg[193], S. Axani[192], H. Bagherpour[184], X. Bai[194], J. P. Barron[195], S. W. Barwick[196], V. Baum[197], R. Bay[198], J. J. Beatty[199,200], J. Becker Tjus[201], E. Bernardini[183], D. Z. Besson[202], G. Binder[198,203], D. Bindig[204], E. Blaufuss[205], S. Blot[183], C. Bohm[187], M. Börner[206], F. Bos[201], D. Bose[207], S. Böser[197], O. Botner[208], E. Bourbeau[186], J. Bourbeau[209], F. Bradascio[183], J. Braun[210], L. Brayeur[210], M. Brenzke[193], H.-P. Bretz[183], S. Bron[188], J. Brostean-Kaiser[183], A. Burgman[208], T. Carver[188], J. Casey[209], M. Casier[210], E. Cheung[205], D. Chirkin[209], A. Christov[188], K. Clark[211], L. Classen[212], S. Coenders[213], G. H. Collin[192], J. M. Conrad[192], D. F. Cowen[191,214], R. Cross[202], M. Day[209], J. P. A. M. de André[216], C. De Clercq[210], J. J. DeLaunay[191], H. Dembinski[217], S. De Ridder[218], P. Desiati[209], K. D. de Vries[210], G. de Wasseige[210], M. de With[219], T. DeYoung[216], J. C. Díaz-Vélez[209], V. di Lorenzo[197], H. Dujmovic[207], J. P. Dumm[187], M. Dunkman[191], E. Dvorak[194], B. Eberhardt[197], T. Ehrhardt[197], B. Eichmann[201], P. Eller[191], P. A. Evenson[217], S. Fahey[209], A. R. Fazely[220], J. Felde[205], K. Filimonov[198], C. Finley[187], S. Flis[187], A. Franckowiak[183], E. Friedman[205], T. Fuchs[206], T. K. Gaisser[217], J. Gallagher[221], L. Gerhardt[205], K. Ghorbani[209], W. Giang[195], T. Glauch[193], T. Glüsenkamp[189], A. Goldschmidt[205], J. G. Gonzalez[217], D. Grant[195], Z. Griffith[209], C. Haack[193], A. Hallgren[208], F. Halzen[209], K. Hanson[209], D. Hebecker[219], D. Heereman[185], K. Helbing[204], R. Hellauer[205], S. Hickford[204], J. Hignight[216], G. C. Hill[182], K. D. Hoffman[205], R. Hoffmann[204], B. Hokanson-Fasig[209], K. Hoshina[209,222], F. Huang[191], M. Huber[213], K. Hultqvist[187], M. Hünnefeld[206], S. In[207], A. Ishihara[223], E. Jacobi[183], G. S. Japaridze[224], M. Jeong[207], K. Jero[209], B. J. P. Jones[225], P. Kalaczynski[193], W. Kang[207], A. Kappes[212], T. Karg[183], A. Karle[209,209], A. Keivani[191], J. L. Kelley[209], A. Kheirandish[209], J. Kim[207], M. Kim[223], T. Kintscher[183], J. Kiryluk[226], T. Kittler[189], S. R. Klein[205,198], G. Kohnen[227], R. Koirala[217], H. Kolanoski[219], L. Köpke[197], C. Kopper[195], S. Kopper[228], J. P. Koschinsky[193], D. J. Koskinen[186], M. Kowalski[219,183], K. Krings[213], M. Kroll[201], G. Krückl[197], J. Kunnen[210], S. Kunwar[183], N. Kurahashi[229], T. Kuwabara[223], A. Kyriacou[182], M. Labare[218], J. L. Lanfranchi[191], M. J. Larson[186], F. Lauber[204], M. Lesiak-Bzdak[226], M. Leuermann[193], Q. R. Liu[209], L. Lu[223], J. Lünemann[210], W. Luszczak[209], J. Madsen[230], G. Maggi[210], K. B. M. Mahn[216], S. Mancina[209], R. Maruyama[231], K. Mase[223], R. Maunu[205], F. McNally[209], K. Meagher[185], M. Medici[186], M. Meier[206], T. Menne[206], G. Merino[209], T. Meures[185], S. Miarecki[205,198], J. Micallef[216], G. Momenté[197], T. Montaruli[188], R. W. Moore[195], M. Moulai[192], R. Nahnhauer[183], P. Nakarmi[228], U. Naumann[204], G. Neer[216], H. Niederhausen[226], S. C. Nowicki[195], D. R. Nygren[203], A. Obertacke Pollmann[204], A. Olivas[205], A. O'Murchadha[185], T. Palczewski[203,198], H. Pandya[217], D. V. Pankova[191], P. Peiffer[197], J. A. Pepper[228], C. Pérez de los Heros[208], D. Pieloth[206], E. Pinat[185], P. B. Price[198], G. T. Przybylski[203], C. Raab[185], L. Rädel[193], M. Rameez[186], K. Rawlins[232], I. C. Rea[213], R. Reimann[193], B. Relethford[229], M. Relich[223], E. Resconi[213], W. Rhode[206], M. Richman[229], S. Robertson[182], M. Rongen[193], C. Rott[207], T. Ruhe[206], D. Ryckbosch[218], D. Rysewyk[216], T. Sälzer[193], S. E. Sanchez Herrera[195], A. Sandrock[206], J. Sandroos[197], M. Santander[228], S. Sarkar[186,233], S. Sarkar[195], K. Satalecka[183], P. Schlunder[206], T. Schmidt[203], A. Schneider[209], S. Schoenen[193], S. Schöneberg[201], L. Schumacher[193], D. Seckel[217], S. Seunarine[230], J. Soedingrekso[206], D. Soldin[204], M. Song[205], G. M. Spiczak[230], C. Spiering[183], J. Stachurska[183], M. Stamatikos[199], T. Stanev[217], A. Stasik[183], J. Stettner[193], A. Steuer[197], T. Stezelberger[203], R. G. Stokstad[203], A. Stössl[223], N. L. Strotjohann[183], T. Stuttard[186], G. W. Sullivan[205], M. Sutherland[199], I. Taboada[234], J. Tatar[203,198], F. Tenholt[201], S. Ter-Antonyan[220], A. Terliuk[183], G. Tešić[191], S. Tilav[217], P. A. Toale[228], M. N. Tobin[209], S. Toscano[210], D. Tosi[209], M. Tselengidou[189], C. F. Tung[234], A. Turcati[213], C. F. Turley[191], B. Ty[209], E. Unger[208], M. Usner[183], J. Vandenbroucke[209], W. Van Driessche[218], N. van Eijndhoven[210], S. Vanheule[218], J. van Santen[183], M. Vehring[193], E. Vogel[193], M. Vraeghe[218], C. Walck[187], A. Wallace[182], M. Wallraff[193], F. D. Wandler[195], N. Wandkowsky[209], A. Waza[193], C. Weaver[195], M. J. Weiss[191], C. Wendt[209], J. Werthebach[206], B. J. Whelan[182], K. Wiebe[197], C. H. Wiebusch[193], L. Wille[209], D. R. Williams[228], L. Wills[229], M. Wolf[209], T. R. Wood[195], E. Woolsey[195], K. Woschnagg[198], D. L. Xu[209], X. W. Xu[220], Y. Xu[226], J. P. Yanez[195], G. Yodh[196], S. Yoshida[223], T. Yuan[209], M. Zoll[187],
(IceCube Collaboration),

A. Balasubramanian[235,236], S. Mate[236], V. Bhalerao[236], D. Bhattacharya[19], A. Vibhute[19], G. C. Dewangan[19], A. R. Rao[119], S. V. Vadawale[237],
(AstroSat Cadmium Zinc Telluride Imager Team),
D. S. Svinkin[238], K. Hurley[239], R. L. Aptekar[238], D. D. Frederiks[238], S. V. Golenetskii[238], A. V. Kozlova[238], A. L. Lysenko[238], Ph. P. Oleynik[238], A. E. Tsvetkova[238], M. V. Ulanov[238], T. Cline[240],
(IPN Collaboration),
T. P. Li[241,82,242], S. L. Xiong[241], S. N. Zhang[241,242], F. J. Lu[241], L. M. Song[241], X. L. Cao[241], Z. Chang[241], G. Chen[241], L. Chen[243], T. X. Chen[241], Y. Chen[241], Y. B. Chen[82], Y. P. Chen[241], W. Cui[241,82], W. W. Cui[241], J. K. Deng[82], Y. W. Dong[241], Y. Y. Du[241], M. X. Fu[82], G. H. Gao[241,242], H. Gao[241,242], M. Gao[241], M. Y. Ge[241], Y. D. Gu[241], J. Guan[241], C. C. Guo[241,242], D. W. Han[241], W. Hu[241], Y. Huang[241], J. Huo[241], S. M. Jia[241], L. H. Jiang[241], W. C. Jiang[241], J. Jin[241], Y. J. Jin[82], B. Li[241], C. K. Li[241], G. Li[241], M. S. Li[241], W. Li[241], X. Li[241], X. B. Li[241], X. F. Li[241], Y. G. Li[241], Z. J. Li[241,242], Z. W. Li[241], X. H. Liang[241], J. Y. Liao[241], C. Z. Liu[241], G. Q. Liu[82], H. W. Liu[241], S. Z. Liu[241], X. J. Liu[241], Y. Liu[241], Y. N. Liu[82], B. Lu[241], X. F. Lu[241], T. Luo[241], X. Ma[241], B. Meng[241], Y. Nang[241,242], J. Y. Nie[241], G. Ou[241], J. L. Qu[241], N. Sai[241,242], L. Sun[241], Y. Tan[241], L. Tao[241], W. H. Tao[241], Y. L. Tuo[241,242], G. F. Wang[241], H. Y. Wang[241], J. Wang[241], W. S. Wang[241], Y. S. Wang[241], X. Y. Wen[241], B. B. Wu[241], M. Wu[241], G. C. Xiao[241,242], H. Xu[241], Y. P. Xu[241], L. L. Yan[241,242], J. W. Yang[241], S. Yang[241], Y. J. Yang[241], A. M. Zhang[241], C. L. Zhang[241], C. M. Zhang[241], F. Zhang[241], H. M. Zhang[241], J. Zhang[241], Q. Zhang[241], S. Zhang[241], T. Zhang[241], W. Zhang[241,242], W. C. Zhang[241], W. Z. Zhang[243], Y. Zhang[241], Y. Zhang[241,242], Y. F. Zhang[241], Y. J. Zhang[241], Z. Zhang[82], Z. L. Zhang[241], H. S. Zhao[241], J. L. Zhao[241], X. F. Zhao[241,242], S. J. Zheng[241], Y. Zhu[241], Y. X. Zhu[241], C. L. Zou[241],
(The Insight-Hxmt Collaboration),
A. Albert[244], M. André[245], M. Anghinolfi[246,247], M. Ardid[247], J.-J. Aubert[248], J. Aublin[249], T. Avgitas[249], B. Baret[249], J. Barrios-Martí[250], S. Basa[251], B. Belhorma[252], V. Bertin[248], S. Biagi[253], R. Bormuth[14,254], S. Bourret[249], M. C. Bouwhuis[14], H. Brânzaş[255], R. Bruijn[14,256], J. Brunner[248], J. Busto[248], A. Capone[257,258], L. Caramete[255], J. Carr[248], S. Celli[257,258,259], R. Cherkaoui El Moursli[260], T. Chiarusi[261], M. Circella[262], J. A. B. Coelho[249], A. Coleiro[249,250], R. Coniglione[253], H. Costantini[248], P. Coyle[248], A. Creusot[249], A. F. Díaz[263], A. Deschamps[264], G. De Bonis[258], C. Distefano[253], I. Di Palma[257,258], A. Domi[246,265], C. Donzaud[249,266], D. Dornic[248], D. Drouhin[244], T. Eberl[189,260,267], I. El Bojaddaini[267], N. El Khayati[260], D. Elsässer[268], A. Enzenhöfer[248], A. Ettahiri[260], F. Fassi[260], I. Felis[247], L. A. Fusco[261,269], P. Gay[270,249], V. Giordano[271], H. Glotin[272,273], T. Grégoire[249], R. Gracia Ruiz[249], K. Graf[189], S. Hallmann[189], H. van Haren[274], A. J. Heijboer[14], Y. Hello[264], J. J. Hernández-Rey[250], J. Hössl[189], J. Hofestädt[189], C. Hugon[246,265], G. Illuminati[250], C. W. James[189], M. de Jong[14,254], M. Jongen[14], M. Kadler[268], O. Kalekin[189], U. Katz[189], D. Kiessling[189], A. Kouchner[249,273], M. Kreter[268], I. Kreykenbohm[275], V. Kulikovskiy[248,276], C. Lachaud[249], R. Lahmann[189], D. Lefèvre[277], E. Leonora[271,278], M. Lotze[250], S. Loucatos[279,249], M. Marcelin[251], A. Margiotta[261,269], A. Marinelli[280,281], J. A. Martínez-Mora[247], R. Mele[282,283], K. Melis[14,256], T. Michael[14], P. Migliozzi[282], A. Moussa[267], S. Navas[284], E. Nezri[251], M. Organokov[285], G. E. Păvălaş[255], C. Pellegrino[261,269], C. Perrina[257,258], P. Piattelli[253], V. Popa[255], T. Pradier[285], L. Quinn[248], C. Racca[244], G. Riccobene[253], A. Sánchez-Losa[262], M. Saldaña[247], I. Salvadori[248], D. F. E. Samtleben[14,254], M. Sanguineti[246,265], P. Sapienza[253], C. Sieger[189], M. Spurio[261,269], Th. Stolarczyk[279], M. Taiuti[246,265], Y. Tayalati[260], A. Trovato[253], D. Turpin[248], C. Tönnis[250], B. Vallage[279,249], V. Van Elewyck[249,273], F. Versari[261,269], D. Vivolo[282,283], A. Vizzoca[257,258], J. Wilms[275], J. D. Zornoza[250], J. Zúñiga[250],
(ANTARES Collaboration),
A. P. Beardmore[286], A. A. Breeveld[287], D. N. Burrows[288], S. B. Cenko[289,290], G. Cusumano[291], A. D'Aì[291], M. de Pasquale[292], S. W. K. Emery[287], P. A. Evans[286], P. Giommi[293], C. Gronwall[288,294], J. A. Kennea[288], H. A. Krimm[295,296], N. P. M. Kuin[287], A. Lien[297,298], F. E. Marshall[287], A. Melandri[299], J. A. Nousek[288], S. R. Oates[300], J. P. Osborne[286], C. Pagani[286], K. L. Page[286], D. M. Palmer[301], M. Perri[302,293], M. H. Siegel[288], B. Sbarufatti[288], G. Tagliaferri[299], A. Tohuvavohu[288,303],
(The Swift Collaboration),
M. Tavani[304,305,306], F. Verrecchia[307,308], A. Bulgarelli[309], Y. Evangelista[304], L. Pacciani[304], M. Feroci[304], C. Pittori[307,308], A. Giuliani[310], E. Del Monte[304], I. Donnarumma[311], A. Argan[304], A. Trois[312], A. Ursi[304], M. Cardillo[304], G. Piano[304], F. Longo[313], F. Lucarelli[307,308], P. Munar-Adrover[314], F. Fuschino[309], C. Labanti[309], M. Marisaldi[315], G. Minervini[304], V. Fioretti[309], N. Parmiggiani[309], F. Gianotti[309], M. Trifoglio[309], G. Di Persio[304], L. A. Antonelli[311], G. Barbiellini[313], P. Caraveo[310], P. W. Cattaneo[316], E. Costa[304], S. Colafrancesco[317], F. D'Amico[311], A. Ferrari[318], A. Morselli[319], F. Paoletti[320], P. Picozza[319], M. Pilia[312], A. Rappoldi[316], P. Soffitta[304], S. Vercellone[321],
(AGILE Team),
R. J. Foley[322], D. A. Coulter[322], C. D. Kilpatrick[322], M. R. Drout[323], A. L. Piro[323], B. J. Shappee[323,324], M. R. Siebert[322], J. D. Simon[323], N. Ulloa[325], D. Kasen[326,327], B. F. Madore[323], A. Murguia-Berthier[322], Y.-C. Pan[322], J. X. Prochaska[322], E. Ramirez-Ruiz[322,328], A. Rest[329,330], C. Rojas-Bravo[322],
(The 1M2H Team),

E. Berger[163], M. Soares-Santos[331,332], J. Annis[332], K. D. Alexander[163], S. Allam[332], E. Balbinot[333], P. Blanchard[163], D. Brout[334], R. E. Butler[335,332], R. Chornock[336], E. R. Cook[337,338], P. Cowperthwaite[163], H. T. Diehl[332], A. Drlica-Wagner[332], M. R. Drout[339], F. Durret[340], T. Eftekhari[163], D. A. Finley[332], W. Fong[341], J. A. Frieman[332], C. L. Fryer[342], J. García-Bellido[343], R. A. Gruendl[344], W. Hartley[345,346], K. Herner[332], R. Kessler[347], H. Lin[332], P. A. A. Lopes[348], A. C. C. Lourenço[348], R. Margutti[349], J. L. Marshall[337], T. Matheson[350], G. E. Medina[351], B. D. Metzger[352], R. R. Muñoz[351], J. Muir[353], M. Nicholl[163], P. Nugent[354], A. Palmese[345], F. Paz-Chinchón[344], E. Quataert[355], M. Sako[334], M. Sauseda[337], D. J. Schlegel[356], D. Scolnic[347], L. F. Secco[334], N. Smith[357], F. Sobreira[358,359], V. A. Villar[163], A. K. Vivas[360], W. Wester[332], P. K. G. Williams[163], B. Yanny[332], A. Zenteno[360], Y. Zhang[332], T. M. C. Abbott[360], M. Banerji[361,362], K. Bechtol[338], A. Benoit-Lévy[363,345,364], E. Bertin[363,364], D. Brooks[345], E. Buckley-Geer[332], D. L. Burke[365,366], D. Capozzi[367], A. Carnero Rosell[359,368], M. Carrasco Kind[369,344], F. J. Castander[370], M. Crocce[370], C. E. Cunha[365], C. B. D'Andrea[334], L. N. da Costa[359,368], C. Davis[365], D. L. DePoy[371], S. Desai[372], J. P. Dietrich[373,374], T. F. Eifler[375,376], E. Fernandez[377], B. Flaugher[332], P. Fosalba[370], E. Gaztanaga[370], D. W. Gerdes[378,379], T. Giannantonio[361,362,380], D. A. Goldstein[381,354], D. Gruen[365,366], J. Gschwend[359,368], G. Gutierrez[332], K. Honscheid[382,383], D. J. James[384], T. Jeltema[385], M. W. G. Johnson[344], M. D. Johnson[344], S. Kent[332,347], E. Krause[365], R. Kron[332,347], K. Kuehn[386], O. Lahav[345], M. Lima[387,359], M. A. G. Maia[359,368], M. March[334], P. Martini[382,388], R. G. McMahon[361,362], F. Menanteau[369,344], C. J. Miller[378,379], R. Miquel[389,377], J. J. Mohr[373,374,390], R. C. Nichol[367], R. L. C. Ogando[359,368], A. A. Plazas[376], A. K. Romer[391], A. Roodman[365,366], E. S. Rykoff[365,366], E. Sanchez[392], V. Scarpine[332], R. Schindler[366], M. Schubnell[379], I. Sevilla-Noarbe[392], E. Sheldon[393], M. Smith[394], R. C. Smith[360], A. Stebbins[332], E. Suchyta[395], M. E. C. Swanson[344], G. Tarle[379], R. C. Thomas[354], M. A. Troxel[382,383], D. L. Tucker[332], V. Vikram[396], A. R. Walker[360], R. H. Wechsler[397,365,366], J. Weller[373,390,380], J. L. Carlin[338], M. S. S. Gill[366], T. S. Li[332], J. Marriner[332], E. Neilsen[332],
(The Dark Energy Camera GW-EM Collaboration and the DES Collaboration),
J. B. Haislip[398], V. V. Kouprianov[398], D. E. Reichart[398], D. J. Sand[399], L. Tartaglia[399,400], S. Valenti[400], S. Yang[400,401,402],
(The DLT40 Collaboration),
S. Benetti[403], E. Brocato[404], S. Campana[405], E. Cappellaro[403], S. Covino[405], P. D'Avanzo[405], V. D'Elia[404,406], F. Getman[407], G. Ghirlanda[405], G. Ghisellini[405], L. Limatola[407], L. Nicastro[408], E. Palazzi[408], E. Pian[408], S. Piranomonte[404], A. Possenti[312], A. Rossi[408], O. S. Salafia[409,405], L. Tomasella[403], L. Amati[408], L. A. Antonelli[404], M. G. Bernardini[410,405], F. Bufano[411], M. Capaccioli[407,412], P. Casella[404], M. Dadina[408], G. De Cesare[408], A. Di Paola[404], G. Giuffrida[404], A. Giunta[404], G. L. Israel[404], M. Lisi[404], E. Maiorano[408], M. Mapelli[403,413], N. Masetti[408,414], A. Pescalli[415,405], L. Pulone[404], R. Salvaterra[416], P. Schipani[407], M. Spera[403], A. Stamerra[146,417], L. Stella[404], V. Testa[404], M. Turatto[403], D. Vergani[408], G. Aresu[312], M. Bachetti[312], F. Buffa[312], M. Burgay[312], M. Buttu[312], T. Caria[312], E. Carretti[312], V. Casasola[418], P. Castangia[312], G. Carboni[312], S. Casu[312], R. Concu[312], A. Corongiu[312], G. L. Deiana[312], E. Egron[312], A. Fara[312], F. Gaudiomonte[312], V. Gusai[312], A. Ladu[312], S. Loru[312], S. Leurini[312], L. Marongiu[312], A. Melis[312], G. Melis[312], Carlo Migoni[312], Sabrina Milia[312], Alessandro Navarrini[312], A. Orlati[312], P. Ortu[312], S. Palmas[312], A. Pellizzoni[312], D. Perrodin[312], T. Pisanu[312], S. Poppi[312], S. Righini[419], A. Saba[312], G. Serra[312], M. Serrau[312], M. Stagni[419], G. Surcis[312], V. Vacca[312], G. P. Vargiu[312], L. K. Hunt[418], Z. P. Jin[420], S. Klose[421], C. Kouveliotou[422,423], P. A. Mazzali[424,425], P. Møller[426], L. Nava[405,427], T. Piran[428], J. Selsing[328], S. D. Vergani[429,405], K. Wiersema[430], K. Toma[431,432], A. B. Higgins[430], C. G. Mundell[433], S. di Serego Alighieri[418], D. Götz[434], W. Gao[435], A. Gomboc[436], L. Kaper[437], S. Kobayashi[438], D. Kopac[439], J. Mao[440], R. L. C. Starling[430], I. Steele[441], A. J. van der Horst[442,423],
(GRAWITA: GRAvitational Wave Inaf TeAm),
F. Acero[443], W. B. Atwood[444], L. Baldini[445], G. Barbiellini[446,447], D. Bastieri[448,449], B. Berenji[450], R. Bellazzini[451], E. Bissaldi[452,453], R. D. Blandford[454], E. D. Bloom[454], R. Bonino[455,456], E. Bottacini[454], J. Bregeon[457], R. Buehler[469], S. Buson[50], R. A. Cameron[454], R. Caputo[459], P. A. Caraveo[180], E. Cavazzuti[460], A. Chekhtman[461], C. C. Cheung[462], J. Chiang[454], S. Ciprini[462,463], J. Cohen-Tanugi[457], L. R. Cominsky[465], D. Costantin[449], A. Cuoco[466,455], F. D'Ammando[466,467], F. de Palma[453,468], S. W. Digel[454], N. Di Lalla[445], M. Di Mauro[454], L. Di Venere[452,453], R. Dubois[454], S. J. Fegan[458], W. B. Focke[454], A. Franckowiak[458], Y. Fukazawa[471], S. Funk[189], P. Fusco[452,453], F. Gargano[453], D. Gasparrini[462,464], N. Giglietto[452,453], F. Giordano[452,453], M. Giroletti[466], T. Glanzman[454], D. Green[472,50], M.-H. Grondin[473], L. Guillemot[474,475], S. Guiriec[50,422], A. K. Harding[50], D. Horan[458], G. Jóhannesson[476,477], T. Kamae[478], S. Kensei[471], M. Kuss[451], G. La Mura[449], L. Latronico[455], M. Lemoine-Goumard[473], F. Longo[446,447], F. Loparco[452,453], M. N. Lovellette[462], P. Lubrano[463], J. D. Magill[472], S. Maldera[455], A. Manfreda[445], M. N. Mazziotta[453], J. E. McEnery[50,472], M. Meyer[454], P. F. Michelson[454], N. Mirabal[50], M. E. Monzani[454], A. Morselli[479], I. V. Moskalenko[454], M. Negro[455,456], E. Nuss[457], R. Ojha[50], N. Omodei[454], M. Orienti[467], E. Orlando[454], M. Palatiello[446,446], V. S. Paliya[480], D. Paneque[481], M. Pesce-Rollins[451], F. Piron[457], T. A. Porter[454], G. Principe[189], S. Rainò[452,453], R. Rando[448,449], M. Razzano[451], S. Razzaque[482], A. Reimer[483,454], O. Reimer[483,454], T. Reposeur[473], L. S. Rochester[454], P. M. Saz Parkinson[444,484,485], C. Sgrò[451], E. J. Siskind[486], F. Spada[451], G. Spandre[451], D. J. Suson[487], M. Takahashi[481], Y. Tanaka[488], J. G. Thayer[454], J. B. Thayer[454], D. J. Thompson[50], L. Tibaldo[489,490], D. F. Torres[491,492], E. Torresi[493], E. Troja[50,472], T. M. Venters[50],

G. Vianello[454], G. Zaharijas[446,447,494],
(The Fermi Large Area Telescope Collaboration),
J. Allison[495,496], K. W. Bannister[497], D. Dobie[495,497,498], D. L. Kaplan[499], E. Lenc[495,498], C. Lynch[495,498], T. Murphy[495,498], E. M. Sadler[495,498],
(ATCA: Australia Telescope Compact Array),
A. Hotan[500], C. W. James[501], S. Oslowski[502], W. Raja[497], R. M. Shannon[497,501], M. Whiting[497],
(ASKAP: Australian SKA Pathfinder),
I. Arcavi[503,504], D. A. Howell[503,504], C. McCully[503,504], G. Hosseinzadeh[503,504], D. Hiramatsu[503,504], D. Poznanski[505], J. Barnes[506], M. Zaltzman[505], S. Vasylyev[503,504], D. Maoz[505],
(Las Cumbres Observatory Group),
J. Cooke[507,508,509], M. Bailes[507,508], C. Wolf[510,509,508], A. T. Deller[507,508,509], C. Lidman[511,509], L. Wang[512,513,514], B. Gendre[515], I. Andreoni[507,508,511,509], K. Ackley[516], T. A. Pritchard[507], M. S. Bessell[510], S.-W. Chang[510,509], A. Möller[510,509], C. A. Onken[510,509], R. A. Scalzo[510,509,517], R. Ridden-Harper[510], R. G. Sharp[510,509], B. E. Tucker[510,509], T. J. Farrell[511], E. Elmer[518], S. Johnston[519,509], V. Venkatraman Krishnan[507,509], E. F. Keane[520,509], J. A. Green[519], A. Jameson[507,509], L. Hu[513,514], B. Ma[521,514], T. Sun[513,514], X. Wu[513,514], X. Wang[522], Z. Shang[521,523,514], Y. Hu[521,514], M. C. B. Ashley[524], X. Yuan[525,514], X. Li[525,514], C. Tao[522], Z. Zhu[526], H. Zhang[527], N. B. Suntzeff[512], J. Zhou[527], J. Yang[513], B. Orange[528], D. Morris[515], A. Cucchiara[515], T. Giblin[529], A. Klotz[530], J. Staff[515], P. Thierry[531], B. P. Schmidt[532,509],
(OzGrav, DWF (Deeper, Wider, Faster program), AST3, and CAASTRO Collaborations),
N. R. Tanvir[533], A. J. Levan[300], Z. Cano[52,534], A. de Ugarte-Postigo[527,534], P. Evans[533], C. González-Fernández[535], J. Greiner[536], J. Hjorth[328], M. Irwin[535], T. Krühler[536], I. Mandel[537], B. Milvang-Jensen[328], P. O'Brien[533], E. Rol[538], S. Rosetti[533], S. Rosswog[539], A. Rowlinson[540,541], D. T. H. Steeghs[300], C. C. Thöne[534], K. Ulaczyk[300], D. Watson[328], S. H. Bruun[328], R. Cutter[300], R. Figuera Jaimes[542], Y. I. Fujii[543,544], A. S. Fruchter[545], B. Gompertz[300], P. Jakobsson[546], G. Hodosan[534], U. G. Jèrgensen[543], T. Kangas[545], D. A. Kann[534], M. Rabus[547,548], S. L. Schrøder[328], E. R. Stanway[300], R. A. M. J. Wijers[540],
(The VINROUGE Collaboration),
V. M. Lipunov[549,550], E. S. Gorbovskoy[550], V. G. Kornilov[549,550], N. V. Tyurina[550], P. V. Balanutsa[550], A. S. Kuznetsov[550], D. M. Vlasenko[549,550], R. C. Podesta[551], C. Lopez[551], F. Podesta[551], H. O. Levato[552], C. Saffe[552], C. C. Mallamaci[553], N. M. Budnev[554], O. A. Gress[554,550], D. A. Kuvshinov[549,550], I. A. Gorbunov[549,550], V. V. Vladimirov[550], D. S. Zimnukhov[549,550], A. V. Gabovich[555], V. V. Yurkov[555], Yu. P. Sergienko[555], R. Rebolo[556], M. Serra-Ricart[556], A. G. Tlatov[557], Yu. V. Ishmuhametova[554],
(MASTER Collaboration),
F. Abe[558], K. Aoki[559], W. Aoki[560], Y. Asakura[558,956], S. Baar[561], S. Barway[562], I. A. Bond[563], M. Doi[564], F. Finet[559], T. Fujiyoshi[559], H. Furusawa[560], S. Honda[561], R. Itoh[565], N. Kanda[566], K. S. Kawabata[567], M. Kawabata[568], J. H. Kim[559], S. Koshida[559], D. Kuroda[569], C.-H. Lee[559], W. Liu[567,570], K. Matsubayashi[569], S. Miyazaki[571], K. Morihana[572], T. Morokuma[564], K. Motohara[564], K. L. Murata[565], H. Nagai[560], H. Nagashima[568], T. Nagayama[573], T. Nakaoka[568], F. Nakata[559], R. Ohsawa[564], T. Ohshima[561], K. Ohta[574], H. Okita[559], T. Saito[561], Y. Saito[565], S. Sako[564,575], Y. Sekiguchi[576], T. Sumi[571], A. Tajitsu[559], J. Takahashi[561], M. Takayama[561], Y. Tamura[572], I. Tanaka[559], M. Tanaka[560], T. Terai[559], N. Tominaga[577,578], P. J. Tristram[579], M. Uemura[567], Y. Utsumi[567], M. S. Yamaguchi[564], N. Yasuda[578], M. Yoshida[559], T. Zenko[574],
(J-GEM),
S. M. Adams[580], J. R. Allison[495,496], G. C. Anupama[581], J. Bally[582], S. Barway[583], E. Bellm[584], N. Blagorodnova[580], C. Cannella[580], P. Chandra[585], D. Chatterjee[586], T. E. Clarke[587], B. E. Cobb[588], D. O. Cook[580], C. Copperwheat[441], K. De[580], S. W. K. Emery[589], P. A. Evans[590], U. Feindt[591], K. Foster[580], O. D. Fox[592], D. A. Frail[593], C. Fremling[580], C. Frohmaier[594,595], J. A. Garcia[580], S. Ghosh[586], S. Giacintucci[587], A. Goobar[591], O. Gottlieb[596], B. W. Grefenstette[580], G. Hallinan[580], F. Harrison[580], M. Heida[580], G. Helou[597], A. Y. Q. Ho[580], A. Horesh[598], K. Hotokezaka[599], W.-H. Ip[600], R. Itoh[601], Bob Jacobs[66], J. E. Jencson[580], D. Kasen[602,603], M. M. Kasliwal[580], N. E. Kassim[587], H. Kim[604], B. S. Kiran[581], N. P. M. Kuin[589], S. R. Kulkarni[580], T. Kupfer[580], R. M. Lau[580], K. Madsen[580], P. A. Mazzali[441,605], A. A. Miller[606,607], H Miyasaka[580], K. Mooley[608], S. T. Myers[593], E. Nakar[596], C.-C. Ngeow[600], P. Nugent[602,354], E. O. Ofek[609], N. Palliyaguru[610], M. Pavana[581], D. A. Perley[611], W. M. Peters[587], S. Pike[580], T. Piran[598], H. Qi[586], R. M. Quimby[612,613], J. Rana[19], S. Rosswog[614], F. Rusu[615], E. M. Sadler[495,616], A. Van Sistine[586], J. Sollerman[614], Y. Xu[580], L. Yan[580,597], Y. Yatsu[601], P.-C. Yu[600], C. Zhang[586], W. Zhao[615],
(GROWTH, JAGWAR, Caltech-NRAO, TTU-NRAO, and NuSTAR Collaborations),
K. C. Chambers[617], M. E. Huber[617], A. S. B. Schultz[617], J. Bulger[617], H. Flewelling[617], E. A. Magnier[617], T. B. Lowe[617], R. J. Wainscoat[617], C. Waters[617], M. Willman[617],
(Pan-STARRS),

K. Ebisawa[618], C. Hanyu[619], S. Harita[620], T. Hashimoto[621], K. Hidaka[619], T. Hori[622], M. Ishikawa[623], N. Isobe[618], W. Iwakiri[624], H. Kawai[625], N. Kawai[620,624], T. Kawamuro[626], T. Kawase[627], Y. Kitaoka[621], K. Makishima[624], M. Matsuoka[624], T. Mihara[624], T. Morita[622], K. Morita[620], S. Nakahira[624], M. Nakajima[627], Y. Nakamura[625], H. Negoro[627], S. Oda[622], A. Sakamaki[627], R. Sasaki[625], M. Serino[621], M. Shidatsu[624], R. Shimomukai[618], Y. Sugawara[618], S. Sugita[620], M. Sugizaki[624], Y. Tachibana[620], Y. Takao[624], A. Tanimoto[622], H. Tomida[618], Y. Tsuboi[625], H. Tsunemi[628], Y. Ueda[622], S. Ueno[618], S. Yamada[622], K. Yamaoka[629], M. Yamauchi[619], F. Yatabe[624], T. Yoneyama[628], T. Yoshii[620],
(The MAXI Team),
D. M. Coward[630], H. Crisp[630], D. Macpherson[630], I. Andreoni[631], R. Laugier[632], K. Noysena[632,633], A. Klotz[633], B. Gendre[632,634], P. Thierry[635], D. Turpin[630],
(TZAC Consortium),
M. Im[636], C. Choi[636], J. Kim[636], Y. Yoon[636], G. Lim[636], S.-K. Lee[636], C.-U. Lee[637], S.-L. Kim[637], S.-W. Ko[637], J. Joe[637], M.-K. Kwon[637], P.-J. Kim[637], S.-K. Lim[637], J.-S. Choi[637],
(KU Collaboration),
J. P. U. Fynbo[328], D. Malesani[328], D. Xu[638],
(Nordic Optical Telescope),
S. J. Smartt[639], A. Jerkstrand[425], E. Kankare[639], S. A. Sim[639], M. Fraser[168], C. Inserra[640], K. Maguire[639], G. Leloudas[328], M. Magee[639], L. J. Shingles[639], K. W. Smith[639], D. R. Young[639], R. Kotak[639], A. Gal-Yam[641], J. D. Lyman[642], D. S. Homan[643], C. Agliozzo[644,645], J. P. Anderson[646], C. R. Angus[640], C. Ashall[611], C. Barbarino[647], F. E. Bauer[648,645,649], M. Berton[650,651], M. T. Botticella[652], M. Bulla[653], G. Cannizzaro[654], R. Cartier[640], A. Cikota[655], P. Clark[639], A. De Cia[655], M. Della Valle[652,656], M. Dennefeld[657], L. Dessart[658], G. Dimitriadis[640], N. Elias-Rosa[659], R. E. Firth[640], A. Flörs[655,425], C. Frohmaier[660], L. Galbany[661], S. González-Gaitán[662], M. Gromadzki[663], C. P. Gutiérrez[640], A. Hamanowicz[655,663], J. Harmanen[664], K. E. Heintz[546,328], M.-S. Hernandez[665], S. T. Hodgkin[666], I. M. Hook[667], L. Izzo[668], P. A. James[611], P. G. Jonker[654,66], W. E. Kerzendorf[655], Z. Kostrzewa-Rutkowska[654,66], M. Kromer[669,670], H. Kuncarayakti[671,664], A. Lawrence[643], I. Manulis[641], S. Mattila[664], O. McBrien[639], A. Müller[672], J. Nordin[673], D. O'Neill[639], F. Onori[654], J. T. Palmerio[674], A. Pastorello[675], F. Patat[655], G. Pignata[644,645], P. Podsiadlowski[676], A. Razza[646,677], T. Reynolds[664], R. Roy[647], A. J. Ruiter[678,532,679], K. A. Rybicki[663], L. Salmon[168], M. L. Pumo[680,675,681], S. J. Prentice[611], I. R. Seitenzahl[678,532], M. Smith[640], J. Sollerman[647], M. Sullivan[640], H. Szegedi[682], F. Taddia[647], S. Taubenberger[655,425], G. Terreran[349,675], B. Van Soelen[682], J. Vos[665], N. A. Walton[666], D. E. Wright[683], Ł. Wyrzykowski[663], O. Yaron[641],
(ePESSTO) ,
T.-W. Chen[684], T. Krühler[684], P. Schady[684], P. Wiseman[684], J. Greiner[684], A. Rau[684], T. Schweyer[684], S. Klose[685], A. Nicuesa Guelbenzu[685],
(GROND),
N. T. Palliyaguru[686],
(Texas Tech University),
M. M. Shara[687,361], T. Williams[688], P. Vaisanen[688,689], S. B. Potter[688], E. Romero Colmenero[688,689], S. Crawford[688,689], D. A. H. Buckley[688], J. Mao[440],
(SALT Group),
M. C. Díaz[690], L. M. Macri[691], D. García Lambas[692], C. Mendes de Oliveira[693], J. L. Nilo Castellón[694,695], T. Ribeiro[696], B. Sánchez[692], W. Schoenell[693,697], L. R. Abramo[698], S. Akras[699], J. S. Alcaniz[699], R. Artola[692], M. Beroiz[690], S. Bonoli[700], J. Cabral[692], R. Camuccio[690], V. Chavushyan[701], P. Coelho[693], C. Colazo[692], M. V. Costa-Duarte[693], H. Cuevas Larenas[695], M. Domínguez Romero[692], D. Dultzin[702], D. Fernández[703], J. García[690], C. Girardini[692], D. R. Gonçalves[704], T. S. Gonçalves[704], S. Gurovich[692], Y. Jiménez-Teja[699], A. Kanaan[697], M. Lares[692], R. Lopes de Oliveira[696,705], O. López-Cruz[701], R. Melia[692], A. Molino[693], N. Padilla[703], T. Peñuela[690,706], V. M. Placco[707,708], C. Quiñones[692], A. Ramírez Rivera[695], V. Renzi[692], L. Riguccini[704], E. Ríos-López[701], H. Rodriguez[692], L. Sampedro[693], M. Schneiter[692], L. Sodré[693], M. Starck[692], S. Torres-Flores[695], M. Tornatore[692], A. Zadrożny[690],
(TOROS: Transient Robotic Observatory of the South Collaboration),
A. J. Castro-Tirado[709,710], J. C. Tello[709], Y.-D. Hu[709], B.-B. Zhang[709], R. Cunniffe[709], A. Castellón[711], D. Hiriart[712], M. D. Caballero-García[713], M. Jelínek[714], P. Kubánek[715], C. Pérez del Pulgar[710], I. H. Park[716], S. Jeong[716], J. M. Castro Cerón[717], S. B. Pandey[718], P. C. Yock[719], R. Querel[720], Y. Fan[721], C. Wang[721],
(The BOOTES Collaboration),
A Beardsley[722], I. S. Brown[499], B. Crosse[501], D. Emrich[501], T. Franzen[501], B. M. Gaensler[723], L. Horsley[501], M. Johnston-Hollitt[724], D. Kenney[501], M. F. Morales[725], D. Pallot[726], M. Sokolowski[501,498,727], K. Steele[501], S. J. Tingay[501,498], C. M. Trott[501,498], M. Walker[501], R. Wayth[501,498], A. Williams[501], C. Wu[726],
(MWA: Murchison Widefield Array),

A. Yoshida[728], T. Sakamoto[728], Y. Kawakubo[728], K. Yamaoka[729], I. Takahashi[730], Y. Asaoka[731], S. Ozawa[731], S. Torii[731], Y. Shimizu[732], T. Tamura[732], W. Ishizaki[733], M. L. Cherry[2], S. Ricciarini[734], A. V. Penacchioni[735], P. S. Marrocchesi[735],
(The CALET Collaboration),
A. S. Pozanenko[736,737,738], A. A. Volnova[736], E. D. Mazaeva[736], P. Yu. Minaev[736], M. A. Krugov[739], A. V. Kusakin[742], I. V. Reva[742], A. S. Moskvitin[740], V. V. Rumyantsev[741], R. Inasaridze[742], E. V. Klunko[743], N. Tungalag[744], S. E. Schmalz[745], O. Burhonov[746],
(IKI-GW Follow-up Collaboration),
H. Abdalla[747], A. Abramowski[748], F. Aharonian[749,750,751], F. Ait Benkhali[749], E. O. Angüner[752], M. Arakawa[753], M. Arrieta[754], P. Aubert[755], M. Backes[756], A. Balzer[757], M. Barnard[747], Y. Becherini[758], J. Becker Tjus[759], D. Berge[760], S. Bernhard[761], K. Bernlöhr[749], R. Blackwell[762], M. Böttcher[747], C. Boisson[754], J. Bolmont[763], S. Bonnefoy[183], P. Bordas[749], J. Bregeon[764], F. Brun[765], P. Brun[766], M. Bryan[757], M. Büchele[189], T. Bulik[767], M. Capasso[768], S. Caroff[470], A. Carosi[755], S. Casanova[752,746], M. Cerruti[763], N. Chakraborty[746], R. C. G. Chaves[764], A. Chen[769], J. Chevalier[755], S. Colafrancesco[769], B. Condon[765], J. Conrad[770], I. D. Davids[756], J. Decock[766], C. Deil[746], J. Devin[764], P. deWilt[762], L. Dirson[745], A. Djannati-Ataï[771], A. Donath[746], L. O'C. Drury[750], K. Dutson[772], J. Dyks[773], T. Edwards[746], K. Egberts[774], G. Emery[763], J.-P. Ernenwein[775], S. Eschbach[189], C. Farnier[770,758], S. Fegan[470], M. V. Fernandes[745], A. Fiasson[755], G. Fontaine[470], S. Funk[189], M. Füssling[183], S. Gabici[771], Y. A. Gallant[764], T. Garrigoux[744], F. Gaté[755], G. Giavitto[183], B. Giebels[470], D. Glawion[776], J. F. Glicenstein[766], D. Gottschall[768], M.-H. Grondin[765], J. Hahn[746], M. Haupt[183], J. Hawkes[762], G. Heinzelmann[745], G. Henri[777], G. Hermann[746], J. A. Hinton[746], W. Hofmann[746], C. Hoischen[774], T. L. Holch[778], M. Holler[761], D. Horns[745], A. Ivascenko[744], H. Iwasaki[753], A. Jacholkowska[763], M. Jamrozy[779], D. Jankowsky[189], F. Jankowsky[776], M. Jingo[769], L. Jouvin[771], I. Jung-Richardt[189], M. A. Kastendieck[745], K. Katarzyński[780], M. Katsuragawa[781,763], D. Khangulyan[753], B. Khélifi[771], J. King[746], S. Klepser[183], D. Klochkov[768], W. Kluźniak[773], Nu. Komin[769], K. Kosack[766], S. Krakau[759], M. Kraus[189], P. P. Krüger[744], H. Laffon[765], G. Lamanna[755], J. Lau[762], J.-P. Lees[755], J. Lefaucheur[754], A. Lemière[771], M. Lemoine-Goumard[765], J.-P. Lenain[763], E. Leser[774], T. Lohse[778], M. Lorentz[766], R. Liu[746], I. Lypova[183], D. Malyshev[768], V. Marandon[746], A. Marcowith[764], C. Mariaud[470], R. Marx[746], G. Maurin[755], N. Maxted[762], M. Mayer[778], P. J. Meintjes[782], M. Meyer[770], A. M. W. Mitchell[746], R. Moderski[773], M. Mohamed[776], L. Mohrmann[189], K. Morå[770], E. Moulin[766], T. Murach[183], S. Nakashima[781], M. de Naurois[470], H. Ndiyavala[744], F. Niederwanger[761], J. Niemiec[752], L. Oakes[778], P. O'Brien[772], H. Odaka[781], S. Ohm[183], M. Ostrowski[779], I. Oya[183], M. Padovani[764], M. Panter[746], R. D. Parsons[746], N. W. Pekeur[744], G. Pelletier[777], C. Perennes[763], P.-O. Petrucci[777], B. Peyaud[766], Q. Piel[755], S. Pita[771], V. Poireau[755], H. Poon[746], D. Prokhorov[758], H. Prokoph[760], G. Pühlhofer[768], M. Punch[771,758], A. Quirrenbach[776], S. Raab[189], R. Rauth[761], A. Reimer[761], O. Reimer[761], M. Renaud[764], R. de los Reyes[746], F. Rieger[746,783], L. Rinchiuso[766], C. Romoli[750], G. Rowell[762], B. Rudak[773], C. B. Rulten[754], V. Sahakian[784,751], S. Saito[753], D. A. Sanchez[755], A. Santangelo[768], M. Sasaki[189], R. Schlickeiser[759], F. Schüssler[766], A. Schulz[183], U. Schwanke[778], S. Schwemmer[776], M. Seglar-Arroyo[766], M. Settimo[763], A. S. Seyffert[744], N. Shafi[769], I. Shilon[189], K. Shiningayamwe[756], R. Simoni[757], H. Sol[754], F. Spanier[744], M. Spir-Jacob[771], Ł. Stawarz[779], R. Steenkamp[756], C. Stegmann[774,183], C. Steppa[774], I. Sushch[744], T. Takahashi[781], J.-P. Tavernet[763], T. Tavernier[771], A. M. Taylor[183], R. Terrier[771], L. Tibaldo[746], D. Tiziani[189], M. Tluczykont[745], C. Trichard[775], M. Tsirou[764], N. Tsuji[753], R. Tuffs[746], Y. Uchiyama[753], D. J. van der Walt[744], C. van Eldik[189], C. van Rensburg[744], B. van Soelen[782], G. Vasileiadis[764], J. Veh[189], C. Venter[744], A. Viana[746], P. Vincent[763], J. Vink[757], F. Voisin[762], H. J. Völk[746], T. Vuillaume[755], Z. Wadiasingh[744], S. J. Wagner[776], P. Wagner[778], R. M. Wagner[770], R. White[746], A. Wierzcholska[752], P. Willmann[189], A. Wörnlein[189], D. Wouters[766], R. Yang[746], D. Zaborov[470], M. Zacharias[744], R. Zanin[746], A. A. Zdziarski[773], A. Zech[754], F. Zefi[470], A. Ziegler[189], J. Zorn[746], N. Żywucka[779],
(H.E.S.S. Collaboration),
R. P. Fender[785], J. W. Broderick[541], A. Rowlinson[786,541], R. A. M. J. Wijers[786], A. J. Stewart[785], S. ter Veen[541], A. Shulevski[541],
(LOFAR Collaboration),
M. Kavic[787], J. H. Simonetti[788], C. League[787], J. Tsai[788], K. S. Obenberger[789], K. Nathaniel[788], G. B. Taylor[790], J. D. Dowell[790], S. L. Liebling[791], J. A. Estes[787], M. Lippert[787], I. Sharma[787], P. Vincent[787], B. Farella[787],
(LWA: Long Wavelength Array),
A. U. Abeysekara[792], A. Albert[793], R. Alfaro[794], C. Alvarez[795], R. Arceo[795], J. C. Arteaga-Velázquez[796], D. Avila Rojas[794], H. A. Ayala Solares[797], A. S. Barber[792], J. Becerra Gonzalez[50], A. Becerril[794], E. Belmont-Moreno[794], S. Y. BenZvi[798], D. Berley[799], A. Bernal[800], J. Braun[801], C. Brisbois[797], K. S. Caballero-Mora[795], T. Capistrán[802], A. Carramiñana[802], S. Casanova[803], M. Castillo[796], U. Cotti[796], J. Cotzomi[804], S. Coutiño de León[802], C. De León[804], E. De la Fuente[805], R. Diaz Hernandez[802], S. Dichiara[800], B. L. Dingus[793], M. A. DuVernois[801], J. C. Díaz-Vélez[805,801], R. W. Ellsworth[806], K. Engel[799], O. Enríquez-Rivera[807], D. W. Fiorino[799], H. Fleischhack[797], N. Fraija[800], J. A. García-González[794], F. Garfias[800], M. Gerhardt[797], A. Gonzõlez Muñoz[794], M. M. González[800], J. A. Goodman[799], Z. Hampel-Arias[801], J. P. Harding[793], S. Hernandez[794], A. Hernandez-Almada[794], B. Hona[797], P. Hüntemeyer[797], A. Iriarte[800], A. Jardin-Blicq[808], V. Joshi[808],

S. Kaufmann[795], D. Kieda[792], A. Lara[807], R. J. Lauer[809], D. Lennarz[810], H. León Vargas[794], J. T. Linnemann[811], A. L. Longinotti[802], G. Luis Raya[812], R. Luna-García[813], R. López-Coto[808], K. Malone[814], S. S. Marinelli[811], O. Martinez[804], I. Martinez-Castellanos[799], J. Martínez-Castro[813], H. Martínez-Huerta[815], J. A. Matthews[809], P. Miranda-Romagnoli[816], E. Moreno[804], M. Mostafá[814], L. Nellen[817], M. Newbold[792], M. U. Nisa[798], R. Noriega-Papaqui[816], R. Pelayo[813], J. Pretz[814], E. G. Pérez-Pérez[812], Z. Ren[809], C. D. Rho[798], C. Rivière[799], D. Rosa-González[802], M. Rosenberg[814], E. Ruiz-Velasco[794], H. Salazar[804], F. Salesa Greus[803], A. Sandoval[794], M. Schneider[818], H. Schoorlemmer[808], G. Sinnis[793], A. J. Smith[799], R. W. Springer[792], P. Surajbali[808], O. Tibolla[795], K. Tollefson[811], I. Torres[802], T. N. Ukwatta[793], T. Weisgarber[801], S. Westerhoff[801], I. G. Wisher[801], J. Wood[801], T. Yapici[811], G. B. Yodh[819], P. W. Younk[793], H. Zhou[793], J. D. Álvarez[796],
(HAWC Collaboration),
A. Aab[66], P. Abreu[820], M. Aglietta[821,822], I. F. M. Albuquerque[823], J. M. Albury[824], I. Allekotte[825], A. Almela[826,827], J. Alvarez Castillo[828], J. Alvarez-Muñiz[829], G. A. Anastasi[830,831], L. Anchordoqui[832], B. Andrada[826], S. Andringa[820], C. Aramo[833], N. Arsene[834], H. Asorey[825,835], P. Assis[820], G. Avila[836,837], A. M. Badescu[838], A. Balaceanu[839], F. Barbato[840,820], R. J. Barreira Luz[820], K. H. Becker[204], J. A. Bellido[824], C. Berat[841], M. E. Bertaina[822,842], X. Bertou[825], P. L. Biermann[843], J. Biteau[844], S. G. Blaess[824], A. Blanco[820], J. Blazek[845], C. Bleve[846,847], M. Boháčová[845], C. Bonifazi[848], N. Borodai[849], A. M. Botti[826,850], J. Brack[851], I. Brancus[839], T. Bretz[852], A. Bridgeman[853], F. L. Briechle[852], P. Buchholz[854], A. Bueno[855], S. Buitink[66], M. Buscemi[856,857], K. S. Caballero-Mora[795], L. Caccianiga[858], A. Cancio[827,826], F. Canfora[14,66], R. Caruso[856,857], A. Castellina[821,822], F. Catalani[859], G. Cataldi[847], L. Cazon[820], A. G. Chavez[860], J. A. Chinellato[861], J. Chudoba[845], R. W. Clay[824], A. C. Cobos Cerutti[862], R. Colalillo[840,833], A. Coleman[863], L. Collica[864], M. R. Coluccia[846,847], R. Conceição[820], G. Consolati[864,865], F. Contreras[836,837], M. J. Cooper[824], S. Coutu[863], C. E. Covault[866], J. Cronin[867,957], S. D'Amico[868,847], B. Daniel[861], S. Dasso[869,870], K. Daumiller[850], B. R. Dawson[824], J. A. Day[824], R. M. de Almeida[871], S. J. de Jong[14,66], G. De Mauro[14,66], J. R. T. de Mello Neto[848,872], I. De Mitri[846,847], J. de Oliveira[871], V. de Souza[873], J. Debatin[853], O. Deligny[844], M. L. Díaz Castro[861], F. Diogo[820], C. Dobrigkeit[861], J. C. D'Olivo[828], Q. Dorosti[854], R. C. Dos Anjos[874], M. T. Dova[875], A. Dundovic[876], J. Ebr[845], R. Engel[850], M. Erdmann[852], M. Erfani[854], C. O. Escobar[877], J. Espadanal[820], A. Etchegoyen[826,827], H. Falcke[14,66,878], J. Farmer[867], G. Farrar[879], A. C. Fauth[861], N. Fazzini[877], F. Feldbusch[880], F. Fenu[822,842], B. Fick[881], J. M. Figueira[826], A. Filipčič[494,882], M. M. Freire[883], T. Fujii[867], A. Fuster[826,827], R. Gaïor[884], B. García[862], F. Gaté[885], H. Gemmeke[880], A. Gherghel-Lascu[839], P. L. Ghia[844], U. Giaccari[848,886], M. Giammarchi[864], M. Giller[887], D. Głas[888], C. Glaser[852], G. Golup[825], M. Gómez Berisso[825], P. F. Gómez Vitale[836,837], N. González[826,850], A. Gorgi[821,822], M. Gottowik[204], A. F. Grillo[831,954], T. D. Grubb[824], F. Guarino[840,833], G. P. Guedes[889], R. Halliday[866], M. R. Hampel[826], P. Hansen[875], D. Harari[825], T. A. Harrison[824], V. M. Harvey[824], A. Haungs[850], T. Hebbeker[852], D. Heck[850], P. Heimann[854], A. E. Herve[853], G. C. Hill[824], C. Hojvat[877], E. W. Holt[850,826], P. Homola[849], J. R. Hörandel[14,66], P. Horvath[890], M. Hrabovský[890], T. Huege[850], J. Hulsman[826,850], A. Insolia[856,857], P. G. Isar[834], I. Jandt[204], J. A. Johnsen[891], M. Josebachuili[826], J. Jurysek[845], A. Kääpä[204], K. H. Kampert[204], B. Keilhauer[850], N. Kemmerich[823], J. Kemp[852], R. M. Kieckhafer[881], H. O. Klages[850], M. Kleifges[880], J. Kleinfeller[836], R. Krause[852], N. Krohm[204], D. Kuempel[204], G. Kukec Mezek[494], N. Kunka[880], A. Kuotb Awad[853], B. L. Lago[892], D. LaHurd[866], R. G. Lang[873], M. Lauscher[852], R. Legumina[887], M. A. Leigui de Oliveira[893], A. Letessier-Selvon[884], I. Lhenry-Yvon[844], K. Link[853], D. Lo Presti[856,857], L. Lopes[820], R. López[894], A. López Casado[829], R. Lorek[866], Q. Luce[844], A. Lucero[826], M. Malacari[867], M. Mallamaci[858,864], D. Mandat[845], P. Mantsch[877], A. G. Mariazzi[875], I. C. Maris[895], G. Marsella[846,847], D. Martello[846,847], H. Martinez[896], O. Martínez Bravo[894], J. J. Masías Meza[870], H. J. Mathes[850], S. Mathys[204], J. Matthews[2], G. Matthiae[897,898], E. Mayotte[204], P. O. Mazur[877], C. Medina[891], G. Medina-Tanco[828], D. Melo[826], A. Menshikov[880], K.-D. Merenda[891], S. Michal[890], M. I. Micheletti[883], L. Middendorf[852], L. Miramonti[858,864], B. Mitrica[839], D. Mockler[853], S. Mollerach[825], F. Montanet[841], C. Morello[821,822], G. Morlino[830,831], A. L. Müller[826,850], G. Müller[852], M. A. Muller[861,899], S. Müller[853,826], R. Mussa[822], I. Naranjo[825], P. H. Nguyen[824], M. Niculescu-Oglinzanu[839], M. Niechciol[854], L. Niemietz[204], T. Niggemann[852], D. Nitz[881], D. Nosek[900], V. Novotny[900], L. Nožka[890], L. A. Núñez[835], F. Oikonomou[863], A. Olinto[867], M. Palatka[845], J. Pallotta[901], P. Papenbreer[204], G. Parente[829], A. Parra[894], T. Paul[832], M. Pech[845], F. Pedreira[829], J. Pękala[849], J. Peña-Rodriguez[835], L. A. S. Pereira[861], M. Perlin[826], L. Perrone[846,847], C. Peters[852], S. Petrera[830,831], J. Phuntsok[863], T. Pierog[850], M. Pimenta[820], V. Pirronello[856,857], M. Platino[826], M. Plum[852], J. Poh[867], C. Porowski[849], R. R. Prado[873], P. Privitera[867], M. Prouza[845], E. J. Quel[901], S. Querchfeld[204], S. Quinn[866], R. Ramos-Pollan[835], J. Rautenberg[204], D. Ravignani[826], J. Ridky[845], F. Riehn[820], M. Risse[854], P. Ristori[901], V. Rizi[831,902], W. Rodrigues de Carvalho[823], G. Rodriguez Fernandez[897,898], J. Rodriguez Rojo[836], M. J. Roncoroni[826], M. Roth[850], E. Roulet[825], A. C. Rovero[869], P. Ruehl[854], S. J. Saffi[824], A. Saftoiu[839], F. Salamida[902,831], H. Salazar[894], A. Saleh[494], G. Salina[898], F. Sánchez[826], P. Sanchez-Lucas[855], E. M. Santos[823], E. Santos[845], F. Sarazin[891], R. Sarmento[820], C. Sarmiento-Cano[826], R. Sato[836], M. Schauer[204], V. Scherini[847], H. Schieler[850], M. Schimp[204], D. Schmidt[850,826], O. Scholten[903,904], P. Schovánek[845], F. G. Schröder[850], S. Schröder[204], A. Schulz[850], J. Schumacher[852], S. J. Sciutto[875], A. Segreto[857,905], A. Shadkam[2], R. C. Shellard[886], G. Sigl[876], G. Silli[826,850], R. Šmída[850], G. R. Snow[906], P. Sommers[863], S. Sonntag[854], J. F. Soriano[832], R. Squartini[836], D. Stanca[839], S. Stanič[494], J. Stasielak[849], P. Stassi[841], M. Stolpovskiy[841], F. Strafella[846,847], A. Streich[853],

F. Suarez[826,827], M. Suarez-Durán[835], T. Sudholz[824], T. Suomijärvi[844], A. D. Supanitsky[869], J. Šupík[890], J. Swain[907], Z. Szadkowski[888], A. Taboada[850], O. A. Taborda[825], C. Timmermans[14,66], C. J. Todero Peixoto[859], L. Tomankova[850], B. Tomé[820], G. Torralba Elipe[829], P. Travnicek[845], M. Trini[494], M. Tueros[875], R. Ulrich[850], M. Unger[850], M. Urban[852], J. F. Valdés Galicia[828], I. Valiño[829], L. Valore[840,833], G. van Aar[66], P. van Bodegom[824], A. M. van den Berg[903], A. van Vliet[66], E. Varela[894], B. Vargas Cárdenas[828], R. A. Vázquez[829], D. Veberič[850], C. Ventura[872], I. D. Vergara Quispe[875], V. Verzi[898], J. Vicha[845], L. Villaseñor[860], S. Vorobiov[494], H. Wahlberg[875], O. Wainberg[826,827], D. Walz[852], A. A. Watson[908], M. Weber[880], A. Weindl[850], M. Wiedeński[888], L. Wiencke[891], H. Wilczyński[849], M. Wirtz[852], D. Wittkowski[204], B. Wundheiler[826], L. Yang[494], A. Yushkov[845], E. Zas[829], D. Zavrtanik[494,882], M. Zavrtanik[494,882], A. Zepeda[896], B. Zimmermann[880], M. Ziolkowski[854], Z. Zong[844], F. Zuccarello[909,857],
(The Pierre Auger Collaboration),
S. Kim[548,910], S. Schulze[911], F. E. Bauer[649,910,912], J. M. Corral-Santana[913], I. de Gregorio-Monsalvo[913,914], J. González-López[910], D. H. Hartmann[915], C. H. Ishwara-Chandra[916], S. Martín[913,914], A. Mehner[913], K. Misra[917], M. J. Michałowski[918], L. Resmi[919],
(ALMA Collaboration),
Z. Paragi[920], I. Agudo[921], T. An[922,923], R. Beswick[924], C. Casadio[925], S. Frey[926], P. Jonker[66,927], M. Kettenis[920], B. Marcote[920], J. Moldon[924], A. Szomoru[920], H. J. Langevelde[920,928], J. Yang[929],
(Euro VLBI Team),
A. Cwiek[727], M. Cwiok[930], H. Czyrkowski[930], R. Dabrowski[930], G. Kasprowicz[931], L. Mankiewicz[932], K. Nawrocki[727], R. Opiela[932], L. W. Piotrowski[933], G. Wrochna[727], M. Zaremba[930], A. F. Żarnecki[930],
(Pi of the Sky Collaboration),
D. Haggard[934], M. Nynka[934], J. J. Ruan[934],
(The Chandra Team at McGill University),
P. A. Bland[935], T. Booler[501], H. A. R. Devillepoix[935], J. S. de Gois[501], P. J. Hancock[501], R. M. Howie[936], J. Paxman[936], E. K. Sansom[935], M. C. Towner[935],
(DFN: Desert Fireball Network),
J. Tonry[617], M. Coughlin[937], C. W. Stubbs[938], L. Denneau[617], A. Heinze[617], B. Stalder[939], H. Weiland[617],
(ATLAS),
R. P. Eatough[940], M. Kramer[940], A. Kraus[940],
(High Time Resolution Universe Survey),
E. Troja[941,942], L. Piro[172], J. Becerra González[943,944], N. R. Butler[722], O. D. Fox[945], H. G. Khandrika[945], A. Kutyrev[941,942], W. H. Lee[946,298], R. Ricci[947], R. E. Ryan Jr.[945], R. Sánchez-Ramírez[172], S. Veilleux[942,290], A. M. Watson[946], M. H. Wieringa[948], J. M. Burgess[949], H. van Eerten[950], C. J. Fontes[951], C. L. Fryer[951], O. Korobkin[951], R. T. Wollaeger[951],
(RIMAS and RATIR),
and
F. Camilo[952], A. R. Foley[952], S. Goedhart[952], S. Makhathini[952], N. Oozeer[952], O. M. Smirnov[952], R. P. Fender[66], and P. A. Woudt[953]
(SKA South Africa/MeerKAT)

[1] LIGO, California Institute of Technology, Pasadena, CA 91125, USA
[2] Louisiana State University, Baton Rouge, LA 70803, USA
[3] Università di Salerno, Fisciano, I-84084 Salerno, Italy
[4] INFN, Sezione di Napoli, Complesso Universitario di Monte S.Angelo, I-80126 Napoli, Italy
[5] University of Florida, Gainesville, FL 32611, USA
[6] OzGrav, School of Physics & Astronomy, Monash University, Clayton, VIC 3800, Australia
[7] LIGO Livingston Observatory, Livingston, LA 70754, USA
[8] Laboratoire d'Annecy-le-Vieux de Physique des Particules (LAPP), Université Savoie Mont Blanc, CNRS/IN2P3, F-74941 Annecy, France
[9] University of Sannio at Benevento, I-82100 Benevento, Italy and INFN, Sezione di Napoli, I-80100 Napoli, Italy
[10] Albert-Einstein-Institut, Max-Planck-Institut für Gravitationsphysik, D-30167 Hannover, Germany
[11] The University of Mississippi, University, MS 38677, USA
[12] NCSA, University of Illinois at Urbana-Champaign, Urbana, IL 61801, USA
[13] University of Cambridge, Cambridge CB2 1TN, UK
[14] Nikhef, Science Park, 1098 XG Amsterdam, The Netherlands
[15] LIGO, Massachusetts Institute of Technology, Cambridge, MA 02139, USA
[16] Instituto Nacional de Pesquisas Espaciais, 12227-010 São José dos Campos, São Paulo, Brazil
[17] Gran Sasso Science Institute (GSSI), I-67100 L'Aquila, Italy
[18] INFN, Laboratori Nazionali del Gran Sasso, I-67100 Assergi, Italy
[19] Inter-University Centre for Astronomy and Astrophysics, Pune 411007, India
[20] International Centre for Theoretical Sciences, Tata Institute of Fundamental Research, Bengaluru 560089, India
[21] University of Wisconsin-Milwaukee, Milwaukee, WI 53201, USA
[22] Leibniz Universität Hannover, D-30167 Hannover, Germany
[23] Università di Pisa, I-56127 Pisa, Italy

[24] INFN, Sezione di Pisa, I-56127 Pisa, Italy
[25] OzGrav, Australian National University, Canberra, ACT 0200, Australia
[26] Laboratoire des Matériaux Avancés (LMA), CNRS/IN2P3, F-69622 Villeurbanne, France
[27] SUPA, University of the West of Scotland, Paisley PA1 2BE, UK
[28] LAL, Univ. Paris-Sud, CNRS/IN2P3, Université Paris-Saclay, F-91898 Orsay, France
[29] California State University Fullerton, Fullerton, CA 92831, USA
[30] European Gravitational Observatory (EGO), I-56021 Cascina, Pisa, Italy
[31] Chennai Mathematical Institute, Chennai 603103, India
[32] Università di Roma Tor Vergata, I-00133 Roma, Italy
[33] INFN, Sezione di Roma Tor Vergata, I-00133 Roma, Italy
[34] Universität Hamburg, D-22761 Hamburg, Germany
[35] INFN, Sezione di Roma, I-00185 Roma, Italy
[36] Cardiff University, Cardiff CF24 3AA, UK
[37] Embry-Riddle Aeronautical University, Prescott, AZ 86301, USA
[38] Albert-Einstein-Institut, Max-Planck-Institut für Gravitationsphysik, D-14476 Potsdam-Golm, Germany
[39] APC, AstroParticule et Cosmologie, Université Paris Diderot, CNRS/IN2P3, CEA/Irfu, Observatoire de Paris, Sorbonne Paris Cité, F-75205 Paris Cedex 13, France
[40] Korea Institute of Science and Technology Information, Daejeon 34141, Korea
[41] West Virginia University, Morgantown, WV 26506, USA
[42] Università di Perugia, I-06123 Perugia, Italy
[43] INFN, Sezione di Perugia, I-06123 Perugia, Italy
[44] Syracuse University, Syracuse, NY 13244, USA
[45] University of Minnesota, Minneapolis, MN 55455, USA
[46] SUPA, University of Glasgow, Glasgow G12 8QQ, UK
[47] LIGO Hanford Observatory, Richland, WA 99352, USA
[48] Caltech CaRT, Pasadena, CA 91125, USA
[49] Wigner RCP, RMKI, Konkoly Thege Miklós út 29-33, H-1121 Budapest, Hungary
[50] NASA Goddard Space Flight Center, Greenbelt, MD 20771, USA
[51] Columbia University, New York, NY 10027, USA
[52] Stanford University, Stanford, CA 94305, USA
[53] Università di Camerino, Dipartimento di Fisica, I-62032 Camerino, Italy
[54] Università di Padova, Dipartimento di Fisica e Astronomia, I-35131 Padova, Italy
[55] INFN, Sezione di Padova, I-35131 Padova, Italy
[56] Institute of Physics, Eötvös University, Pázmány P. s. 1/A, H-1117 Budapest, Hungary
[57] Nicolaus Copernicus Astronomical Center, Polish Academy of Sciences, 00-716, Warsaw, Poland
[58] Rochester Institute of Technology, Rochester, NY 14623, USA
[59] University of Birmingham, Birmingham B15 2TT, UK
[60] INFN, Sezione di Genova, I-16146 Genova, Italy
[61] RRCAT, Indore MP 452013, India
[62] Faculty of Physics, Lomonosov Moscow State University, Moscow 119991, Russia
[63] SUPA, University of Strathclyde, Glasgow G1 1XQ, UK
[64] The Pennsylvania State University, University Park, PA 16802, USA
[65] OzGrav, University of Western Australia, Crawley, WA 6009, Australia
[66] Institute of Mathematics, Astrophysics and Particle Physics, Radboud University, 6525 AJ Nijmegen, The Netherlands
[67] Artemis, Université Côte d'Azur, Observatoire Côte d'Azur, CNRS, CS 34229, F-06304 Nice Cedex 4, France
[68] Institut FOTON, CNRS, Université de Rennes 1, F-35042 Rennes, France
[69] Washington State University, Pullman, WA 99164, USA
[70] University of Oregon, Eugene, OR 97403, USA
[71] Laboratoire Kastler Brossel, UPMC-Sorbonne Universités, CNRS, ENS-PSL Research University, Collège de France, F-75005 Paris, France
[72] Carleton College, Northfield, MN 55057, USA
[73] OzGrav, University of Adelaide, Adelaide, SA 5005, Australia
[74] Astronomical Observatory Warsaw University, 00-478 Warsaw, Poland
[75] VU University Amsterdam, 1081 HV Amsterdam, The Netherlands
[76] University of Maryland, College Park, MD 20742, USA
[77] Center for Relativistic Astrophysics, Georgia Institute of Technology, Atlanta, GA 30332, USA
[78] Université Claude Bernard Lyon 1, F-69622 Villeurbanne, France
[79] Università di Napoli "Federico II," Complesso Universitario di Monte S. Angelo, I-80126 Napoli, Italy
[80] Dipartimento di Fisica, Università degli Studi di Genova, I-16146 Genova, Italy
[81] RESCEU, University of Tokyo, Tokyo, 113-0033, Japan
[82] Tsinghua University, Beijing 100084, China
[83] Texas Tech University, Lubbock, TX 79409, USA
[84] Kenyon College, Gambier, OH 43022, USA
[85] Departamento de Astronomía y Astrofísica, Universitat de València, E-46100 Burjassot, València, Spain
[86] Museo Storico della Fisica e Centro Studi e Ricerche Enrico Fermi, I-00184 Roma, Italy
[87] National Tsing Hua University, Hsinchu City, 30013 Taiwan, Republic of China
[88] Charles Sturt University, Wagga Wagga, NSW 2678, Australia
[89] Center for Interdisciplinary Exploration & Research in Astrophysics (CIERA), Northwestern University, Evanston, IL 60208, USA
[90] Canadian Institute for Theoretical Astrophysics, University of Toronto, Toronto, ON M5S 3H8, Canada
[91] University of Chicago, Chicago, IL 60637, USA
[92] Pusan National University, Busan 46241, Korea
[93] The Chinese University of Hong Kong, Shatin, NT, Hong Kong
[94] INAF, Osservatorio Astronomico di Padova, I-35122 Padova, Italy
[95] INFN, Trento Institute for Fundamental Physics and Applications, I-38123 Povo, Trento, Italy
[96] OzGrav, University of Melbourne, Parkville, VIC 3010, Australia
[97] Università di Roma "La Sapienza," I-00185 Roma, Italy

[98] Université Libre de Bruxelles, Brussels 1050, Belgium
[99] Sonoma State University, Rohnert Park, CA 94928, USA
[100] Departamento de Matemáticas, Universitat de València, E-46100 Burjassot, València, Spain
[101] Montana State University, Bozeman, MT 59717, USA
[102] Universitat de les Illes Balears, IAC3—IEEC, E-07122 Palma de Mallorca, Spain
[103] The University of Texas Rio Grande Valley, Brownsville, TX 78520, USA
[104] Bellevue College, Bellevue, WA 98007, USA
[105] Institute for Plasma Research, Bhat, Gandhinagar 382428, India
[106] The University of Sheffield, Sheffield S10 2TN, UK
[107] Dipartimento di Scienze Matematiche, Fisiche e Informatiche, Università di Parma, I-43124 Parma, Italy
[108] INFN, Sezione di Milano Bicocca, Gruppo Collegato di Parma, I-43124 Parma, Italy
[109] California State University, Los Angeles, 5151 State University Drive, Los Angeles, CA 90032, USA
[110] Università di Trento, Dipartimento di Fisica, I-38123 Povo, Trento, Italy
[111] Montclair State University, Montclair, NJ 07043, USA
[112] National Astronomical Observatory of Japan, 2-21-1 Osawa, Mitaka, Tokyo 181-8588, Japan
[113] Observatori Astronòmic, Universitat de València, E-46980 Paterna, València, Spain
[114] School of Mathematics, University of Edinburgh, Edinburgh EH9 3FD, UK
[115] University and Institute of Advanced Research, Koba Institutional Area, Gandhinagar Gujarat 382007, India
[116] IISER-TVM, CET Campus, Trivandrum Kerala 695016, India
[117] University of Szeged, Dóm tér 9, H-6720 Szeged, Hungary
[118] University of Michigan, Ann Arbor, MI 48109, USA
[119] Tata Institute of Fundamental Research, Mumbai 400005, India
[120] INAF, Osservatorio Astronomico di Capodimonte, I-80131, Napoli, Italy
[121] Università degli Studi di Urbino "Carlo Bo," I-61029 Urbino, Italy
[122] INFN, Sezione di Firenze, I-50019 Sesto Fiorentino, Firenze, Italy
[123] Physik-Institut, University of Zurich, Winterthurerstrasse 190, CH-8057 Zurich, Switzerland
[124] American University, Washington, DC 20016, USA
[125] University of Białystok, 15-424 Białystok, Poland
[126] University of Southampton, Southampton SO17 1BJ, UK
[127] University of Washington Bothell, 18115 Campus Way NE, Bothell, WA 98011, USA
[128] Institute of Applied Physics, Nizhny Novgorod, 603950, Russia
[129] Korea Astronomy and Space Science Institute, Daejeon 34055, Korea
[130] Inje University Gimhae, South Gyeongsang 50834, Korea
[131] National Institute for Mathematical Sciences, Daejeon 34047, Korea
[132] NCBJ, 05-400 Świerk-Otwock, Poland
[133] Institute of Mathematics, Polish Academy of Sciences, 00656 Warsaw, Poland
[134] Hillsdale College, Hillsdale, MI 49242, USA
[135] Hanyang University, Seoul 04763, Korea
[136] Seoul National University, Seoul 08826, Korea
[137] NASA Marshall Space Flight Center, Huntsville, AL 35812, USA
[138] ESPCI, CNRS, F-75005 Paris, France
[139] Southern University and A&M College, Baton Rouge, LA 70813, USA
[140] College of William and Mary, Williamsburg, VA 23187, USA
[141] Centre Scientifique de Monaco, 8 quai Antoine Ier, MC-98000, Monaco
[142] Indian Institute of Technology Madras, Chennai 600036, India
[143] IISER-Kolkata, Mohanpur, West Bengal 741252, India
[144] Whitman College, 345 Boyer Avenue, Walla Walla, WA 99362 USA
[145] Indian Institute of Technology Bombay, Powai, Mumbai, Maharashtra 400076, India
[146] Scuola Normale Superiore, Piazza dei Cavalieri 7, I-56126 Pisa, Italy
[147] Université de Lyon, F-69361 Lyon, France
[148] Hobart and William Smith Colleges, Geneva, NY 14456, USA
[149] OzGrav, Swinburne University of Technology, Hawthorn, VIC 3122, Australia
[150] Janusz Gil Institute of Astronomy, University of Zielona Góra, 65-265 Zielona Góra, Poland
[151] University of Washington, Seattle, WA 98195, USA
[152] King's College London, University of London, London WC2R 2LS, UK
[153] Indian Institute of Technology, Gandhinagar Ahmedabad Gujarat 382424, India
[154] Indian Institute of Technology Hyderabad, Sangareddy, Khandi, Telangana 502285, India
[155] International Institute of Physics, Universidade Federal do Rio Grande do Norte, Natal RN 59078-970, Brazil
[156] Andrews University, Berrien Springs, MI 49104, USA
[157] Università di Siena, I-53100 Siena, Italy
[158] Trinity University, San Antonio, TX 78212, USA
[159] Abilene Christian University, Abilene, TX 79699, USA
[160] Colorado State University, Fort Collins, CO 80523, USA
[161] INFN Sezione di Bari, I-70126 Bari, Italy
[162] Politecnico di Bari, I-70126 Bari BA, Italy
[163] Harvard-Smithsonian Center for Astrophysics, 60 Garden Street, Cambridge, MA 02138, USA
[164] University of Alabama in Huntsville, Huntsville, AL 35899, USA
[165] Universities Space Research Association, Huntsville, AL 35805, USA
[166] Jacobs Technology, Inc., Huntsville, AL 35806, USA
[167] Los Alamos National Laboratory, Los Alamos, NM 87545, USA
[168] School of Physics, O'Brien Centre for Science North, University College Dublin, Belfield, Dublin 4, Ireland
[169] Max-Planck-Institut für extraterrestrische Physik, D-85748 Garching, Germany
[170] ISDC, Department of Astronomy, University of Geneva, Chemin d'Écogia, 16 CH-1290 Versoix, Switzerland
[171] European Space Research and Technology Centre (ESA/ESTEC), Keplerlaan 1, 2201 AZ Noordwijk, The Netherlands
[172] INAF, Istituto di Astrofisica e Planetologia Spaziali, via Fosso del Cavaliere 100, I-00133 Rome, Italy

[173] DTU Space, National Space Institute Elektrovej, Building 327 DK-2800 Kongens Lyngby Denmark
[174] Centro de Astrobiología (CAB-CSIC/INTA, ESAC Campus), Camino bajo del Castillo S/N, E-28692 Villanueva de la Cañada, Madrid, Spain
[175] IRAP, Université de Toulouse, CNRS, UPS, CNES, 9 Av. Roche, F-31028 Toulouse, France
[176] APC, AstroParticule et Cosmologie, Université Paris Diderot, CNRS/IN2P3, CEA/Irfu, Observatoire de Paris Sorbonne Paris Cité, 10 rue Alice Domont et Léonie Duquet, F-75205 Paris Cedex 13, France.
[177] DSM/Irfu/Service d'Astrophysique, Bat. 709 Orme des Merisiers CEA Saclay, F-91191 Gif-sur-Yvette Cedex, France
[178] Space Research Institute of Russian Academy of Sciences, Profsoyuznaya 84/32, Moscow, 117997, Russia
[179] Moscow Institute of Physics and Technology, Institutskiy per. 9, Dolgoprudny, Moscow Region, 141700, Russia
[180] INAF-Istituto di Astrofisica Spaziale e Fisica Cosmica Milano, via E. Bassini 15, I-20133 Milano, Italy
[181] Max Planck Institute for Astrophysics, Karl-Schwarzschild-Str. 1, Garching b. Munchen D-85741, Germany
[182] Department of Physics, University of Adelaide, Adelaide, 5005, Australia
[183] DESY, D-15738 Zeuthen, Germany
[184] Dept. of Physics and Astronomy, University of Canterbury, Private Bag 4800, Christchurch, New Zealand
[185] Université Libre de Bruxelles, Science Faculty CP230, B-1050 Brussels, Belgium
[186] Niels Bohr Institute, University of Copenhagen, DK-2100 Copenhagen, Denmark
[187] Oskar Klein Centre and Dept. of Physics, Stockholm University, SE-10691 Stockholm, Sweden
[188] Département de physique nucléaire et corpusculaire, Université de Genève, CH-1211 Genève, Switzerland
[189] Friedrich-Alexander-Universität Erlangen-Nürnberg, Erlangen Centre for Astroparticle Physics, Erwin-Rommel-Str. 1, D-91058 Erlangen, Germany
[190] Department of Physics, Marquette University, Milwaukee, WI, 53201, USA
[191] Dept. of Physics, Pennsylvania State University, University Park, PA 16802, USA
[192] Dept. of Physics, Massachusetts Institute of Technology, Cambridge, MA 02139, USA
[193] III. Physikalisches Institut, RWTH Aachen University, D-52056 Aachen, Germany
[194] Physics Department, South Dakota School of Mines and Technology, Rapid City, SD 57701, USA
[195] Dept. of Physics, University of Alberta, Edmonton, AB T6G 2E1, Canada
[196] Dept. of Physics and Astronomy, University of California, Irvine, CA 92697, USA
[197] Institute of Physics, University of Mainz, Staudinger Weg 7, D-55099 Mainz, Germany
[198] Dept. of Physics, University of California, Berkeley, CA 94720, USA
[199] Dept. of Physics and Center for Cosmology and Astro-Particle Physics, Ohio State University, Columbus, OH 43210, USA
[200] Dept. of Astronomy, Ohio State University, Columbus, OH 43210, USA
[201] Fakultät für Physik & Astronomie, Ruhr-Universität Bochum, D-44780 Bochum, Germany
[202] Dept. of Physics and Astronomy, University of Kansas, Lawrence, KS 66045, USA
[203] Lawrence Berkeley National Laboratory, Berkeley, CA 94720, USA
[204] Bergische Universität Wuppertal, Department of Physics, Wuppertal, Germany
[205] Dept. of Physics, University of Maryland, College Park, MD 20742, USA
[206] Dept. of Physics, TU Dortmund University, D-44221 Dortmund, Germany
[207] Dept. of Physics, Sungkyunkwan University, Suwon 440-746, Korea
[208] Dept. of Physics and Astronomy, Uppsala University, Box 516, S-75120 Uppsala, Sweden
[209] Dept. of Physics and Wisconsin IceCube Particle Astrophysics Center, University of Wisconsin, Madison, WI 53706, USA
[210] Vrije Universiteit Brussel (VUB), Dienst ELEM, B-1050 Brussels, Belgium
[211] SNOLAB, 1039 Regional Road 24, Creighton Mine 9, Lively, ON P3Y 1N2, Canada
[212] Institut für Kernphysik, Westfälische Wilhelms-Universität Münster, D-48149 Münster, Germany
[213] Physik-department, Technische Universität München, D-85748 Garching, Germany
[214] Dept. of Astronomy and Astrophysics, Pennsylvania State University, University Park, PA 16802, USA
[215] Dept. of Physics and Astronomy, University of Rochester, Rochester, NY 14627, USA
[216] Dept. of Physics and Astronomy, Michigan State University, East Lansing, MI 48824, USA
[217] Bartol Research Institute and Dept. of Physics and Astronomy, University of Delaware, Newark, DE 19716, USA
[218] Dept. of Physics and Astronomy, University of Gent, B-9000 Gent, Belgium
[219] Institut für Physik, Humboldt-Universität zu Berlin, D-12489 Berlin, Germany
[220] Dept. of Physics, Southern University, Baton Rouge, LA 70813, USA
[221] Dept. of Astronomy, University of Wisconsin, Madison, WI 53706, USA
[222] Earthquake Research Institute, University of Tokyo, Bunkyo, Tokyo 113-0032, Japan
[223] Dept. of Physics and Institute for Global Prominent Research, Chiba University, Chiba 263-8522, Japan
[224] CTSPS, Clark-Atlanta University, Atlanta, GA 30314, USA
[225] Dept. of Physics, University of Texas at Arlington, 502 Yates Street, Science Hall Room 108, Box 19059, Arlington, TX 76019, USA
[226] Dept. of Physics and Astronomy, Stony Brook University, Stony Brook, NY 11794-3800, USA
[227] Université de Mons, B-7000 Mons, Belgium
[228] Dept. of Physics and Astronomy, University of Alabama, Tuscaloosa, AL 35487, USA
[229] Dept. of Physics, Drexel University, 3141 Chestnut Street, Philadelphia, PA 19104, USA
[230] Dept. of Physics, University of Wisconsin, River Falls, WI 54022, USA
[231] Dept. of Physics, Yale University, New Haven, CT 06520, USA
[232] Dept. of Physics and Astronomy, University of Alaska Anchorage, 3211 Providence Drive, Anchorage, AK 99508, USA
[233] Dept. of Physics, University of Oxford, 1 Keble Road, Oxford OX1 3NP, UK
[234] School of Physics and Center for Relativistic Astrophysics, Georgia Institute of Technology, Atlanta, GA 30332, USA
[235] Indian Institute of Science Education and Research, Dr. Homi Bhabha Road, Pashan, Pune 411008, India
[236] Department of Physics, Indian Institute of Technology Bombay, Mumbai 400076, India
[237] Physical Research Laboratory, Ahmedabad, India
[238] Ioffe Institute, Politekhnicheskaya 26, St. Petersburg 194021, Russia
[239] University of California-Berkeley, Space Sciences Lab, 7 Gauss Way, Berkeley, CA 94720, USA
[240] Emeritus, NASA Goddard Space Flight Center, Greenbelt, MD 20771, USA
[241] Key Laboratory Of Particle Astrophysics, Institute Of High Energy Physics, Chinese Academy Of Sciences, Beijing 100049, China
[242] University Of Chinese Academy Of Sciences, Chinese Academy Of Sciences, Beijing 100049, China
[243] Beijing Normal University, Beijing 100088, China
[244] GRPHE, Université de Haute Alsace, Institut universitaire de technologie de Colmar, 34 rue du Grillenbreit BP 50568, F-68008 Colmar, France
[245] Technical University of Catalonia, Laboratory of Applied Bioacoustics, Rambla Exposició, E-08800 Vilanova i la Geltrú, Barcelona, Spain
[246] INFN—Sezione di Genova, Via Dodecaneso 33, I-16146 Genova, Italy

[247] Institut d'Investigació per a la Gestió Integrada de les Zones Costaneres (IGIC), Universitat Politècnica de València. C/ Paranimf 1, E-46730 Gandia, Spain
[248] Aix Marseille Univ, CNRS/IN2P3, CPPM, Marseille, France
[249] APC, Univ Paris Diderot, CNRS/IN2P3, CEA/Irfu, Obs de Paris, Sorbonne Paris Cité, France
[250] IFIC, Instituto de Física Corpuscular (CSIC—Universitat de València), c/ Catedrático José Beltrán, 2 E-46980 Paterna, Valencia, Spain
[251] LAM—Laboratoire d'Astrophysique de Marseille, Pôle de l'Étoile Site de Château-Gombert, rue Frédéric Joliot-Curie 38, F-13388 Marseille Cedex 13, France
[252] National Center for Energy Sciences and Nuclear Techniques, B.P. 1382, R. P. 10001 Rabat, Morocco
[253] INFN—Laboratori Nazionali del Sud (LNS), Via S. Sofia 62, I-95123 Catania, Italy
[254] Huygens-Kamerlingh Onnes Laboratorium, Universiteit Leiden, The Netherlands
[255] Institute for Space Science, RO-077125 Bucharest, Măgurele, Romania
[256] Universiteit van Amsterdam, Instituut voor Hoge-Energie Fysica, Science Park 105, 1098 XG Amsterdam, The Netherlands
[257] INFN—Sezione di Roma, P.le Aldo Moro 2, I-00185 Roma, Italy
[258] Dipartimento di Fisica dell'Università La Sapienza, P.le Aldo Moro 2, I-00185 Roma, Italy
[259] Gran Sasso Science Institute, Viale Francesco Crispi 7, I-00167 L'Aquila, Italy
[260] University Mohammed V in Rabat, Faculty of Sciences, 4 av. Ibn Battouta, B.P. 1014, 10000, Rabat, Morocco
[261] INFN—Sezione di Bologna, Viale Berti-Pichat 6/2, I-40127 Bologna, Italy
[262] INFN—Sezione di Bari, Via E. Orabona 4, I-70126 Bari, Italy
[263] Department of Computer Architecture and Technology/CITIC, University of Granada, E-18071 Granada, Spain
[264] Géoazur, UCA, CNRS, IRD, Observatoire de la Côte d'Azur, Sophia Antipolis, France
[265] Dipartimento di Fisica dell'Università, Via Dodecaneso 33, I-16146 Genova, Italy
[266] Université Paris-Sud, F-91405 Orsay Cedex, France
[267] University Mohammed I, Laboratory of Physics of Matter and Radiations, B.P. 717, Oujda 6000, Morocco
[268] Institut für Theoretische Physik und Astrophysik, Universität Würzburg, Emil-Fischer Str. 31, D-97074 Würzburg, Germany
[269] Dipartimento di Fisica e Astronomia dell'Università, Viale Berti Pichat 6/2, I-40127 Bologna, Italy
[270] Laboratoire de Physique Corpusculaire, Clermont Université, Université Blaise Pascal, CNRS/IN2P3, BP 10448, F-63000 Clermont-Ferrand, France
[271] INFN—Sezione di Catania, Viale Andrea Doria 6, I-95125 Catania, Italy
[272] LSIS, Aix Marseille Université CNRS ENSAM LSIS UMR 7296 F-13397 Marseille, France; Université de Toulon CNRS LSIS UMR 7296, F-83957 La Garde, France
[273] Institut Universitaire de France, F-75005 Paris, France
[274] Royal Netherlands Institute for Sea Research (NIOZ) and Utrecht University, Landsdiep 4, 1797 SZ 't Horntje (Texel), The Netherlands
[275] Dr. Remeis-Sternwarte and ECAP, Universität Erlangen-Nürnberg, Sternwartstr. 7, D-96049 Bamberg, Germany
[276] Moscow State University, Skobeltsyn Institute of Nuclear Physics, Leninskie gory, 119991 Moscow, Russia
[277] Mediterranean Institute of Oceanography (MIO), Aix-Marseille University, F-13288, Marseille, Cedex 9, France; Université du Sud Toulon-Var, CNRS-INSU/IRD UM 110, 83957, La Garde Cedex, France
[278] Dipartimento di Fisica ed Astronomia dell'Università, Viale Andrea Doria 6, I-95125 Catania, Italy
[279] Direction des Sciences de la Matière, Institut de recherche sur les lois fondamentales de l'Univers, Service de Physique des Particules, CEA Saclay, F-91191 Gif-sur-Yvette Cedex, France
[280] INFN—Sezione di Pisa, Largo B. Pontecorvo 3, I-56127 Pisa, Italy
[281] Dipartimento di Fisica dell'Università, Largo B. Pontecorvo 3, I-56127 Pisa, Italy
[282] INFN—Sezione di Napoli, Via Cintia I-80126 Napoli, Italy
[283] Dipartimento di Fisica dell'Università Federico II di Napoli, Via Cintia I-80126, Napoli, Italy
[284] Dpto. de Física Teórica y del Cosmos & C.A.F.P.E., University of Granada, E-18071 Granada, Spain
[285] Université de Strasbourg, CNRS, IPHC UMR 7178, F-67000 Strasbourg, France
[286] University of Leicester, X-ray and Observational Astronomy Research Group, Leicester Institute for Space and Earth Observation, Department of Physics & Astronomy, University Road, Leicester, LE1 7RH, UK
[287] University College London, Mullard Space Science Laboratory, Holmbury St. Mary, Dorking, RH5 6NT, UK
[288] Department of Astronomy and Astrophysics, The Pennsylvania State University, University Park, PA 16802, USA
[289] Astrophysics Science Division, NASA Goddard Space Flight Center, Greenbelt, MD 20771 USA
[290] Joint Space-Science Institute, University of Maryland, College Park, MD 20742, USA
[291] Istituto Nazionale di Astrofisica – Istituto di Astrofisica Spaziale e Fisica Cosmica Palermo, Via Ugo La Malfa 153, I-90146, Palermo, Italy
[292] Department of Astronomy and Space Sciences, University of Istanbul, Beyzıt 34119, Istanbul, Turkey
[293] Space Science Data Center—Agenzia Spaziale Italiana, I-00133 Roma, Italy
[294] Institute for Gravitation and the Cosmos, The Pennsylvania State University, University Park, PA 16802, USA
[295] Universities Space Research Association, 7178 Columbia Gateway Drive, Columbia, MD 21046, USA
[296] National Science Foundation, 2415 Eisenhower Avenue, Alexandria, VA 22314, USA
[297] Center for Research and Exploration in Space Science and Technology (CRESST) and NASA Goddard Space Flight Center, Greenbelt MD, 20771 USA
[298] Department of Physics, University of Maryland, Baltimore County, 1000 Hilltop Circle, Baltimore, MD 21250, USA
[299] Istituto Nazionale di Astrofisica – Osservatorio Astronomico di Brera, Via Bianchi 46, I-23807 Merate, Italy
[300] Department of Physics, University of Warwick, Coventry CV4 7AL, UK
[301] Los Alamos National Laboratory, B244, Los Alamos, NM, 87545, USA
[302] Istituto Nazionale di Astrofisica – Osservatorio Astronomico di Roma, Via Frascati 33, I-00040 Monteporzio Catone, Italy
[303] Department of Physics and Astronomy, University of Maryland, College Park, MD 20742-4111, USA
[304] INAF-IAPS, via del Fosso del Cavaliere 100, I-00133 Roma, Italy
[305] Dip. di Fisica, Univ. di Roma "Tor Vergata," via della Ricerca Scientifica 1, I-00133 Roma, Italy
[306] Gran Sasso Science Institute, viale Francesco Crispi 7, I-67100 L'Aquila, Italy
[307] INAF-OAR, via Frascati 33, I-00078 Monte Porzio Catone (Roma), Italy
[308] ASI Space Science Data Center (SSDC), via del Politecnico, I-00133 Roma, Italy
[309] INAF-IASF-Bologna, via Gobetti 101, I-40129 Bologna, Italy
[310] INAF-IASF Milano, via E.Bassini 15, I-20133 Milano, Italy
[311] Agenzia Spaziale Italiana, via del Politecnico, I-00133 Roma, Italy
[312] INAF, Osservatorio Astronomico di Cagliari, Via della Scienza 5, I-09047 Selargius (CA), Italy
[313] Dip. di Fisica, Università di Trieste and INFN, via Valerio 2, I-34127 Trieste, Italy
[314] Unitat de Física de les Radiacions, Departament de Física, and CERES-IEEC, Universitat Autònoma de Barcelona, E-08193 Bellaterra, Spain
[315] Birkeland Centre for Space Science, Department of Physics and Technology, University of Bergen, Bergen, Norway
[316] INFN-Pavia, via Bassi 6, I-27100 Pavia, Italy
[317] University of Witwatersrand, Johannesburg, South Africa

[318] CIFS, c/o Physics Department, University of Turin, via P. Giuria 1, I-10125, Torino, Italy
[319] INFN Roma Tor Vergata, via della Ricerca Scientifica 1, I-00133 Roma, Italy
[320] East Windsor RSD, 25A Leshin Lane, Hightstown, NJ 08520, USA
[321] Osservatorio Astronomico di Brera, via Emilio Bianchi 46, I-23807 Merate (LC), Italy
[322] Department of Astronomy and Astrophysics, University of California, Santa Cruz, CA 95064, USA
[323] The Observatories of the Carnegie Institution for Science, 813 Santa Barbara Street, Pasadena, CA 91101, USA
[324] Institute for Astronomy, University of Hawai'i, 2680 Woodlawn Drive, Honolulu, HI 96822, USA
[325] Departamento de Física y Astronomía, Universidad de La Serena, La Serena, Chile
[326] Nuclear Science Division, Lawrence Berkeley National Laboratory, Berkeley, CA 94720, USA
[327] Departments of Physics and Astronomy, University of California, Berkeley, CA 94720, USA
[328] Dark Cosmology Centre, Niels Bohr Institute, University of Copenhagen, Juliane Maries Vej 30, DK-2100 Copenhagen Ø, Denmark
[329] Space Telescope Science Institute, 3700 San Martin Drive, Baltimore, MD 21218, USA
[330] Department of Physics and Astronomy, The Johns Hopkins University, 3400 North Charles Street, Baltimore, MD 21218, USA
[331] Department of Physics, Brandeis University, Waltham, MA, USA
[332] Fermi National Accelerator Laboratory, P. O. Box 500, Batavia, IL 60510, USA
[333] Department of Physics, University of Surrey, Guildford GU2 7XH, UK
[334] Department of Physics and Astronomy, University of Pennsylvania, Philadelphia, PA 19104, USA
[335] Department of Astronomy, Indiana University, 727 E. Third Street, Bloomington, IN 47405, USA
[336] Astrophysical Institute, Department of Physics and Astronomy, 251B Clippinger Lab, Ohio University, Athens, OH 45701, USA
[337] George P. and Cynthia Woods Mitchell Institute for Fundamental Physics and Astronomy, and Department of Physics and Astronomy, Texas A&M University, College Station, TX 77843, USA
[338] LSST, 933 North Cherry Avenue, Tucson, AZ 85721, USA
[339] The Observatories of the Carnegie Institution for Science, 813 Santa Barbara St., Pasadena, CA 91101, USA
[340] Institut d'Astrophysique de Paris (UMR7095: CNRS & UPMC), 98 bis Bd Arago, F-75014, Paris, France
[341] Center for Interdisciplinary Exploration and Research in Astrophysics (CIERA) and Department of Physics and Astronomy, Northwestern University, Evanston, IL 60208, USA
[342] Center for Theoretical Astrophysics, Los Alamos National Laboratory, Los Alamos, NM 87544, USA
[343] Instituto de Fisica Teorica UAM/CSIC, Universidad Autonoma de Madrid, E-28049 Madrid, Spain
[344] National Center for Supercomputing Applications, 1205 West Clark Street, Urbana, IL 61801, USA
[345] Department of Physics & Astronomy, University College London, Gower Street, London WC1E 6BT, UK
[346] Department of Physics, ETH Zurich, Wolfgang-Pauli-Strasse 16, CH-8093 Zurich, Switzerland
[347] Kavli Institute for Cosmological Physics, University of Chicago, Chicago, IL 60637, USA
[348] Observatòrio do Valongo, Universidade Federal do Rio de Janeiro, Ladeira do Pedro Antônio 43, Rio de Janeiro, RJ, 20080-090, Brazil
[349] Center for Interdisciplinary Exploration and Research in Astrophysics (CIERA) and Department of Physics and Astronomy, Northwestern University, Evanston, IL 60208, USA
[350] National Optical Astronomy Observatory, 950 North Cherry Avenue, Tucson, AZ 85719, USA
[351] Departamento de Astronomonía, Universidad de Chile, Camino del Observatorio 1515, Las Condes, Santiago, Chile
[352] Department of Physics and Columbia Astrophysics Laboratory, Columbia University, New York, NY 10027, USA
[353] Department of Physics, University of Michigan, 450 Church Street, Ann Arbor, MI 48109-1040, USA
[354] Lawrence Berkeley National Laboratory, 1 Cyclotron Road, Berkeley, CA 94720, USA
[355] Department of Astronomy & Theoretical Astrophysics Center, University of California, Berkeley, CA 94720-3411, USA
[356] Physics Division, Lawrence Berkeley National Laboratory, Berkeley, CA 94720-8160, USA
[357] Steward Observatory, University of Arizona, 933 N. Cherry Avenue, Tucson, AZ 85721, USA
[358] Instituto de Física Gleb Wataghin, Universidade Estadual de Campinas, Campinas, SP—13083-859, Brazil
[359] Laboratório Interinstitucional de e-Astronomia—LIneA, Rua Gal. José Cristino 77, Rio de Janeiro, RJ—20921-400, Brazil
[360] Cerro Tololo Inter-American Observatory, National Optical Astronomy Observatory, Casilla 603, La Serena, Chile
[361] Institute of Astronomy, University of Cambridge Madingley Road, Cambridge CB3 0HA, UK
[362] Kavli Institute for Cosmology, University of Cambridge, Madingley Road, Cambridge CB3 0HA, UK
[363] CNRS, UMR 7095, Institut d'Astrophysique de Paris, F-75014, Paris, France
[364] Sorbonne Universités, UPMC Univ Paris 06, UMR 7095, Institut d'Astrophysique de Paris, F-75014, Paris, France
[365] Kavli Institute for Particle Astrophysics & Cosmology, P. O. Box 2450, Stanford University, Stanford, CA 94305, USA
[366] SLAC National Accelerator Laboratory, Menlo Park, CA 94025, USA
[367] Institute of Cosmology & Gravitation, University of Portsmouth, Portsmouth PO1 3FX, UK
[368] Observatório Nacional, Rua Gal. José Cristino 77, Rio de Janeiro, RJ—20921-400, Brazil
[369] Department of Astronomy, University of Illinois, 1002 W. Green Street, Urbana, IL 61801, USA
[370] Institute of Space Sciences, IEEC-CSIC, Campus UAB, Carrer de Can Magrans, s/n, E-08193 Barcelona, Spain
[371] George P. and Cynthia Woods Mitchell Institute for Fundamental Physics and Astronomy, and Department of Physics and Astronomy, Texas A&M University, College Station, TX 77843, USA
[372] Department of Physics, IIT Hyderabad, Kandi, Telangana 502285, India
[373] Excellence Cluster Universe, Boltzmannstr. 2, D-85748 Garching, Germany
[374] Faculty of Physics, Ludwig-Maximilians-Universität, Scheinerstr. 1, D-81679 Munich, Germany
[375] Department of Physics, California Institute of Technology, Pasadena, CA 91125, USA
[376] Jet Propulsion Laboratory, California Institute of Technology, 4800 Oak Grove Drive, Pasadena, CA 91109, USA
[377] Institut de Física d'Altes Energies (IFAE), The Barcelona Institute of Science and Technology, Campus UAB, 08193 Bellaterra (Barcelona), Spain
[378] Department of Astronomy, University of Michigan, Ann Arbor, MI 48109, USA
[379] Department of Physics, University of Michigan, Ann Arbor, MI 48109, USA
[380] Universitäts-Sternwarte, Fakultät für Physik, Ludwig-Maximilians Universität München, Scheinerstr. 1, D-81679 München, Germany
[381] Department of Astronomy, University of California, Berkeley, 501 Campbell Hall, Berkeley, CA 94720, USA
[382] Center for Cosmology and Astro-Particle Physics, The Ohio State University, Columbus, OH 43210, USA
[383] Department of Physics, The Ohio State University, Columbus, OH 43210, USA
[384] Astronomy Department, University of Washington, Box 351580, Seattle, WA 98195, USA
[385] Santa Cruz Institute for Particle Physics, Santa Cruz, CA 95064, USA
[386] Australian Astronomical Observatory, North Ryde, NSW 2113, Australia
[387] Departamento de Física Matemática, Instituto de Física, Universidade de São Paulo, CP 66318, São Paulo, SP—05314-970, Brazil
[388] Department of Astronomy, The Ohio State University, Columbus, OH 43210, USA

[389] Institució Catalana de Recerca i Estudis Avançats, E-08010 Barcelona, Spain
[390] Max Planck Institute for Extraterrestrial Physics, Giessenbachstrasse, D-85748 Garching, Germany
[391] Department of Physics and Astronomy, Pevensey Building, University of Sussex, Brighton BN1 9QH, UK
[392] Centro de Investigaciones Energéticas, Medioambientales y Tecnológicas (CIEMAT), Madrid, Spain
[393] Brookhaven National Laboratory, Building 510, Upton, NY 11973, USA
[394] School of Physics and Astronomy, University of Southampton, Southampton SO17 1BJ, UK
[395] Computer Science and Mathematics Division, Oak Ridge National Laboratory, Oak Ridge, TN 37831, USA
[396] Argonne National Laboratory, 9700 South Cass Avenue, Lemont, IL 60439, USA
[397] Department of Physics, Stanford University, 382 Via Pueblo Mall, Stanford, CA 94305, USA
[398] Department of Physics and Astronomy, University of North Carolina at Chapel Hill, Chapel Hill, NC 27599, USA
[399] Department of Astronomy and Steward Observatory, University of Arizona, 933 N Cherry Avenue, Tucson, AZ 85719, USA
[400] Department of Physics, University of California, 1 Shields Avenue, Davis, CA 95616-5270, USA
[401] Department of Physics and Astronomy, University of Padova, Via 8 Febbraio, I-35122 Padova, Italy
[402] INAF—Osservatorio Astronomico di Padova, Vicolo della Osservatorio 5, I-35122 Padova, Italy
[403] INAF—Osservatorio Astronomico di Padova, Vicolo dell'Osservatorio 5, I-35122 Padova, Italy
[404] INAF—Osservatorio Astronomico di Roma, Via di Frascati, 33, I-00078 Monteporzio Catone, Italy
[405] INAF—Osservatorio Astronomico di Brera, Via E. Bianchi 46, I-23807 Merate (LC), Italy
[406] Space Science Data Center, ASI, Via del Politecnico, s.n.c., I-00133, Roma, Italy
[407] INAF—Osservatorio Astronomico di Capodimonte, salita Moiariello 16, I-80131, Napoli, Italy
[408] INAF—Istituto di Astrofisica Spaziale e Fisica Cosmica di Bologna, Via Gobetti 101, I-40129 Bologna, Italy
[409] Dipartimento di Fisica "G. Occhialini," Università degli Studi di Milano-Bicocca, P.za della Scienza 3, I-20126 Milano, Italy
[410] Laboratoire Univers et Particules de Montpellier, Université Montpellier 2, 34095, Montpellier, France
[411] INAF—Osservatorio Astronomico di Catania, Via S. Sofia 78, I-95123, Catania, Italy
[412] Department of physics, University of Naples Federico II, Corso Umberto I, 40, I-80138 Napoli, Italy
[413] Institute for Astrophysics and Particle Physics, University of Innsbruck, Technikerstrasse 25/8, A-6020 Innsbruck, Austria
[414] Departamento de Ciencias Fısicas, Universidad Andrés Bello, Fernández Concha 700, Las Condes, Santiago, Chile
[415] Università degli Studi dell'Insubria, via Valleggio 11, I-22100, Como, Italy
[416] INAF—Istituto di Astrofisica Spaziale e Fisica Cosmica di Milano, via E. Bassini 15, I-20133 Milano, Italy
[417] INAF—Osservatorio Astrofisico di Torino, Pino Torinese, Italy
[418] INAF—Osservatorio Astrofisico di Arcetri, Largo Enrico Fermi 5, I-50125, Florence, Italy
[419] INAF—Istituto di Radioastronomia di Bologna, Bologna Italy
[420] Key Laboratory of dark Matter and Space Astronomy, Purple Mountain Observatory, Chinese Academy of Science, Nanjing 210008, China
[421] Thüringer Landessternwarte Tautenburg, Sternwarte 5, D-07778 Tautenburg, Germany
[422] Department of Physics, The George Washington University, Corcoran Hall, Washington, DC 20052, USA
[423] Astronomy, Physics, and Statistics Institute of Sciences (APSIS)
[424] Astrophysics Research Institute, Liverpool John Moores University, Liverpool Science Park, IC2, 146 Brownlow Hill, Liverpool L3 5RF, UK
[425] Max-Planck-Institut für Astrophysik, Karl-Schwarzschild-Str. 1, D-85748 Garching bei München, Germany
[426] European Southern Observatory, Karl-Schwarzschild-Strasse 2, D-85748 Garching bei München, Germany
[427] INAF—Osservatorio Astronomico di Trieste, Via G.B. Tiepolo 11, I-34143 Trieste, Italy
[428] Racah Institute of Physics, The Hebrew University of Jerusalem, Jerusalem 91904, Israel
[429] GEPI, Observatoire de Paris, PSL Research University, CNRS, Place Jules Janssen, F-92190, Meudon, France
[430] Department of Physics and Astronomy, University of Leicester, Leicester LE1 7RH, UK
[431] Frontier Research Institute for Interdisciplinary Sciences, Tohoku University, Sendai 980-8578, Japan
[432] Astronomical Institute, Tohoku University, Sendai 980-8578, Japan
[433] Department of Physics, University of Bath Claverton Down, Bath, BA2 7AY, UK
[434] CEA Saclay—DRF/Irfu/Département d'Astrophysique, F-91191 Gif-sur-Yvette, France
[435] Department of Physics and Institute of Theoretical Physics, Nanjing Normal University, Nanjing 210046, China
[436] Center for Astrophysics and Cosmology (CAC), University of Nova Gorica, Nova Gorica, Slovenia
[437] Anton Pannekoek Institute, University of Amsterdam, Science Park 904, 1098XH Amsterdam, The Netherlands
[438] Astrophysics Research Institute, Liverpool John Moores University, ic2, Liverpool Science Park, 146 Brownlow Hill, Liverpool L3 5RF, UK
[439] Faculty of Mathematics and Physics, University of Ljubljana, Jadranska 19, 1000 Ljubljana, Slovenia
[440] Yunnan Observatories, Chinese Academy of Sciences, 650011 Kunming, Yunnan Province, China
[441] Astrophysics Research Institute, Liverpool John Moores University, Liverpool, L3 5RF, UK
[442] Department of Physics, The George Washington University, 725 21st Street NW, Washington, DC 20052, USA
[443] Laboratoire AIM, CEA-IRFU/CNRS/Université Paris Diderot, Service d'Astrophysique, CEA Saclay, F-91191 Gif-sur-Yvette, France
[444] Santa Cruz Institute for Particle Physics, Department of Physics and Department of Astronomy and Astrophysics, University of California at Santa Cruz, Santa Cruz, CA 95064, USA
[445] Università di Pisa and Istituto Nazionale di Fisica Nucleare, Sezione di Pisa, I-56127 Pisa, Italy
[446] Istituto Nazionale di Fisica Nucleare, Sezione di Trieste, and Università di Trieste, I-34127 Trieste, Italy
[447] Dipartimento di Fisica, Università di Trieste, I-34127 Trieste, Italy
[448] Istituto Nazionale di Fisica Nucleare, Sezione di Padova, I-35131 Padova, Italy
[449] Dipartimento di Fisica e Astronomia "G. Galilei," Università di Padova, I-35131 Padova, Italy
[450] California State University, Los Angeles, Department of Physics and Astronomy, Los Angeles, CA 90032, USA
[451] Istituto Nazionale di Fisica Nucleare, Sezione di Pisa, I-56127 Pisa, Italy
[452] Dipartimento di Fisica "M. Merlin" dell'Università e del Politecnico di Bari, I-70126 Bari, Italy
[453] Istituto Nazionale di Fisica Nucleare, Sezione di Bari, I-70126 Bari, Italy
[454] W. W. Hansen Experimental Physics Laboratory, Kavli Institute for Particle Astrophysics and Cosmology, Department of Physics and SLAC National Accelerator Laboratory, Stanford University, Stanford, CA 94305, USA
[455] Istituto Nazionale di Fisica Nucleare, Sezione di Torino, I-10125 Torino, Italy
[456] Dipartimento di Fisica, Università degli Studi di Torino, I-10125 Torino, Italy
[457] Laboratoire Univers et Particules de Montpellier, Université Montpellier, CNRS/IN2P3, F-34095 Montpellier, France
[458] Deutsches Elektronen Synchrotron DESY, D-15738 Zeuthen, Germany
[459] Center for Research and Exploration in Space Science and Technology (CRESST) and NASA Goddard Space Flight Center, Greenbelt, MD 20771, USA
[460] Italian Space Agency, Via del Politecnico, snc, I-00133 Roma, Italy
[461] College of Science, George Mason University, Fairfax, VA 22030; Resident at Naval Research Laboratory, Washington, DC 20375, USA

[462] Space Science Division, Naval Research Laboratory, Washington, DC 20375-5352, USA
[463] Space Science Data Center—Agenzia Spaziale Italiana, Via del Politecnico, snc, I-00133, Roma, Italy
[464] Istituto Nazionale di Fisica Nucleare, Sezione di Perugia, I-06123 Perugia, Italy
[465] Department of Physics and Astronomy, Sonoma State University, Rohnert Park, CA 94928-3609, USA
[466] RWTH Aachen University, Institute for Theoretical Particle Physics and Cosmology (TTK), D-52056 Aachen, Germany
[467] INAF Istituto di Radioastronomia, I-40129 Bologna, Italy
[468] Dipartimento di Astronomia, Università di Bologna, I-40127 Bologna, Italy
[469] Università Telematica Pegaso, Piazza Trieste e Trento, 48, I-80132 Napoli, Italy
[470] Laboratoire Leprince-Ringuet, École polytechnique, CNRS/IN2P3, F-91128 Palaiseau, France
[471] Department of Physical Sciences, Hiroshima University, Higashi-Hiroshima, Hiroshima 739-8526, Japan
[472] Department of Physics and Department of Astronomy, University of Maryland, College Park, MD 20742, USA
[473] Centre d'Études Nucléaires de Bordeaux Gradignan, IN2P3/CNRS, Université Bordeaux 1, BP120, F-33175 Gradignan Cedex, France
[474] Laboratoire de Physique et Chimie de l'Environnement et de l'Espace, Université d'Orléans/CNRS, F-45071 Orléans Cedex 02, France
[475] Station de radioastronomie de Nançay, Observatoire de Paris, CNRS/INSU, F-18330 Nançay, France
[476] Science Institute, University of Iceland, IS-107 Reykjavik, Iceland
[477] Nordita, Roslagstullsbacken 23, 106 91 Stockholm, Sweden
[478] Department of Physics, Graduate School of Science, University of Tokyo, 7-3-1 Hongo, Bunkyo-ku, Tokyo 113-0033, Japan
[479] Istituto Nazionale di Fisica Nucleare, Sezione di Roma "Tor Vergata," I-00133 Roma, Italy
[480] Department of Physics and Astronomy, Clemson University, Kinard Lab of Physics, Clemson, SC 29634-0978, USA
[481] Max-Planck-Institut für Physik, D-80805 München, Germany
[482] Department of Physics, University of Johannesburg, PO Box 524, Auckland Park 2006, South Africa
[483] Institut für Astro- und Teilchenphysik and Institut für Theoretische Physik, Leopold-Franzens-Universität Innsbruck, A-6020 Innsbruck, Austria
[484] Department of Physics, The University of Hong Kong, Pokfulam Road, Hong Kong, China
[485] Laboratory for Space Research, The University of Hong Kong, Hong Kong, China
[486] NYCB Real-Time Computing Inc., Lattingtown, NY 11560-1025, USA
[487] Purdue University Northwest, Hammond, IN 46323, USA
[488] Hiroshima Astrophysical Science Center, Hiroshima University, Higashi-Hiroshima, Hiroshima 739-8526, Japan
[489] CNRS, IRAP, F-31028 Toulouse cedex 4, France
[490] GAHEC, Universit de Toulouse, UPS-OMP, IRAP, F-31400 Toulouse, France
[491] Institute of Space Sciences (CSICIEEC), Campus UAB, Carrer de Magrans s/n, E-08193 Barcelona, Spain
[492] Institució Catalana de Recerca i Estudis Avançats (ICREA), E-08010 Barcelona, Spain
[493] INAF-Istituto di Astrofisica Spaziale e Fisica Cosmica Bologna, via P. Gobetti 101, I-40129 Bologna, Italy
[494] Centre for Astrophysics and Cosmology, University of Nova Gorica, Vipavska 11c, 5270 Ajdovščina, Slovenia
[495] Sydney Institute for Astronomy, School of Physics, The University of Sydney, Sydney, NSW 2006, Australia
[496] ARC Centre of Excellence for All-sky Astrophysics in 3 Dimensions (ASTRO 3D)
[497] ATNF, CSIRO Astronomy and Space Science, PO Box 76, Epping, NSW 1710, Australia
[498] ARC Centre of Excellence for All-sky Astrophysics (CAASTRO)
[499] University of Wisconsin–Milwaukee, Milwaukee, WI 53201, USA
[500] ATNF, CSIRO Astronomy and Space Science, 26 Dick Perry Avenue, Kensington, WA 6152, Australia
[501] International Centre for Radio Astronomy Research, Curtin University, Bentley, WA 6102, Australia
[502] Centre for Astrophysics and Supercomputing, Swinburne University of Technology, Mail H30, PO Box 218, VIC 3122, Australia
[503] Department of Physics, University of California, Santa Barbara, CA 93106-9530, USA
[504] Las Cumbres Observatory, 6740 Cortona Drive, Suite 102, Goleta, CA 93117-5575, USA
[505] School of Physics and Astronomy, Tel Aviv University, Tel Aviv 69978, Israel
[506] Columbia Astrophysics Laboratory, Columbia University, New York, NY, 10027, USA
[507] Centre for Astrophysics and Supercomputing, Swinburne University of Technology, PO Box 218, H29, Hawthorn, VIC 3122, Australia
[508] The Australian Research Council Centre of Excellence for Gravitational Wave Discovery (OzGrav), Australia
[509] The Australian Research Council Centre of Excellence for All-Sky Astrophysics (CAASTRO), Australia
[510] Research School of Astronomy and Astrophysics, The Australian National University, Canberra, ACT 2611, Australia
[511] Australian Astronomical Observatory, 105 Delhi Road, North Ryde, NSW 2113, Australia
[512] George P. and Cynthia Woods Mitchell Institute for Fundamental Physics & Astronomy, Texas A. & M. University, Department of Physics and Astronomy, 4242 TAMU, College Station, TX 77843, USA
[513] Purple Mountain Observatory, Chinese Academy of Sciences, Nanjing 210008, China
[514] Chinese Center for Antarctic Astronomy, Nanjing 210008, China
[515] The University of the Virgin Islands, 2 John Brewer's Bay, St. Thomas 00802, USVI
[516] Monash Centre for Astrophysics, Monash University, VIC 3800, Australia
[517] Centre for Translational Data Science, University of Sydney, Sydney, NSW 2006, Australia
[518] School of Physics and Astronomy, University of Nottingham, Nottingham, UK
[519] CSIRO Astronomy & Space Science, Australia Telescope National Facility, P.O. Box 76, Epping, NSW 1710, Australia
[520] SKA Organisation, Jodrell Bank Observatory, SK11 9DL, UK
[521] National Astronomical Observatories, Chinese Academy of Sciences, Beijing 100012, China
[522] Physics Department and Tsinghua Center for Astrophysics (THCA), Tsinghua University, Beijing, 100084, China
[523] Tianjin Normal University, Tianjin 300074, China
[524] School of Physics, University of New South Wales, NSW 2052, Australia
[525] Nanjing Institute of Astronomical Optics and Technology, Nanjing 210042, China
[526] Department of Astronomy, Beijing Normal University, Beijing 100875, China
[527] School of Astronomy and Space Science and Key Laboratory of Modern Astronomy and Astrophysics in Ministry of Education, Nanjing University, Nanjing 210093, China
[528] Orangewave Innovation Science, 2113 Old Highway 52, Moncks Corner, SC 29461, USA
[529] Department of Physics, 2354 Fairchild Drive, U.S. Air Force Academy, CO 80840, USA
[530] Universite de Toulouse, IRAP 14 Av. Edouard Belin, F-31000 Toulouse France
[531] Auragne Observatory, France
[532] Research School of Astronomy and Astrophysics, The Australian National University, Canberra, ACT 2611, Australia
[533] Department of Physics and Astronomy, University of Leicester, University Road, Leicester, LE1 7RH, UK
[534] Instituto de Astrofísica de Andalucía (IAA-CSIC), Glorieta de la Astronomía s/n, E-18008 Granada, Spain

[535] Institute of Astronomy, University of Cambridge, Madingley Road, Cambridge, CB3 0HA, UK
[536] Max-Planck-Institut für extraterrestrische Physik, Giessenbachstr. 1, D-85740 Garching, Germany
[537] Birmingham Institute for Gravitational Wave Astronomy and School of Physics and Astronomy, University of Birmingham, Birmingham B15 2TT, UK
[538] School of Physics and Astronomy and Monash Centre for Astrophysics, Monash University, VIC 3800, Australia
[539] The Oskar Klein Centre, Department of Astronomy, AlbaNova, Stockholm University, SE-106 91 Stockholm, Sweden
[540] Anton Pannekoek Institute, University of Amsterdam, Science Park 904, 1098 XH Amsterdam, the Netherlands
[541] ASTRON, the Netherlands Institute for Radio Astronomy, Postbus 2, 7990 AA Dwingeloo, the Netherlands
[542] SUPA, School of Physics & Astronomy, University of St Andrews, North Haugh, St Andrews KY16 9SS, UK
[543] Niels Bohr Institute & Centre for Star and Planet Formation, University of Copenhagen Øster Voldgade 5, DK-1350—Copenhagen, Denmark
[544] Institute for Advanced Research, Nagoya University, Furo-cho, Chikusa-ku, Nagoya 464-8601, Japan
[545] Space Telescope Science Institute, 3700 San Martin Drive, Baltimore, MD 21218, USA
[546] Centre for Astrophysics and Cosmology, Science Institute, University of Iceland, Dunhagi 5, 107 Reykjavík, Iceland
[547] Instituto de Astrofísica, Pontificia Universidad Católica de Chile, Av. Vicuña Mackenna 4860, 7820436 Macul, Santiago, Chile
[548] Max-Planck-Institut für Astronomie Königstuhl 17, D-69117 Heidelberg, Germany
[549] Lomonosov Moscow State University, Physics Department, Vorobievy gory, 1 Moscow, 119991, Russia
[550] Lomonosov Moscow State University,SAI, Universitetsky prospekt, 13 Moscow, 119234, Russia
[551] Observatorio Astronomico Felix Aguilar (OAFA), Avda Benavides s/n, Rivadavia, El Leonsito, Argentina
[552] Instituto de Ciencias Astronomicas de la Tierra y del Espacio, Casilla de Correo 49, 5400 San Juan, Argentina
[553] Universidad Nacional de San Juan, Av. Ignacio de la Roza 391, San Juan, 5400, Argentina
[554] Irkutsk State University Applied Physics Institute, 20, Boulevard, 664003, Irkutsk, Russia
[555] Blagoveschenk State Pedagogical University, Lenin str., 104, Blagoveschensk, 675000, Russia
[556] Instituto de Astrofsica de Canarias, C/Via Lctea, s/n E-38205, La Laguna, Tenerife, Spain
[557] Kislovodsk Solar Station, Pulkovo Observatory RAS, Gagarina str. 100, Kislovodsk, 357700, Russia
[558] Institute for Space-Earth Environmental Research, Nagoya, 464-8601, Japan
[559] Subaru Telescope, Hilo, HI 96720, USA
[560] National Astronomical Observatory of Japan, Mitaka, Tokyo 181-8588, Japan
[561] University of Hyogo, Sayo 679-5313, Japan
[562] South African Astronomical Observatory, Cape Town, South Africa
[563] Massey University, Auckland 0745, New Zealand
[564] Institute of Astronomy, Graduate School of Science, Mitaka 181-0015, Japan
[565] Tokyo Institute of Technology, Tokyo 152-8551, Japan
[566] Osaka City University, Osaka 558-8585, Japan
[567] Hiroshima Astrophysical Science Center, Higashi-Hiroshima 739-8526, Japan
[568] Hiroshima University, Higashi-Hiroshima, 739-8526, Japan
[569] Okayama Astrophysical Observatory, Asakuchi 719-0232, Japan
[570] Purple Mountain Observatory, Nanjing 210008, China
[571] Osaka University, Toyonaka 560-0043, Japan
[572] Nagoya University, Nagoya 464-8602, Japan
[573] Kagoshima University, Kogoshima 890-0065, Japan
[574] Kyoto University, Kyoto 606-8502, Japan
[575] Precursory Research for Embryonic Science and Technology, Mitaka, Tokyo 181-0015, Japan
[576] Toho University, Funabashi 274-8510, Japan
[577] Konan University, Kobe 658-8501, Japan
[578] Kavli Institute for the Physics and Mathematics of the Universe (WPI), Kashiwa 277-8583, Japan
[579] University of Canterbury, Mt John Observatory, Lake Tekapo 7945, New Zealand
[580] Division of Physics, Math and Astronomy, California Institute of Technology, Pasadena, CA 91125, USA
[581] Indian Institute of Astrophysics, Bangalore-560034, India
[582] University of Colorado, Boulder, CO 80309, USA
[583] South African Astronomical Observatory (SAAO), Cape Town 7935, South Africa
[584] Department of Astronomy, University of Washington, Seattle, WA 98195, USA
[585] National Center for Radio Astrophysics, Tata Institute of Fundamental Research, Pune University Campus, Ganeshkhind Pune 411007, India
[586] Department of Physics, University of Wisconsin, Milwaukee, WI 53201, USA
[587] Remote Sensing Division, Naval Research Laboratory, Code 7213, Washington, DC 20375, USA
[588] Department of Physics, George Washington University,Washington, DC 20052, USA
[589] University College London, Mullard Space Science Laboratory, RH5 6NT, UK
[590] X-ray and Observational Astronomy Research Group, Leicester Institute for Space and Earth Observation, Department of Physics & Astronomy, University of Leicester, Leicester LE1 7RH, UK
[591] The Oskar Klein Centre, Department of Physics, Stockholm, University, AlbaNova, SE-106 91 Stockholm, Sweden
[592] Space Telescope Science Institute, Baltimore, MD 21218, USA
[593] National Radio Astronomy Observatory, Socorro, NM, USA
[594] Department of Physics and Astronomy, University of Southampton, Southampton, Hampshire SO17 1BJ, UK
[595] Institute of Cosmology and Gravitation, University of Portsmouth, Portsmouth PO1 3FX, UK
[596] The Raymond and Beverly Sackler School of Physics and Astronomy, Tel Aviv University, Tel Aviv 69978, Israel
[597] Infrared Processing and Analysis Center, California Institute of Technology, Pasadena, CA 91125, USA
[598] Racah Institute of Physics, The Hebrew University of Jerusalem, Jerusalem, 91904, Israel
[599] Center for Computational Astrophysics, Simons Foundation, New York, NY 10010, USA
[600] Graduate Institute of Astronomy, National Central University, Taoyuan City 32001, Taiwan
[601] Department of Physics, Tokyo Institute of Technology, Tokyo 152-8551, Japan
[602] Department of Astronomy, University of California, Berkeley, CA 94720-3411, USA
[603] Department of Physics, University of California, Berkeley, CA 94720, USA
[604] Gemini Observatory, Casilla 603, La Serena, Chile
[605] Max-Planck Institute for Astrophysics, Garching, Germany
[606] Center for Interdisciplinary Exploration and Research in Astrophysics (CIERA), Department of Physics and Astronomy, Northwestern University, Evanston, IL 60208, USA
[607] The Adler Planetarium, Chicago, IL 60605, USA

[608] Astrophysics, Department of Physics, University of Oxford, Oxford OX1 3RH, UK
[609] Department of Particle Physics & Astrophysics, Weizmann Institute of Science, Rehovot 7610001, Israel
[610] Department of Physics and Astronomy, Texas Tech University, Lubbock, TX 79409-1051, USA
[611] Astrophysics Research Institute, Liverpool John Moores University, IC2, Liverpool Science Park, 146 Brownlow Hill, Liverpool L3 5RF, UK
[612] Department of Astronomy, San Diego State University, CA 92182, USA
[613] Kavli Institute for the Physics and Mathematics of the Universe (WPI), The University of Tokyo Institutes for Advanced Study, The University of Tokyo, Kashiwa, Chiba 277-8583, Japan
[614] The Oskar Klein Centre, Department of Astronomy, Stockholm University, AlbaNova, SE-106 91 Stockholm, Sweden
[615] University of California Merced, Merced, CA, USA
[616] Australian Research Council Centre of Excellence for All-sky Astrophysics (CAASTRO), Sydney Institute for Astronomy, School of Physics, The University of Sydney, Sydney, NSW 2006, Australia
[617] Institute for Astronomy, University of Hawaii, 2680 Woodlawn Drive, Honolulu, Hawaii 96822, USA
[618] ISAS/JAXA, Sagamihara, Kanagawa 229-8510, Japan
[619] University of Miyazaki, Miyazaki, Miyazaki 889-2192, Japan
[620] Tokyo Institute of Technology, Meguro-ku, Tokyo 152-8551, Japan
[621] Aoyama Gakuin University, Sagamihara, Kanagawa 229-8558, Japan
[622] Kyoto University, Kyoto, Kyoto, 606-8502, Japan
[623] JAXA, Tsukuba, Ibaraki 305-8505, Japan
[624] RIKEN, Wako, Saitama, 351-0198, Japan
[625] Chuo University, Bunkyo-ku, Tokyo 112-8551, Japan
[626] National Astronomical Observatory of Japan, Mitaka, Tokyo 181-8588, Japan
[627] Nihon University, Chiyoda-ku, Tokyo 101-8308, Japan
[628] Osaka University, Toyonaka, Osaka 560-0043, Japan
[629] Nagoya University, Nagoya, Aichi 464-8601, Japan
[630] The University of Western Australia, 35, Stirling Highway, Perth, WA 6009, Australia
[631] Swinburne University, John Street, Hawthorn, VIC 3122, Australia
[632] ARTEMIS (UCA, CNRS, OCA), boulevard de l'Observatoire, CS 34229, F-06304 Nice, France
[633] IRAP (CNRS, UPS), 14 avenue Edouard Belin, F-31029 Toulouse, France
[634] The University of the Virgin Islands, 2 John Brewer's Bay, St Thomas 00802, USVI
[635] The Auragne Observatory, F-31190 Auragne, France
[636] Center of the Exploration of the Origin of the Universe, Astronomy Program, Dept. of Physics & Astronomy, Seoul National University, 1 Gwanak-rho, Gwanak-gu, Seoul 08826, Korea
[637] Korea Astronomy and Space Science Institute, 776 Daedeokdae-ro, Yuseong-gu, Daejeon 34055, Korea
[638] CAS Key Laboratory of Space Astronomy and Technology, National Astronomical Observatories, Chinese Academy of Sciences, Beijing 100012, China
[639] Astrophysics Research Centre, School of Mathematics and Physics, Queens University Belfast, Belfast BT7 1NN, UK
[640] Department of Physics and Astronomy, University of Southampton, Southampton SO17 1BJ, UK
[641] Department of Particle Physics and Astrophysics, Weizmann Institute of Science, Rehovot 76100, Israel
[642] Department of Physics, University of Warwick, Coventry CV4 7AL, UK
[643] Institute for Astronomy, SUPA (Scottish Universities Physics Alliance), University of Edinburgh, Royal Observatory, Blackford Hill, Edinburgh EH9 3HJ, UK
[644] Departamento de Ciencias Fisicas, Universidad Andres Bello, Avda. Republica 252, Santiago, 8320000, Chile
[645] Millennium Institute of Astrophysics (MAS), Nuncio Monseñor Sótero Sanz 100, Providencia, Santiago, Chile
[646] European Southern Observatory, Alonso de Córdova 3107, Casilla 19, Santiago, Chile
[647] The Oskar Klein Centre, Department of Astronomy, Stockholm University, AlbaNova, SE-10691 Stockholm, Sweden
[648] Instituto de Astrofísica and Centro de Astroingeniería, Facultad de Física, Pontificia Universidad Católica de Chile, Casilla 306, Santiago 22, Chile
[649] Space Science Institute, 4750 Walnut Street, Suite 205, Boulder, CO 80301, USA
[650] Dipartimento di Fisica e Astronomia "G. Galilei," Università di Padova, Vicolo dell'Osservatorio 3, I-35122, Padova, Italy
[651] INAF—Osservatorio Astronomico di Brera, via E. Bianchi 46, I-23807 Merate (LC), Italy
[652] INAF—Osservatorio Astronomico di Capodimonte, via Salita Moiariello 16, I-80131 Napoli, Italy
[653] The Oskar Klein Centre, Department of Physics, Stockholm University, AlbaNova, SE-10691 Stockholm, Sweden
[654] SRON, Netherlands Institute for Space Research, Sorbonnelaan 2, NL-3584 CA Utrecht, The Netherlands
[655] European Southern Observatory, Karl-Schwarzschild-Str. 2, D-85748 Garching b. München, Germany
[656] ICRANet-Pescara, Piazza della Repubblica 10, I-65122 Pescara, Italy
[657] IAP/CNRS and Université Pierre et Marie Curie, Paris, France
[658] Unidad Mixta Internacional Franco-Chilena de Astronomía (CNRS UMI 3386), Departamento de Astronomía, Universidad de Chile, Camino El Observatorio 1515, Las Condes, Santiago, Chile
[659] Istituto Nazionale di Astrofisica, Viale del Parco Mellini 84, I-00136 Roma, Italy
[660] Institute of Cosmology and Gravitation, Dennis Sciama Building, University of Portsmouth, Burnaby Road, Portsmouth PO1 3FX, UK
[661] PITT PACC, Department of Physics and Astronomy, University of Pittsburgh, Pittsburgh, PA 15260, USA
[662] CENTRA, Instituto Superior Técnico, Universidade de Lisboa, Portugal
[663] Warsaw University Astronomical Observatory, Al. Ujazdowskie 4, 00-478 Warszawa, Poland
[664] Tuorla Observatory, Department of Physics and Astronomy, University of Turku, Väisäläntie 20, FI-21500 Piikkiö, Finland
[665] Instituto de Física y Astronomía, Universidad de Valparaiso, Gran Bretaña 1111, Playa Ancha, Valparaíso 2360102, Chile
[666] Institute of Astronomy, University of Cambridge, Madingley Road, Cambridge, CB3 0HA, UK
[667] Department of Physics, Lancaster University, Lancaster LA1 4YB, UK
[668] Instituto de Astrofísica de Andalucía (IAA-CSIC), Glorieta de la Astronomía s/n, E-18008, Granada, Spain
[669] Zentrum für Astronomie der Universität Heidelberg, Institut für Theoretische Astrophysik, Philosophenweg 12, D-69120 Heidelberg, Germany
[670] Heidelberger Institut für Theoretische Studien, Schloss-Wolfsbrunnenweg 35, D-69118 Heidelberg, Germany
[671] Finnish Centre for Astronomy with ESO (FINCA), University of Turku, Väisäläntie 20, 21500 Piikkiö, Finland
[672] Max Planck Institute for Astronomy, Königstuhl 17, D-69117 Heidelberg, Germany
[673] Institut fur Physik, Humboldt-Universitat zu Berlin, Newtonstr. 15, D-12489 Berlin, Germany
[674] Sorbonne Universités, UPMC Univ. Paris 6 and CNRS, UMR 7095, Institut d'Astrophysique de Paris, 98 bis bd Arago, F-75014 Paris, France
[675] INAF-Osservatorio Astronomico di Padova, Vicolo dell'Osservatorio 5, I-35122 Padova, Italy
[676] Department of Astrophysics, University of Oxford, Oxford OX1 3RH, UK
[677] Department of Astronomy, Universidad de Chile, Camino El Observatorio 1515, Las Condes, Santiago de Chile, Chile

[678] School of Physical, Environmental, and Mathematical Sciences, University of New South Wales, Australian Defence Force Academy, Canberra, ACT 2600, Australia
[679] ARC Centre of Excellence for All-sky Astrophysics (CAASTRO), Canberra, ACT 2611, Australia
[680] Università degli studi di Catania, DFA & DIEEI, Via Santa Sofia 64, I-95123 Catania, Italy
[681] INFN—Laboratori Nazionali del Sud, Via Santa Sofia 62, I-95123 Catania, Italy
[682] Department of Physics, University of the Free State, Bloemfontein, 9300 South Africa
[683] School of Physics and Astronomy, University of Minnesota, 116 Church Street SE, Minneapolis, MN 55455-0149, USA
[684] Max-Planck-Institut für Extraterrestrische Physik, Giessenbachstraße 1, D-85748, Garching, Germany
[685] Thüringer Landessternwarte Tautenburg, Sternwarte 5, D-07778 Tautenburg, Germany
[686] Texas Tech University, Lubbock, TX 79409, USA
[687] Department of Astrophysics, American Museum of Natural History, Central Park West and 79th Street, New York, NY 10024, USA
[688] South African Astronomical Observatory, PO Box 9, 7935 Observatory, South Africa
[689] Southern African Large Telescope Foundation, P.O. Box 9, 7935 Observatory, South Africa.
[690] Center for Gravitational Wave Astronomy and Department of Physics & Astronomy, University of Texas—Río Grande Valley, Brownsville, TX, USA
[691] George P. and Cynthia W. Mitchell Institute for Fundamental Physics & Astronomy, Department of Physics & Astronomy, Texas A&M University, College Station, TX, USA
[692] IATE-OAC, Universidad Nacional de Córdoba-CONICET, Córdoba, Argentina
[693] Instituto de Astronomia, Geofísica e Ciências Atmosféricas da U. de São Paulo, São Paulo, SP, Brazil
[694] Instituto de Investigación Multidisciplinario en Ciencia y Tecnología, Universidad de La Serena, La Serena, Chile
[695] Departamento de Física y Astronomía, Universidad de La Serena, La Serena, Chile
[696] Departamento de Física, Universidade Federal de Sergipe, São Cristóvão, SE, Brazil
[697] Departamento de Física, Universidade Federal de Santa Catarina, Florianópolis, SC, Brazil
[698] Departamento de Física Matemática, Instituto de Física, Universidade de São Paulo, São Paulo, SP, Brazil
[699] Departamento de Astronomia, Observatório Nacional, Rio de Janeiro, RJ, Brazil
[700] Centro de Estudios de Física del Cosmos de Aragón, E-44001 Teruel, Spain
[701] Instituto Nacional de Astrofísica, Óptica y Electrónica, Tonantzintla, Puebla, México
[702] Instituto de Astronomía, Universidad Nacional Autónoma de México, Ciudad de México, México
[703] Instituto de Astrofísica, Pontificia Universidad Católica de Chile, Santiago, Chile
[704] Observatorio do Valongo, Universidade Federal do Rio de Janeiro, Río de Janeiro, RJ, Brazil
[705] X-ray Astrophysics Laboratory and CRESST, NASA Goddard Space Flight Center, Greenbelt, MD, USA
[706] Ludwig Maximilian Universität Munich, Faculty of Physics, Munich, Germany
[707] Department of Physics, University of Notre Dame, Notre Dame, IN, USA
[708] Joint Institute for Nuclear Astrophysics—Center for the Evolution of the Elements, USA
[709] Instituto de Astrofísica de Andalucía del Consejo Superior de Investigaciones Científicas (IAA-CSIC), Granada, Apdo. 03004, E-18080 Granada, Spain
[710] Departamento de Ingeniería de Sistemas y Automática, Escuela de Ingenierías (Unidad Asociada al IAA-CSIC), Universidad de Málaga, Dr. Pedro Ortiz Ramos, E-29071 Málaga, Spain
[711] Departamento de Álgebra, Geometría y Topología, Facultad de Ciencias, Universidad de Málaga, Málaga, Campus de Teatinos, E-29071 sn, Málaga, Spain
[712] Instituto de Astronomía, Universidad Nacional Autónoma de México, Apdo. Postal 870, 2800 Ensenada, Baja California, México
[713] Astronomical Institute, Academy of Sciences of the Czech Republic, Boční II 1401, CZ-141 00 Prague, Czech Republic
[714] Astronomical Institute, Academy of Sciences of the Czech Republic, 251 65 Ondřejov, Czech Republic
[715] Institute of Physics of the Czech Academy of Sciences, Na Slovance 1999/2, 182 21 Praha 8, Czech Republic
[716] Department of Physics, Sungkyunkwan University, 2066, Seobu-ro, Jangan-gu, Suwon, Gyeonggi-do, 16419, Korea
[717] ISDEFE for ESA, ESAC, E-28692 Villanueva de la Cañada (Madrid), Spain
[718] Aryabhatta Research Institute of Observational Sciences, Manora Peak, Nainital 263 002, India
[719] Department of Physics, University of Auckland, Private Bag 92019, Auckland, New Zealand
[720] National Institute of Water and Atmospheric Research (NIWA), Lauder, New Zealand
[721] Yunnan Astronomical Observatory, CAS, Kunming 650011, Yunnan, China
[722] School of Earth and Space Exploration, Arizona State University, Tempe, AZ 85287, USA
[723] Dunlap Institute for Astronomy and Astrophysics, University of Toronto, Toronto, ON M5S 3H4, Canada
[724] Peripety Scientific Ltd., PO Box 11355 Manners Street, Wellington, 6142, New Zealand
[725] Department of Physics, University of Washington, Seattle, WA 98195, USA
[726] International Centre for Radio Astronomy Research, University of Western Australia, Crawley, WA 6009, Australia
[727] National Centre for Nuclear Research, 00-681 Warsaw, Poland
[728] Aoyama Gakuin University, 5-10-1 Fuchinobe, Chuo, Sagamihara, Kanagawa 252-5258, Japan
[729] Nagoya University, Furo, Chikusa, Nagoya 464-8601, Japan
[730] Kavli Institute for the Physics and Mathematics of the Universe, The University of Tokyo, 5-1-5 Kashiwanoha, Kashiwa 277-8583, Japan
[731] Waseda University, 3-4-1 Okubo, Shinjuku, Tokyo 169-8555, Japan
[732] Kanagawa University, 3-27-1 Rokkakubashi, Kanagawa, Yokohama, Kanagawa 221-8686, Japan
[733] Institute for Cosmic Ray Research, The University of Tokyo, 5-1-5 Kashiwa-no-Ha, Kashiwa, Chiba 277-8582, Japan
[734] Institute of Applied Physics (IFAC), National Research Council (CNR), Via Madonna del Piano, 10, I-50019 Sesto, Fiorentino, Italy
[735] University of Siena, Rettorato, via Banchi di Sotto 55, I-53100 Siena, Italy
[736] Space Research Institute, Moscow, 117997, Russia
[737] National Research University Higher School of Economics, Moscow, 101000, Russia
[738] National Research Nuclear University MEPhI, Moscow, 115409, Russia
[739] Fesenkov Astrophysical Institute, Almaty, 050020, Kazakhstan
[740] Special Astrophysical Observatory of Russian Academy of Sciences, Nizhniy Arkhyz, 369167, Russia
[741] Crimean Astrophysical Observatory, Nauchny, Crimea 298409
[742] Kharadze Abastumani Astrophysical Observatory, Ilia State University, Tbilisi, 0162, Georgia
[743] Institute of Solar Terrestrial Physics, Irkutsk, 664033 Russia
[744] Institute of Astronomy and Geophysics, Mongolian Academy of Sciences, 13343, Ulaanbaatar, Mongolia
[745] Keldysh Institute of Applied Mathematics, Russian Academy of Sciences, Miusskaya 4, 125047, Moscow, Russia
[746] Ulugh Beg Astronomical Institute, Astronomicheskaya st., 33, Tashkent, 100052, Uzbekistan
[747] Centre for Space Research, North-West University, Potchefstroom 2520, South Africa
[748] Universität Hamburg, Institut für Experimentalphysik, Luruper Chaussee 149, D 22761 Hamburg, Germany
[749] Max-Planck-Institut für Kernphysik, P.O. Box 103980, D-69029 Heidelberg, Germany

[750] Dublin Institute for Advanced Studies, 31 Fitzwilliam Place, Dublin 2, Ireland
[751] National Academy of Sciences of the Republic of Armenia, Marshall Baghramian Avenue, 24, 0019 Yerevan, Republic of Armenia
[752] Instytut Fizyki Jądrowej PAN, ul. Radzikowskiego 152, 31-342 Kraków, Poland
[753] Department of Physics, Rikkyo University, 3-34-1 Nishi-Ikebukuro, Toshima-ku, Tokyo 171-8501, Japan
[754] LUTH, Observatoire de Paris, PSL Research University, CNRS, Université Paris Diderot, 5 Place Jules Janssen, F-92190 Meudon, France
[755] Laboratoire d'Annecy-le-Vieux de Physique des Particules, Université Savoie Mont-Blanc, CNRS/IN2P3, F-74941 Annecy-le-Vieux, France
[756] University of Namibia, Department of Physics, Private Bag 13301, Windhoek, Namibia
[757] GRAPPA, Anton Pannekoek Institute for Astronomy, University of Amsterdam, Science Park 904, 1098 XH Amsterdam, The Netherlands
[758] Department of Physics and Electrical Engineering, Linnaeus University, 351 95 Växjö, Sweden
[759] Institut für Theoretische Physik, Lehrstuhl IV: Weltraum und Astrophysik, Ruhr-Universität Bochum, D 44780 Bochum, Germany
[760] GRAPPA, Anton Pannekoek Institute for Astronomy and Institute of High-Energy Physics, University of Amsterdam, Science Park 904, 1098 XH Amsterdam, The Netherlands
[761] Institut für Astro- und Teilchenphysik, Leopold-Franzens-Universität Innsbruck, A-6020 Innsbruck, Austria
[762] School of Physical Sciences, University of Adelaide, Adelaide 5005, Australia
[763] Sorbonne Universités, UPMC Université Paris 06, Université Paris Diderot, Sorbonne Paris Cité, CNRS, Laboratoire de Physique Nucléaire et de Hautes Energies (LPNHE), 4 place Jussieu, F-75252, Paris Cedex 5, France
[764] Laboratoire Univers et Particules de Montpellier, Université Montpellier, CNRS/IN2P3, CC 72, Place Eugène Bataillon, F-34095 Montpellier Cedex 5, France
[765] Université Bordeaux, CNRS/IN2P3, Centre d'Études Nucléaires de Bordeaux Gradignan, F-33175 Gradignan, France
[766] IRFU, CEA, Université Paris-Saclay, F-91191 Gif-sur-Yvette, France
[767] Astronomical Observatory, The University of Warsaw, Al. Ujazdowskie 4, 00-478 Warsaw, Poland
[768] Institut für Astronomie und Astrophysik, Universität Tübingen, Sand 1, D 72076 Tübingen, Germany
[769] School of Physics, University of the Witwatersrand, 1 Jan Smuts Avenue, Braamfontein, Johannesburg 2050, South Africa
[770] Oskar Klein Centre, Department of Physics, Stockholm University, Albanova University Center, SE-10691 Stockholm, Sweden
[771] APC, AstroParticule et Cosmologie, Université Paris Diderot, CNRS/IN2P3, CEA/Irfu, Observatoire de Paris, Sorbonne Paris Cité, 10, rue Alice Domon et Léonie Duquet, 75205 Paris Cedex 13, France
[772] Department of Physics and Astronomy, The University of Leicester, University Road, Leicester LE1 7RH, UK
[773] Nicolaus Copernicus Astronomical Center, Polish Academy of Sciences, ul. Bartycka 18, 00-716 Warsaw, Poland
[774] Institut für Physik und Astronomie, Universität Potsdam, Karl-Liebknecht-Strasse 24/25, D 14476 Potsdam, Germany
[775] Aix Marseille Université, CNRS/IN2P3, CPPM, Marseille, France
[776] Landessternwarte, Universität Heidelberg, Königstuhl, D 69117 Heidelberg, Germany
[777] Univ. Grenoble Alpes, CNRS, IPAG, F-38000 Grenoble, France
[778] Institut für Physik, Humboldt-Universität zu Berlin, Newtonstr. 15, D 12489 Berlin, Germany
[779] Obserwatorium Astronomiczne, Uniwersytet Jagielloński, ul. Orla 171, 30-244 Kraków, Poland
[780] Centre for Astronomy, Faculty of Physics, Astronomy and Informatics, Nicolaus Copernicus University, Grudziadzka 5, 87-100 Torun, Poland
[781] Japan Aerpspace Exploration Agency (JAXA), Institute of Space and Astronautical Science (ISAS), 3-1-1 Yoshinodai, Chuo-ku, Sagamihara, Kanagawa 229-8510, Japan
[782] Department of Physics, University of the Free State, PO Box 339, Bloemfontein 9300, South Africa
[783] Heisenberg Fellow (DFG), ITA Universität Heidelberg, Germany
[784] Yerevan Physics Institute, 2 Alikhanian Brothers Street, 375036 Yerevan, Armenia
[785] Astrophysics, Department of Physics, University of Oxford, Keble Road, Oxford OX1 3RH, UK
[786] Anton Pannekoek Institute, University of Amsterdam, Science Park 904, 1098 XH Amsterdam, The Netherlands
[787] Long Island University, New York, NY 11201, USA
[788] Virginia Tech, Blacksburg, VA 24061, USA
[789] Air Force Research Laboratory, NM 87117, USA
[790] University of New Mexico, Albuquerque, NM 87131, USA
[791] Long Island University, Brookville, NY 11548, USA
[792] Department of Physics and Astronomy, University of Utah, Salt Lake City, UT, USA
[793] Physics Division, Los Alamos National Laboratory, Los Alamos, NM, USA
[794] Instituto de Física, Universidad Nacional Autónoma de México, Ciudad de México, México
[795] Universidad Autónoma de Chiapas, Tuxtla Gutiérrez, Chiapas, México
[796] Universidad Michoacana de San Nicolas de Hidalgo, Morelia, Mexico
[797] Department of Physics, Michigan Technological University, Houghton, MI, USA
[798] Department of Physics & Astronomy, University of Rochester, Rochester, NY, USA
[799] Department of Physics, University of Maryland, College Park, MD, USA
[800] Instituto de Astronomía, Universidad Nacional Autónoma de México, Ciudad de México, México
[801] Department of Physics, University of Wisconsin-Madison, Madison, WI, USA
[802] Instituto Nacional de Astrofísica, Optica y Electrónica, Puebla, Mexico
[803] Instytut Fizyki Jadrowej im Henryka Niewodniczanskiego Polskiej Akademii Nauk, IFJ-PAN, Krakow, Poland
[804] Facultad de Ciencias Físico Matematicas, Benemérita Universidad Autónoma de Puebla, Puebla, Mexico
[805] Departamento de Física, Centro Universitario de Ciencias Exactase Ingenierias, Universidad de Guadalajara, Guadalajara, Mexico
[806] School of Physics, Astronomy, and Computational Sciences, George Mason University, Fairfax, VA, USA
[807] Instituto de Geofísica, Universidad Nacional Autónoma de México, Ciudad de México, México
[808] Max-Planck Institute for Nuclear Physics, 69117 Heidelberg, Germany
[809] Department of Physics and Astronomy, University of New Mexico, Albuquerque, NM, USA
[810] School of Physics and Center for Relativistic Astrophysics—Georgia Institute of Technology, Atlanta, GA, USA 30332
[811] Department of Physics and Astronomy, Michigan State University, East Lansing, MI, USA
[812] Universidad Politecnica de Pachuca, Pachuca, Hgo, Mexico
[813] Centro de Investigación en Computación, Instituto Politécnico Nacional, México City, México.
[814] Department of Physics, Pennsylvania State University, University Park, PA, USA
[815] Physics Department, Centro de Investigacion y de Estudios Avanzados del IPN, México City, DF, México
[816] Universidad Autónoma del Estado de Hidalgo, Pachuca, Mexico
[817] Instituto de Ciencias Nucleares, Universidad Nacional Autónoma de Mexico, Ciudad de México, México
[818] Santa Cruz Institute for Particle Physics, University of California, Santa Cruz, Santa Cruz, CA, USA
[819] Department of Physics and Astronomy, University of California, Irvine, Irvine, CA, USA

[820] Laboratório de Instrumentação e Física Experimental de Partículas – LIP and Instituto Superior Técnico – IST, Universidade de Lisboa – UL, Lisboa, Portugal
[821] Osservatorio Astrofisico di Torino (INAF), Torino, Italy
[822] INFN, Sezione di Torino, Torino, Italy
[823] Universidade de São Paulo, Instituto de Física, São Paulo, SP, Brazil
[824] University of Adelaide, Adelaide, SA, Australia
[825] Centro Atómico Bariloche and Instituto Balseiro (CNEA-UNCuyo-CONICET), San Carlos de Bariloche, Argentina
[826] Instituto de Tecnologías en Detección y Astropartículas (CNEA, CONICET, UNSAM), Buenos Aires, Argentina
[827] Universidad Tecnológica Nacional, Facultad Regional Buenos Aires, Buenos Aires, Argentina
[828] Universidad Nacional Autónoma de México, México, D.F., México
[829] Universidad de Santiago de Compostela, Santiago de Compostela, Spain
[830] Gran Sasso Science Institute (INFN), L'Aquila, Italy
[831] INFN Laboratori Nazionali del Gran Sasso, Assergi (L'Aquila), Italy
[832] Department of Physics and Astronomy, Lehman College, City University of New York, New York, NY, USA
[833] INFN, Sezione di Napoli, Napoli, Italy
[834] Institute of Space Science, Bucharest-Magurele, Romania
[835] Universidad Industrial de Santander, Bucaramanga, Colombia
[836] Observatorio Pierre Auger, Malargüe, Argentina
[837] Observatorio Pierre Auger and Comisión Nacional de Energía Atómica, Malargüe, Argentina
[838] University Politehnica of Bucharest, Bucharest, Romania
[839] "Horia Hulubei" National Institute for Physics and Nuclear Engineering, Bucharest-Magurele, Romania
[840] Università di Napoli "Federico II", Dipartimento di Fisica "Ettore Pancini," Napoli, Italy
[841] Laboratoire de Physique Subatomique et de Cosmologie (LPSC), Université Grenoble-Alpes, CNRS/IN2P3, Grenoble, France
[842] Università Torino, Dipartimento di Fisica, Torino, Italy
[843] Max-Planck-Institut für Radioastronomie, Bonn, Germany
[844] Institut de Physique Nucléaire d'Orsay (IPNO), Université Paris-Sud, Univ. Paris/Saclay, CNRS-IN2P3, Orsay, France
[845] Institute of Physics of the Czech Academy of Sciences, Prague, Czech Republic
[846] Università del Salento, Dipartimento di Matematica e Fisica "E. De Giorgi," Lecce, Italy
[847] INFN, Sezione di Lecce, Lecce, Italy
[848] Universidade Federal do Rio de Janeiro, Instituto de Física, Rio de Janeiro, RJ, Brazil
[849] Institute of Nuclear Physics PAN, Krakow, Poland
[850] Karlsruhe Institute of Technology, Institut für Kernphysik, Karlsruhe, Germany
[851] Colorado State University, Fort Collins, CO 80523
[852] RWTH Aachen University, III. Physikalisches Institut A, Aachen, Germany
[853] Karlsruhe Institute of Technology, Institut für Experimentelle Kernphysik (IEKP), Karlsruhe, Germany
[854] Universität Siegen, Fachbereich 7 Physik – Experimentelle Teilchenphysik, Siegen, Germany
[855] Universidad de Granada and C.A.F.P.E., Granada, Spain
[856] Università di Catania, Dipartimento di Fisica e Astronomia, Catania, Italy
[857] INFN, Sezione di Catania, Catania, Italy
[858] Università di Milano, Dipartimento di Fisica, Milano, Italy
[859] Universidade de São Paulo, Escola de Engenharia de Lorena, Lorena, SP, Brazil
[860] Universidad Michoacana de San Nicolás de Hidalgo, Morelia, Michoacán, México
[861] Universidade Estadual de Campinas, IFGW, Campinas, SP, Brazil
[862] Instituto de Tecnologías en Detección y Astropartículas (CNEA, CONICET, UNSAM), and Universidad Tecnológica Nacional – Facultad Regional Mendoza (CONICET/CNEA), Mendoza, Argentina
[863] Pennsylvania State University, University Park, PA, USA
[864] INFN, Sezione di Milano, Milano, Italy
[865] Politecnico di Milano, Dipartimento di Scienze e Tecnologie Aerospaziali, Milano, Italy
[866] Case Western Reserve University, Cleveland, OH, USA
[867] University of Chicago, Enrico Fermi Institute, Chicago, IL, USA
[868] Università del Salento, Dipartimento di Ingegneria, Lecce, Italy
[869] Instituto de Astronomía y Física del Espacio (IAFE, CONICET-UBA), Buenos Aires, Argentina
[870] Departamento de Física and Departamento de Ciencias de la Atmósfera y los Océanos, FCEyN, Universidad de Buenos Aires and CONICET, Buenos Aires, Argentina
[871] Universidade Federal Fluminense, EEIMVR, Volta Redonda, RJ, Brazil
[872] Universidade Federal do Rio de Janeiro (UFRJ), Observatório do Valongo, Rio de Janeiro, RJ, Brazil
[873] Universidade de São Paulo, Instituto de Física de São Carlos, São Carlos, SP, Brazil
[874] Universidade Federal do Paraná, Setor Palotina, Palotina, Brazil
[875] IFLP, Universidad Nacional de La Plata and CONICET, La Plata, Argentina
[876] Universität Hamburg, II. Institut für Theoretische Physik, Hamburg, Germany
[877] Fermi National Accelerator Laboratory, USA
[878] Stichting Astronomisch Onderzoek in Nederland (ASTRON), Dwingeloo, The Netherlands
[879] New York University, New York, NY, USA
[880] Karlsruhe Institute of Technology, Institut für Prozessdatenverarbeitung und Elektronik, Karlsruhe, Germany
[881] Michigan Technological University, Houghton, MI, USA
[882] Experimental Particle Physics Department, J. Stefan Institute, Ljubljana, Slovenia
[883] Instituto de Física de Rosario (IFIR) – CONICET/U.N.R. and Facultad de Ciencias Bioquímicas y Farmacéuticas U.N.R., Rosario, Argentina
[884] Laboratoire de Physique Nucléaire et de Hautes Energies (LPNHE), Universités Paris 6 et Paris 7, CNRS-IN2P3, Paris, France
[885] SUBATECH, École des Mines de Nantes, CNRS-IN2P3, Université de Nantes, France
[886] Centro Brasileiro de Pesquisas Fisicas, Rio de Janeiro, RJ, Brazil
[887] University of Łódź, Faculty of Astrophysics, Łódź, Poland
[888] University of Łódź, Faculty of High-Energy Astrophysics,Łódź, Poland
[889] Universidade Estadual de Feira de Santana, Feira de Santana, Brazil
[890] Palacky University, RCPTM, Olomouc, Czech Republic
[891] Colorado School of Mines, Golden, CO, USA
[892] Centro Federal de Educação Tecnológica Celso Suckow da Fonseca, Nova Friburgo, Brazil

[893] Universidade Federal do ABC, Santo André, SP, Brazil
[894] Benemérita Universidad Autónoma de Puebla, Puebla, México
[895] Université Libre de Bruxelles (ULB), Brussels, Belgium
[896] Centro de Investigación y de Estudios Avanzados del IPN (CINVESTAV), México, D.F., México
[897] Università di Roma "Tor Vergata," Dipartimento di Fisica, Roma, Italy
[898] INFN, Sezione di Roma "Tor Vergata", Roma, Italy
[899] Also at Universidade Federal de Alfenas, Brasília, Brazil
[900] Charles University, Faculty of Mathematics and Physics, Institute of Particle and Nuclear Physics, Prague, Czech Republic
[901] Centro de Investigaciones en Láseres y Aplicaciones, CITEDEF and CONICET, Villa Martelli, Argentina
[902] Università dell'Aquila, Dipartimento di Scienze Fisiche e Chimiche, L'Aquila, Italy
[903] KVI – Center for Advanced Radiation Technology, University of Groningen, Groningen, The Netherlands
[904] Also at Vrije Universiteit Brussels, Brussels, Belgium
[905] INAF – Istituto di Astrofisica Spaziale e Fisica Cosmica di Palermo, Palermo, Italy
[906] University of Nebraska, Lincoln, NE, USA
[907] Northeastern University, Boston, MA, USA
[908] School of Physics and Astronomy, University of Leeds, Leeds, UK
[909] Universitá di Catania, Dipartimento di Fisica e Astronomia, Catania, Italy
[910] Instituto de Astrofísica and Centro de Astroingeniería, Facultad de Física, Pontificia Universidad Católica de Chile, Casilla 306, Santiago 22, Chile
[911] Department of Particle Physics and Astrophysics, Weizmann Institute of Science, Rehovot 761000, Israel
[912] Millennium Institute of Astrophysics (MAS), Nuncio Monseñor Sótero Sanz 100, Providencia, Santiago, Chile
[913] European Southern Observatory, Alonso de Córdova 3107, Vitacura, Santiago 763-0355, Chile
[914] Joint ALMA Observatory, Alonso de Córdova 3107, Vitacura, Santiago 763-0355, Chile
[915] Department of Physics & Astronomy, Clemson University, Clemson, SC 29634, USA
[916] National Center for Radio Astrophysics, Pune 411007, India
[917] ARIES, Manora Peak, Nainital 263 001, India
[918] Astronomical Observatory Institute, Faculty of Physics, Adam Mickiewicz University, ul. Słoneczna 36, 60-286 Poznań, Poland
[919] Indian Institute of Space Science & Technology, Trivandrum 695547, India
[920] Joint Institute for VLBI ERIC (JIVE), 7991 PD Dwingeloo, The Netherlands
[921] Instituto de Astrofísica de Andalucía-CSIC, Granada, Spain
[922] Shanghai Astronomical Observatory (ShAO), Key Laboratory of Radio Astronomy, CAS, Shanghai 200030 China
[923] Guilin University of Electronic Technology (GUET), Guilin 541004, China
[924] JBCA, The University of Manchester, Manchester M13 9PL, UK
[925] Max Planck Institut für Radioastronomie, D-53121, Bonn, Germany
[926] Konkoly Observatory, MTA CSFK, H-1121 Budapest, Hungary
[927] SRON Netherlands Institute for Space Research, 3584 CA Utrecht, The Netherlands
[928] Leiden Observatory, Leiden University, 2300 RA Leiden, The Netherlands.
[929] Onsala Space Observatory, 439 92 Onsala, Sweden
[930] University of Warsaw, Faculty of Physics, 02-093 Warsaw, Poland
[931] Warsaw University of Technology, Institute of Electronic Systems, 00-665 Warsaw, Poland
[932] Center for Theoretical Physics, Polish Academy of Sciences, 02-668 Warsaw, Poland
[933] RIKEN, Wako, 351-0198 Saitama, Japan
[934] McGill Space Institute and Department of Physics, McGill University, 3600 rue University, Montreal, QC H3A 2T8, Canada
[935] Department of Applied Geology, Curtin University, GPO Box U1987, Perth, WA 6845, Australia
[936] Department of Mechanical Engineering, Curtin University, GPO Box U1987, Perth, WA 6845, Australia
[937] LIGO Laboratory West Bridge, California Institute of Technology, MC 100-36, Room 257, Pasadena, CA 91125
[938] Department of Physics, Harvard University, Cambridge, MA 02138, USA
[939] LSST, 950 N. Cherry Avenue, Tucson, AZ 85719, USA
[940] Max-Planck-Institut für Radioastronomie, Auf dem Hügel 69, D-53177 Bonn, Germany
[941] NASA Goddard Space Flight Center, 8800 Greenbelt Road, Greenbelt, MD 20771, USA
[942] Department of Astronomy, University of Maryland, College Park, MD 20742-4111, USA
[943] Inst. de Astrofísica de Canarias, E-38200 La Laguna, Tenerife, Spain
[944] Universidad de La Laguna, Dpto. Astrofísica, E-38206 La Laguna, Tenerife, Spain
[945] Space Telescope Science Institute, Baltimore MD, 21218
[946] Instituto de Astronomía, Universidad Nacional Autónoma de México, Apartado Postal 70-264, 04510 México, CDMX, Mexico
[947] INFN—Istituto di Radioastronomia, Via Gobetti 101, I-40129, Italy
[948] CSIRO Astronomy and Space Science, P.O. Box 76, Epping, NSW 1710, Australia
[949] Max-Planck-Institut für extraterrestrische Physik, Giessenbachstrasse, D-85748 Garching, Germany
[950] Department of Physics, University of Bath, Claverton Down, Bath BA2 7AY, UK
[951] Center for Theoretical Astrophysics, Los Alamos National Laboratory, Los Alamos, NM 87545, USA
[952] SKA South Africa, Pinelands, 7405, South Africa
[953] Department of Astronomy, Astrophysics, Cosmology and Gravity Centre, University of Cape Town, Private Bag X3 Rondebosch, 7701 South Africa
[954] Deceased 2017 February.
[955] Deceased 2016 December.
[956] Deceased 2017 August 18.
[957] Deceased 2016 August.